\documentclass[fleqn,usenatbib]{mnras}

\usepackage{newtxtext,newtxmath}

\usepackage[T1]{fontenc}

\DeclareRobustCommand{\VAN}[3]{#2}
\let\VANthebibliography\thebibliography
\def\thebibliography{\DeclareRobustCommand{\VAN}[3]{##3}\VANthebibliography}

\usepackage{graphicx}	
\usepackage{amsmath}	
 
\usepackage{amssymb}	
\usepackage{longtable}
\usepackage{float}

\title[Parsec-scale Properties of 8 FR 0s]{Parsec-scale Properties of eight Fanaroff–Riley Type 0 Radio Galaxies}

\author[Cheng et al.]{Xiaopeng Cheng,$^{1,2}$\thanks{E-mail: xcheng@kasi.re.kr (KASI)}
Tao An,$^{1,3}$\thanks{E-mail: antao@shao.ac.cn}
Bong Won Sohn,$^{2}$
Xiaoyu Hong,$^{1,3,4}$
Ailing Wang$^{1,4}$
\\
$^{1}$ Shanghai Astronomical Observatory, Chinese Academy of Sciences, Nandan Road 80, Shanghai 200030, China \\
$^{2}$ Korea Astronomy and Space Science Institute, 776 Daedeok-daero, Yuseong-gu, Daejeon 34055, Korea \\
$^{3}$ Key Laboratory of Radio Astronomy, Chinese Academy of Sciences, Nanjing 210008, China\\
$^{4}$ {University of Chinese Academy of Sciences, 19A Yuquanlu, Beijing 100049, China}
}

\date{Accepted XXX. Received YYY; in original form ZZZ}

\pubyear{2021}

\begin{document}
\label{firstpage}
\pagerange{\pageref{firstpage}--\pageref{lastpage}}
\maketitle

\begin{abstract}
We report the high-resolution radio observations of eight Fanaroff–Riley type 0 radio galaxies (FR 0s), selected from the published FR 0 sample. These observations were carried out with the Very Long Baseline Array (VLBA) and European VLBI Network (EVN) at frequencies of 5 and 8 GHz with a highest resolution of $\sim$0.6 milli-arcsec (mas).
All eight sources show compact structures on projected physical sizes of 0.3 - 10 parsec.
Six sources show a two-sided structure and two sources show a one-sided jet structure.
J1025+1022 shows an X-shaped jet structure, which could result from a reorientation of the jet axis due to a restart of the central engine or a projection of a highly curved inner jet, but more studies are needed to examine these scenarios.
Proper motions for 22 jet components of the eight sources are determined to be between $-0.08\, c$ and $0.51\, c$. 
Although most of the sources exhibit flat spectra, other observed characteristics, such as, low-amplitude flux density variations, low jet proper motion speeds, and symmetric two-sided jet structures, tend to support that the pc-scale FR~0 jets are mildly relativistic with lower bulk Lorentz factors and larger viewing angles.
\end{abstract}




\section{Introduction}

The first observations of radio galaxies were made in the 1950s \citep{1953Natur.172..996J}, and their radio emission structure is often far beyond the physical extent of their host galaxies \citep{1956ApJ...124..416B}.
The development of the radio interferometers has allowed us to identify radio sources with radio galaxies and quasars \citep[e.g.][]{1959MmRAS..68...37E} and to image the large-scale jet structures in detail \citep[e.g.][]{1968MNRAS.138..259M}.
Radio galaxies were first classified into two distinct populations, namely Fanaroff-Riley (FR) type I and II, according to their luminosity and large-scale morphology \citep{1974MNRAS.167P..31F}. 
FR Is have relatively prominent inner jets and hotspots but diffuse emission at their edges (edge-darken) \citep[e.g., 3C 31, ][]{2008MNRAS.386..657L}.
On the contrary, FR II are more powerful and exhibit two symmetric edge-brightening lobes spanning a scale of several hundred kiloparsecs \citep[][]{1996A&ARv...7....1C}.
Powerful radio galaxies have two primary evolutionary tracks, corresponding to the formation of high-jet-power sources (FR IIs) and low-jet-power sources (FR Is). Each radio source starts as a Compact Symmetric Object (CSO), and grows into a Medium-sized Symmetric Object (MSO), and then a fraction of MSOs may successfully evolve into Large Symmetric Objects (LSOs) including FR II and I; the end product, either FR II or I, depends on the initial jet power, the duration of nuclear activity, re-currency of the radio activity, and the interactions between the jet and the interstellar medium \citep{2010MNRAS.408.2261K,2012ApJ...760...77A}.
This FR classification reveals a morphological sequence that is related to the ability of a jet to transport the momentum and energy to a large distance \citep[][]{2012ApJ...760...77A}.

Recent radio survey observations have found that most detected sources are associated with low-redshift early-type galaxies, and their typical observational characteristic in radio band is the lack of large-scale extended structure  \citep{2012MNRAS.421.1569B,2014MNRAS.438..796S}. In general, their Faint Images of the Radio Sky at Twenty-Centimeters (FIRST) made with the Very Large Array \citep[VLA;][]{1995ApJ...450..559B} show unresolved images with a resolution of 5\arcsec\ (corresponding to a projected linear size of $\lesssim$5 kpc) \citep[][]{2018A&A...609A...1B}.
The deficit of extended radio emission and high core dominance of these sources distinguish them from the classical FR I and FR II galaxies. 
These sources seem to a stand-alone class, and therefore named as a separate class, FR type 0 radio galaxies (FR 0s) \citep{2011AIPC.1381..180G}. 
FR 0s have similar host galaxy and central engine properties with FR Is, but with lower radio power and higher core dominance. 
FR 0s are the dominant radio source population in the local Universe, and the number density of FR 0s is 5 times higher than that of FR Is at $z < 0.05$. Therefore FR 0s are important for understanding the overall nature of extragalactic radio sources. 
Currently, the study of FR 0s is very limited to the available observations in the radio band, and in particular, high resolution radio observations are lacking.
The physical nature of the absence of extended emission in FR 0s and the relation to other AGN classes (e.g., FR classes, blazars, radio-quiet AGN) are still subjects of debate \citep[][]{2014MNRAS.438..796S, 2015A&A...576A..38B,2018A&A...609A...1B}.

Recently, a larger sample of 18,286 sources \citep{2012MNRAS.421.1569B} was cross-matched with the Sloan Digital Sky Survey \citep[SDSS, ][]{2000AJ....120.1579Y,2009ApJS..182..543A}, NRAO VLA Sky Surve \citep[NVSS, ][]{1998AJ....115.1693C}, and FIRST \citep[][]{1995ApJ...450..559B}
 datasets with 104 FR 0s identified \citep[FR0CAT,][]{2018A&A...609A...1B}.
The comparison between the FR 0 and FR I galaxies suggests that FR 0 jets may have lower bulk Lorentz factors than FR I jets, and that 
FR 0s have shorter active periods that may be  recurrent \citep{2018A&A...609A...1B}.  However, FR 0s are not just a down-sized version of FR Is. The difference in the observational characteristics between these two classes might be dominated by an intrinsic physical mechanism that reduces the production ability of extended radio emission in FR 0s. 
The low luminosity and high number density suggest that FR 0s may represent short-lived AGN activity and the short duty cycles are not long enough for a galaxy to develop large-scale radio jets \citep{2016AN....337..105S,2018A&A...609A...1B,2019Galax...7...76B}.
\citet{2020A&A...642A.107C} explored the recurrent behaviour of FR 0s based on the observations made by the Low-Frequency ARray \citep[LOFAR, ][]{2013A&A...556A...2V}  at 150 MHz. They identified 66 FR 0s from the available LOFAR data, and investigated their low-frequency radio properties. Of the 66 FR 0s, 54 are unresolved with sizes of $\lesssim$3--6 kpc, while the remaining twelve sources are in the range of 15--40 kpc.
The spectral slopes between 150 MHz and 1.4 GHz show that 20 per cent of their FR 0 sample sources have steep spectra ($\alpha$\footnote{Spectral index $\alpha$ is defined as $\rm S_{\nu}$ $\propto$ $\rm \nu^{\alpha}$ .}  $<-0.5$).
These observations suggest that FR 0s and FR Is may represent the two extremes of the jetted source population with a continuous distribution of size and radio luminosity, while FR 0s lie at the lower end of the size and radio power \citep{2020A&A...642A.107C}.

In addition, \citet{2019MNRAS.482.2294B} randomly selected 18 FR 0s from the FR0CAT and made new observations with the Karl G. Jansky VLA at 1.5, 4.5 and 7.5 GHz with a highest resolution of $\sim0.3\arcsec$. Most of these sources remain unresolved, suggesting a compact radio structure on the sub-kpc scales and confirming the general lack of extended radio emission and high core dominance of FR 0s.
The size distribution favors the possibility that FR 0s can not be all young radio galaxies that will eventually become FR I/II \citep{2019MNRAS.482.2294B}.

We have studied the radio properties of 14 FR 0s on pc scales from very long baseline interferometry (VLBI) imaging observations \citep{2018ApJ...863..155C}. 
The diverse distribution of low brightness temperatures, slow structure and flux density variations, compact structures, steep spectra and moderate jet speeds implies that the VLBI-detected FR 0s might be  mixed with GHz-peaked spectrum (GPS) and compact steep-spectrum (CSS) sources \citep{2018ApJ...863..155C}. 
\citet{2014MNRAS.438..796S} also found that FR 0s is a heterogeneous population, in addition to the genuine FR 0s, this population is also mixed with GPS,  CSS as well as a minority of beamed sources.
Young radio galaxies, GPS and CSS radio galaxies (see recent review in \citet[][]{2021A&ARv..29....3O}), must be present in FR 0s \citep{2014MNRAS.438..796S}, but their proportion is not yet well constrained. 
GPS and CSS galaxies represent an early stage in the evolution of extended radio sources \citep{2012ApJ...760...77A} and carry important information about the onset of nuclear activity.
The overall radio properties of the VLBI-detected FR 0s \citep{2018ApJ...863..155C} are consistent with what expected for young radio sources, but there are still some open questions to be addressed.
Since young radio galaxies have extremely compact radio structures, only VLBI imaging observations are able to explore the pc-scale sources structure.

In addition to revealing the pc-scale morphology, VLBI observations can determine the jet/hotspot kinematics and spectral properties providing additional important information for discerning the nature of FR 0s.
Following our previous results \citep{2018ApJ...863..155C}, we selected and observed eight FR 0s, which have only one epoch VLBI data, with the Very Long Baseline Array (VLBA) and European VLBI Network (EVN) at 5 and 8 GHz, to study the detailed jet structure, spectral distribution and jet kinematic properties on parsec scales.
Of particular interest is the  determination of the advanced speed of the hotspot or jet knots, which can be diagnostic of whether the radio structure is still growing.

In this paper, we present radio images of eight FR 0s at 5 and 8 GHz and report the spectral index distribution and jet proper motions. 
Section \ref{2} describes the sample selection and observations. 
The observational results are presented in Section \ref{3} and discussed in Section \ref{4}. 
Section \ref{5} gives a summary. 
Throughout the paper, we assume a cosmological model with $\rm H_{0}$ = 73 km\,s$^{-1}$\,Mpc$^{-1}$, $\Omega_{\rm M} = 0.27$, and $\Omega_{\Lambda} = 0.73$. 
At a redshift of $z = 0.01$, an angular size of 1 mas corresponds to a  projected linear size of 0.196 pc, and the conversion from angular velocity to projected linear speed is 1 mas yr$^{-1}$ = $0.65\, c$.

\section{The Sample and Observations}
\label{2} 

\subsection{Sample selection}
\label{sec:sample} 

Cross-matching of the FR 0s catalogue \citep{2018A&A...609A...1B} with the mJy Imaging VLBA Exploration at 20 cm \citep[mJIVE-20][]{2014AJ....147...14D} catalogue and Astrogeo database \footnote{VLBA calibrator survey data base is maintained by Leonid Petrov, \url{http://astrogeo.org/}.} identified 14 FR 0s \citep {2018ApJ...863..155C}. 
Among the 14 sources: (i) three sources have only compact and unresolved cores at 8.4 GHz; (ii) three sources have multi-epoch data with extended jets of 1--6 mas in length, for which we have obtained the jet proper motions and source spectral indices in \citet{2018ApJ...863..155C}; (iii) the remaining eight sources have either core-jet structure or core-double-hotspot structure, but only one epoch of VLBI data. 
Therefore, we selected the remaining eight sources, whose main properties are listed in Table \ref{general properties}, and observed with the VLBA and EVN, so that we have a total of 11 sources for our jet proper motion analysis.
We should note that the FR~0 sources selected for this paper is actually the fraction with the highest radio flux density. As we have discussed in \citet{2018ApJ...863..155C}, the remaining weaker sources in the FR0CAT may not have similar radio power or morphological characteristics to these 14 sources, and thus the conclusions of this study may not be universally applicable to the entire FR~0 sample. This would require a systematic deep VLBI imaging of a large sample to help clarify the physical nature of the radio emission of FR 0s on pc scales.

Six of the selected sources show a two-sided jet structure. 
Thus, another motivation of our observations is to identify the core position of the sample and to classify the radio source structure.

\subsection{Observations}

We obtained 10.5 h of observations with the EVN and VLBA, respectively.
Based on the same motivation, the schedule of the EVN and VLBA observations are almost identical.
We observed the eight sources in total flux density mode at 5 and 8 GHz. 
To ensure the success of the observations and to optimise the $uv$ coverage of each source, we split these sources into two groups. 
The flux densities of these sources are not variable on weeks time scale, so the observations were scheduled to be performed within one week. 
Each source was observed three to four separate scans, with an effective on-source time of approximately 30 min.

The VLBA observations are summarized in Table \ref{observations log} and were carried out from 2018 August 4 to August 12 (project codes of BC241A to BC241D; PI: X.-P. Cheng). 
All 10 VLBA antennas were used for BC241C. The other three sessions used 8 or 9 antennas due to the maintenance of some antennas. 
The observed data were recorded with two consecutive 128 MHz channels, sampled in 2 bits, at an aggregate data rate of 2 giga bits per second (Gbps). 
We used 3C~345 and 4C~39.25 as the finger searching calibrators. 
The raw data were correlated at the Array Operations Centre in Socorro (New Mexico, U.S.) using the DiFX software correlator \citep{2007PASP..119..318D,2011PASP..123..275D} with an averaging time of 2 s, 256 frequency channels per IF, and uniform weighting.

The EVN observations are summarized in Table \ref{observations log} and were carried out  at 5 and 8 GHz frequency bands from 2019 February 24 to March 7 (project codes of EC063A to EC063D; PI: X.-P. Cheng). 
Fifteen antennas were used with 2-bit sampling and an aggregate data rate of 1 Gbps (dual-polarization, 8 consecutive 16 MHz frequency channels). Twelve antennas were used at 8 GHz with the same data rate. 
We used 3C~345 and 4C~39.25 as the calibration sources. 
The correlation was performed at the Joint Institute for VLBI ERIC (JIVE), Dwingeloo, the Netherlands, using the EVN software correlator \citep[SFXC, ][]{2015ExA....39..259K} .

The correlated VLBA and EVN data were transferred to the computer clusters at the China SKA Regional Centre \citep{2019NatAs...3.1030A}, where the post-processing including calibration and imaging analysis was performed.

\subsection{Data Reduction and Model Fitting}
\label{2.3}

The data were calibrated with the U.S. National Radio Astronomy Observatory (NRAO) Astronomical Imaging Processing Software (\textsc{AIPS}) package \citep{2003ASSL..285..109G} following the standard procedure. 
Firstly, we imported the data using the task FITLD and inspected the data quality.
The voltage offsets of the samplers were corrected by auto-correlation. 
A-priori amplitude calibration was performed using the information from  the gain curve (GC), system temperature (TY), and weather (WX) tables  for all antennas to correct for the atmospheric opacity. 
The ionospheric delay was corrected using total electron content measurements monitored by the global positioning system (GPS). 
The phase contributions from the antenna parallactic angles were removed before the subsequent phase corrections were applied. 
The station Pie Town (PT) and Effelsberg (EF) were chosen as the reference antennas during the VLBA and EVN data calibration process. 
The fringe-fitting and bandpass calibration were performed with 3C~345 and 4C~39.25. 
Finally, the data in each sub-band were averaged and exported into single-source files. 
To remove residual amplitude and phase errors, we carried out the hydrib mapping \citep{1984ARA&A..22...97P} in \textsc{Difmap} software package \citep{1997ASPC..125...77S}. Several iterations of self-calibration and imaging were carried out and the final images were made with natural weighting. 
We found that the overall antennas gain correction factors determined were small, typically within 10 per cent. 

A number of circular Gaussian components were fitted to model the brightness distribution of each source in the visibility domain in \textsc{Difmap}. 
The fitted parameters of the gaussian models are presented in Table \ref{modelfit}. 
Typical flux density uncertainties are less than 10 per cent and are mainly contributed by the visibility amplitude calibration errors and the antenna gain calibration errors. 
The uncertainty in the fitted component size is typically less than 15 per cent of the deconvolved size of the fitted Gaussian model. 
We followed \citet{2009AJ....138.1874L} to estimate the errors in the best-fit positions of the Gaussian jet components: \textit{i.e.}, $\sim$20 per cent of the component size convolved with the synthesised beam size.
Since the analysis of the proper motions of the 8 sources contains data obtained from our own EVN and VLBA, as well as the archived VLBA data from the Astrogeo database, to reduce errors due to difference in data quality (in practice, the data quality difference is not much),  we used the same model fitting uncertainties for all the data.

\section{Results}
\label{3}

\subsection{Radio Morphology on Parsec-scale}

In the first and second columns of Fig.~\ref{fig:VLBI}, we show the naturally weighted total intensity EVN images of the 8 sources at 5 and 8 GHz.
The typical image size displayed is 15 mas $\times$ 15 mas and the dimension of the restoring beam in each image is given in Table \ref{image result}.
The VLBA images show similar morphology to the EVN images, so we only present them in Appendix \ref{A}.
The elliptical Gaussian restoring beam is indicated in the bottom-left corner of each image in Figures~\ref{fig:VLBI} and \ref{image:VLBA}. 
The parameters associated with the images are listed in Table ~\ref{image result}. 
The quasi-simultaneous dual-frequency VLBI observations allow estimation of the spectral index (listed in the last column of Table 3), which we calculated using the integrated flux densities on the VLBI images. We found that seven sources show a flat spectrum ($-0.5 < \alpha < 0.5$), with only J1037+4335 exhibiting a slightly steeper spectrum. J1213+5044 shows a change of the spectral index from $\alpha = 0.19$ in 2018 August to $\alpha = -0.61$ in 2019 February, which may be related to a variation in the opacity in the core region due to a change in the emission structure \ref{3.4}).

Six sources (J0906+4124, J0909+1928, J1025+1022, J1205+2031, J1213+5044 and J1230+4700) display a two-sided structure on parsec scales, in agreement with the previous results \citep{2018ApJ...863..155C}.
Two sources (J1037+4335 and J1246+1153) exhibit a faint one-sided jet structure. In our new observations of J1037+4335, we do not detect the SW component, but detect three new jet components (J1 -- J3), which suggest a core-jet structure extending to the northwest (NW) direction.
J1246+1153 shows a core-jet structure with the jet extending to the NW direction.

The identification of the core is crucial for kinematic studies, as it provides the stationary reference point for all epochs.
In our sample, all sources have a prominent core, which greatly simplifies the identification of the reference component.
With the high-resolution data at two frequencies, we can easily get the spectral index of individual components to help the core identification which corresponds to a flat-spectrum component.
In most cases, the identification of jet features across epochs proves to be relatively straightforward due to the slow evolution of the source structure with time.
We also used an additional check on the continuity of flux density and size to confirm the cross-identification. 
To avoid misidentification of the component and incorrect estimates of apparent jet speeds, we discuss the detailed characteristics of the radio morphology and components identification for each individual source in Appendix \ref{3.4}.

\subsection{Jet Proper Motion}
\label{3.2}

In order to quantitatively study the jet kinematics, we used a non-acceleration, two-dimensional vector fit to the jet components position with time, using the core as a reference \citep{2016AJ....152...12L}. 
Combined with our previous studies \citep{2018ApJ...863..155C}, data from three epochs over a maximum baseline of $\sim14$ years are available for the jet proper motion measurement. 
One sources (J1037+4335) was unresolved in the previous observations , so its proper motion is determined only based on two epochs of 2018 and 2019. 
The other seven sources have data from three epochs that allow a linear regression fitting of the jet proper motions.
By model-fitting to our VLBI data, we were able to identify 22 jet components of 8 sources across the different epochs.

Column 4 in Fig.~\ref{fig:VLBI} shows the radial distance of the jet component versus the observing time. 
The slope of the linear regression fitting to the temporal change of jet distance gives the proper motion speeds. 
The corresponding values of the angular speed, apparent speed and observing frequencies are presented in Table \ref{proper motion}.

We note that 13 jet components in six sources (J0906+4124, J1037+4335, J1205+2031, J1213+5044, J1230+4700 and J1246+1153) basically move along a constant radial direction. 
Since the jet components move away from the core, their flux densities tend to decrease in general and their sizes increase. 
In J0909+2918, we find that the north component (N1) moves along a constant direction, but the south jet components (S1 and S2) exhibit non-radial motion as they bend at 5 mas. 
During our observation period, S1 shows little motion and little change in position angle. 
The trajectory of S2 appears a slight eastward bending and a change in the position angle from 140$\degr$ to 155$\degr$. 
In J1025+1022, W3 follows a straight trajectory toward the west direction to the core, but the position angles of W2 and E2 vary significantly from $\pm$90$\degr$ to 130$\degr$ along the NW-SE direction. 
However, each of these three jet components seems to move in ballistic trajectories.

Combining with proper motion results of three FR 0s reported in \citet{2018ApJ...863..155C}, we obtained jet proper motions of a total of 24 jet components in 11 FR 0 sources to study the jet kinematics. 
The derived jet speeds range from $-0.08\, c$ to $0.51\, c$, suggesting stationary motion to mildly relativistic speeds. The only exception of jet proper motion speed exceeding $1,c$ is associated with the J1 component in J1037+4335, which has only two epochs data separated by $<$7 months. Follow-up VLBI monitoring of this source would place strict constraint on its jet proper motion.

\subsection{Spectral Indices}
\label{3.3}

These sources present diverse distribution of spectral indices from $\rm \alpha \sim -0.72$  to  0.29:  seven sources (J0906+4124, J0909+1928, J1025+1022, J1205+2031, J1213+5044, J1230+4700 and J1246+1153) show flat spectrum and one source (J1037+4335 ) show steep spectrum ($\alpha < -0.5$).

Column 3 of Fig.~\ref{fig:VLBI} shows the spectral index maps for all sources between 5 and 8 GHz. 
We also created the spectral index maps using the VIMAP program \citep{2014JKAS...47..195K} from the 5 and 8 GHz EVN data by aligning the maps on the core positions. 
The 8-GHz images were created with the same pixel size (0.05 mas) and restoring beam size as the 5-GHz maps. 
First, we excluded the core region with an elliptical mask approximately twice the size of the restoring beam. 
Then, VIMAP calculated the two-dimensional  cross-correlation product to determine the shift between the two images. 
Due to the relatively closeness of the two frequencies, the typical shift is about 0.05 mas due to the frequency-dependent opacity effect. 
After correcting this relative offset, we computed the spectral index maps and superimposed them (in coloured scale) on the 5 GHz total intensity contours.  
The spectral index maps confirm the flat-spectrum core as the brightest component 
and reveal twin-jet structure for 6 sources and one-sided jet structure for 2 sources. 
The outer jet region has a steeper spectrum with a spectral index of $\alpha < -0.5$, typical for optically-thin AGN jets \citep{2014AJ....147..143H}.

\section{Discussion}
\label{4}

Several low frequency observations have confirmed the absence of extended radio emission from FR 0s, which is the main difference with other FR classification sources  \citep{2019MNRAS.482.2294B,2020A&A...642A.107C}.
Therefore, the physical properties of FR 0s and their possible relation to large-scale FR Is depend on high-resolution observations on pc scales.
Our previous study has shown that FR 0s detected with VLBI could not be a homogeneous population of radio sources \citep{2018ApJ...863..155C}.
Thanks to the new EVN and VLBA data, we obtain the spectral index information for 12 FR 0s and jet kinematic properties for 11 FR 0 sources.
The VLBI data allow us to study the physical properties of FR 0s at the onset phase of their long-term evolution.

The cores other than J1230+4700 occupy more than 50 per cent of the integrated VLBI flux density, and the core dominance is not directly related to the radio structure: that is, sources with a one-sided jet do not show a higher core flux density fraction than those with two-sided jets, although the sample in this paper is too small to draw statistically significant conclusions yet. Although most of the cores exhibit flat radio spectra and core dominance, it is still not sufficient to conclude that these cores have strong Doppler boosting: most jets exhibit two-sided symmetric structures, indicating rather large viewing angles; the  variability of the cores is considerably low, below 30 per cent for all radio cores except for J1213+5044 which has a  variability of 40 per cent at 8 GHz; as mentioned earlier, FR~0 sources have lower jet bulk Lorentz factors than FR Is. These characteristics distinguish these FR 0s from quasars which contain highly relativistic jets and prominent variability. The observed flat radio spectrum may be due to the fact that the nuclear region is still active, with a constant injection of fresh relativistic electrons. This also suggests that FR~0s (at least the brightest fraction in radio) are still growing. The number density of FR~0s in local Universe is much higher than that of FR~Is \citep{2018A&A...609A...1B}, implying that most FR~0s cannot grow into extended sources with sizes of a few hundred kilo-parsec to Mega-parsec.
The successful evolution of an FR~0 galaxy into an extended source depends not only on the intrinsic physical properties of the central engines (the duration and persistence/intermittency of the nuclear activity, initial jet power), but also on the environmental factors such as the density gradient of the ambient medium in the host galaxy, jet-ISM interactions along the jet path and within the terminal lobes (see \citet{2012ApJ...760...77A} and references therein).

Interestingly, J1025+1022 shows an X-shaped jet on pc scales. This relatively large jet misalignment ($\rm \sim 40 \degr$) has been observed in other nearby radio galaxies, e.g., 3C~84 \citep{2018NatAs...2..472G} and Mrk~231 \citep{2021MNRAS.tmp..609W}. If the X-shaped jet of J1025+1022 is associated with re-started AGN activity, then the change in the jet direction could be the result of a change in the direction of the spin axis of the black hole \citep[e.g.,][]{1978Natur.276..588E,2002MNRAS.330..609D}.
The continued evolution depends mainly on the continuity of the nuclear activity and the persistence of the jet \citep{2012ApJ...760...77A}. Or the jet misalignment could simply be a projection of a helically twisted jet. 
Detailed follow-up studies can help to clarify these physical pictures.

Four sources (J0909+1928, J1205+2031, J1230+4700 and J1246+1153) show  flat spectra. But their jet velocities, ranging from $\sim0.04\,c$ to about $0.4\,c$, are small compared with the typical flat-spectrum quasars \citep{2016AJ....152...12L}.
From our previous study, 4 other sources (J0910+1841, J0943+3614, J1604+1744, J1606+1814) also show flat spectra but small jet speed \citep{2018ApJ...863..155C}.
This suggests that the bulk Lorentz factor $\Gamma$ of the jet is low for most of the VLBI-detected FR 0s.
It is important to note that the VLBI sample we used in this study is at the high flux density end of the overall FR~0 population; since the jets of these sources are found to be mildly relativistic (lower Lorentz factor), it is reasonable to believe that the jets of the majority of FR~0 sources have an even lower Lorentz factor. This is consistent with the conclusion obtained by other groups through other methods \citep[e.g.,][]{2014MNRAS.438..796S,2018A&A...609A...1B,2020A&A...642A.107C}.
These sources do not show any stationary hotspots at the ends of their jets, but the lack of terminal hotspots cannot be used to rule out the possibility that FR 0s represent short-lived episodes of AGN activity that do not last long enough for a galaxy to develop large-scale radio jets.
If this scenario is correct, the non-detection of extended relic emission of past radio activity indicates a rapid radiative or adiabatic loss leading to a shorter lifetime of the large-scale lobes.

There is no obvious trend of increasing/decreasing apparent component speed with increasing core distance in these 8 sources, indicating no  clear acceleration or deceleration in the pc-scale radio jets.
The mildly relativistic jets and the lack of extended radio emission indicate relatively lower magnetic field strength on the pc scales and lower radiative efficiency in the vicinity of the SMBHs.
The low jet speeds are more likely a consequence of a lower BH mass and spin than that in FR I/IIs \citep{2018A&A...609A...1B,2019ApJ...871..259G}.
Therefore, the extended jets should be dissipative and lie below the surface-brightness threshold of existing large-scale radio surveys \citep{2016AN....337..105S}.
Follow-up studies with lower-frequency observations \citep[e.g.,][]{2020A&A...642A.107C} should be performed on larger samples and should also be expanded to lower radio power FR 0s.

We must point out that in most of our FR 0 sources, neither evidence of deceleration along the jet nor double-double morphology (manifesting re-current AGN activity) has been observed.
The present results cannot distinguish between these two possibilities: short-lived recurrent sources and low jet-speed sources associated with  low BH mass and spin.
Obviously, we can draw a conclusion that the jets are accelerated in the sub-parsec regions from the central engine and remain mildly relativistic on pc scales.
Since the value of the BH parameter in FR 0s is probably lower than that of FR I/IIs, the deceleration of the jet flow may occur at large distances ($>$10 pc), which is beyond the region that the present VLBI images can show. EVN + eMERLIN observations could provide intermediate resolutions to reveal the radio emission on tens of pc scales \citep{2018MNRAS.476.3478B}.

\section{Summary}
\label{5}

We have observed 8 FR 0s with the VLBA and EVN at 5 and 8 GHz, and studied the inner jet morphology, spectral indices and proper motions. 
All eight sources show compact structures.
We identified a total of 33 jet components from these sources.
The focus of this paper is on jet kinematics and spectral property studies.
Combining these results with our earlier VLBI measurements \citep{2018ApJ...863..155C}, we find that:

\begin{trivlist}

\item{1.} Six sources (J0906+4124, J0909+1928, J1025+1022, J1205+2031, J1213+5044 and J1230+4700) show two-sided structure and two sources (J1037+4335 and J1246+1153) exhibit a one-sided jet structure.
We first detect three new jet components in J1037+4335, which suggests a core-jet structure.
The X-shaped jetted source J1025+1022 is supposed to be a restarted source.

\item{2.} Proper motions of 22 jet components in eight sources are determined, between $-0.08\, c$ and $0.51\, c$. Most of the sources have a two-sided jet structure indicating a relatively large viewing angle of the jet; the low-level variability and low jet proper motion speeds exclude strong beaming effect and suggest sub-relativistic or mildly relativistic jet flow. These observations tend to support low bulk Lorentz factors for the FR~0 jets.

\item{3.} The spectral indices for these sources are distributed over a relatively wide range from $-0.72$ to 0.29. 
Most of these sources have flat spectra, indicating that the nuclear region is in an active state with a continuous injection of fresh relativistic particles. The spectrum of J1037+4335 is slightly steeper. The spectral index of J1213+5044 has changed significantly between 2018 and 2019. No correlation is found between the spectral index and the radio structure as well as the core dominance. The sample in this paper is relatively small and the analysis of these correlations requires more data.
We created spectral index maps for all eight sources, revealing the spectral index distribution of the core and jet.

\end{trivlist}

We expect to further study the jet kinematics at high spatial resolution in a larger FR 0 sample to enrich our understanding of the radio source nature of FR~0s and the possible evolutionary relation to other radio galaxy classes.
Upcoming next-generation radio facilities, such as the Square Kilometre Array (SKA) and the next generation Very Large Array \citep[ngVLA,][]{2018ASPC..517....3M}, 
will provide a huge increase in sensitivity \citep{2015aska.confE.173K}, up to several orders of magnitude, spanning the whole radio frequency band, and offer a promising opportunity to discover a large population of FR~0s.
Although black-hole spins are difficult to measure observationally, the large-scale environment is associated with BH spins and host galaxy dynamics, which would reflect the intrinsic differences between FR~0s and FR~Is and shed light on the low-power radio galaxy evolution.

\section*{Acknowledgements}
We thank the anonymous referee for his/her constructive comments.
This work was supported by the SKA pre-research funding granted by the National Key R\&D Programme of China (2018YFA0404602, 2018YFA0404603) and the Chinese Academy of Sciences (CAS, 114231KYSB20170003).
X.-P. Cheng and B.-W. Sohn were supported by Brain Pool Program through the National Research Foundation of Korea (NRF) funded by the Ministry of Science and ICT (2019H1D3A1A01102564). 
The European VLBI Network (EVN) is a joint facility of independent European, African, Asian, and North American radio astronomy institutes. Scientific results from data presented in this publication are derived from the following EVN project code: EC063. 
The VLBA observations were sponsored by Shanghai Astronomical Observatory through an MoU with the NRAO (Project code: BC241). 
The Very Long Baseline Array is a facility of the National Science Foundation operated under cooperative agreement by Associated Universities, Inc.

\section*{Data Availability}
The correlation data used in this article are available in the EVN data archive (\url{http://www.jive.nl/select-experiment}) and VLBA data archive (\url{https://archive.nrao.edu/archive/archiveproject.jsp}). The calibrated visibility data underlying this article can be requested from the corresponding authors.




\bibliographystyle{mnras}
\bibliography{mnras} 




\begin{figure*}
\centering
 \includegraphics[height=3.8cm]{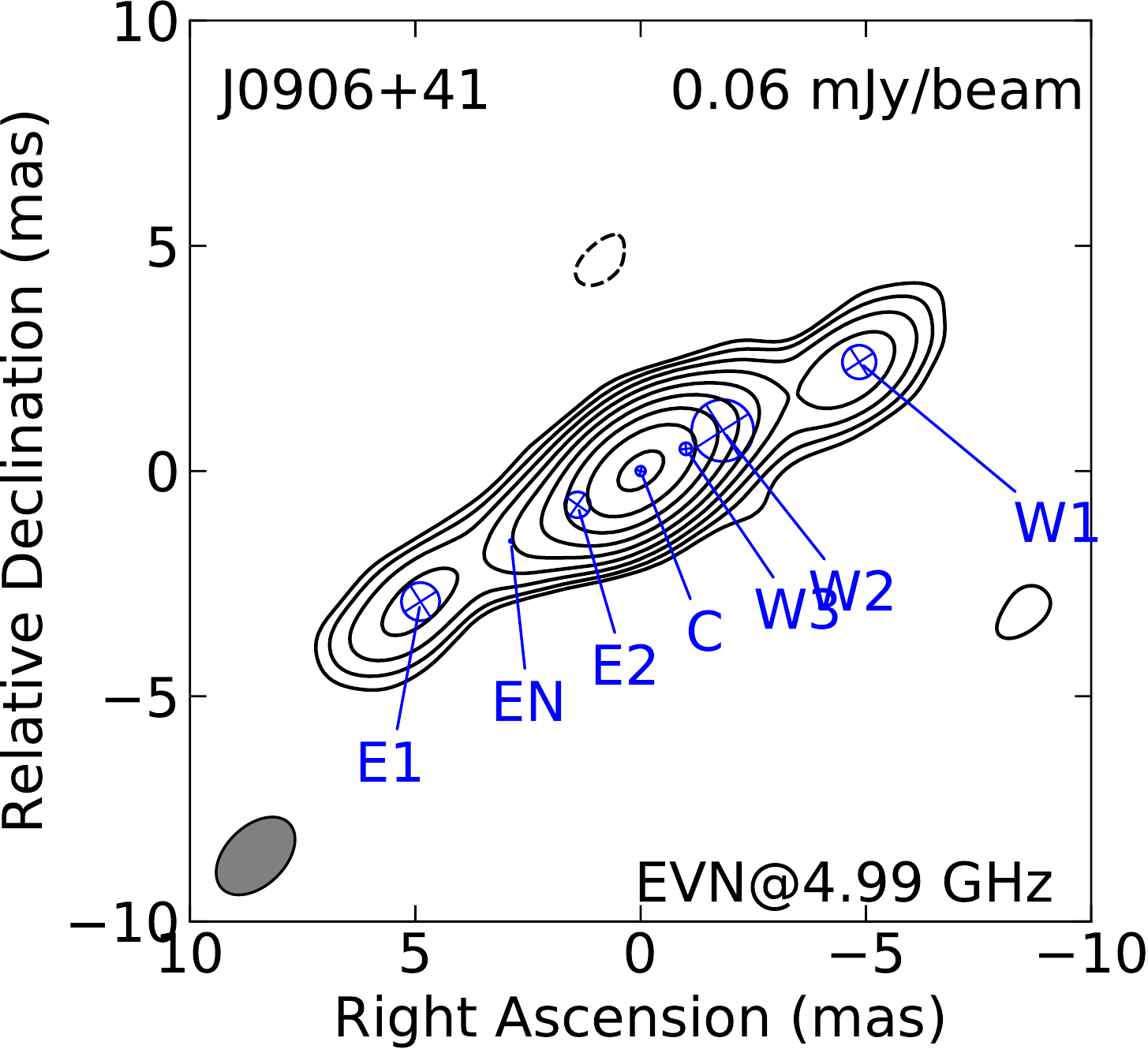}
 \includegraphics[height=3.8cm]{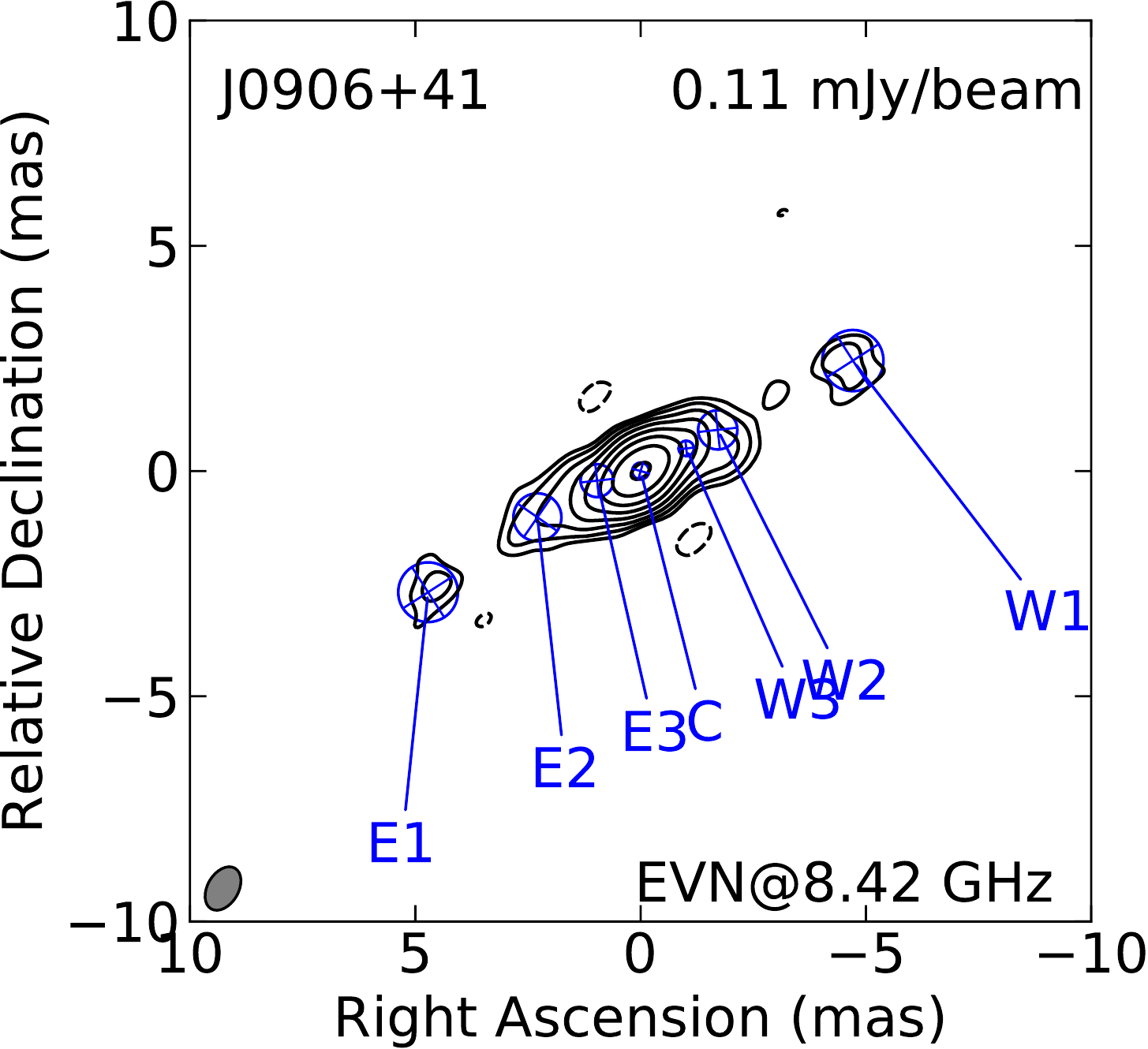}
 \includegraphics[height=3.8cm]{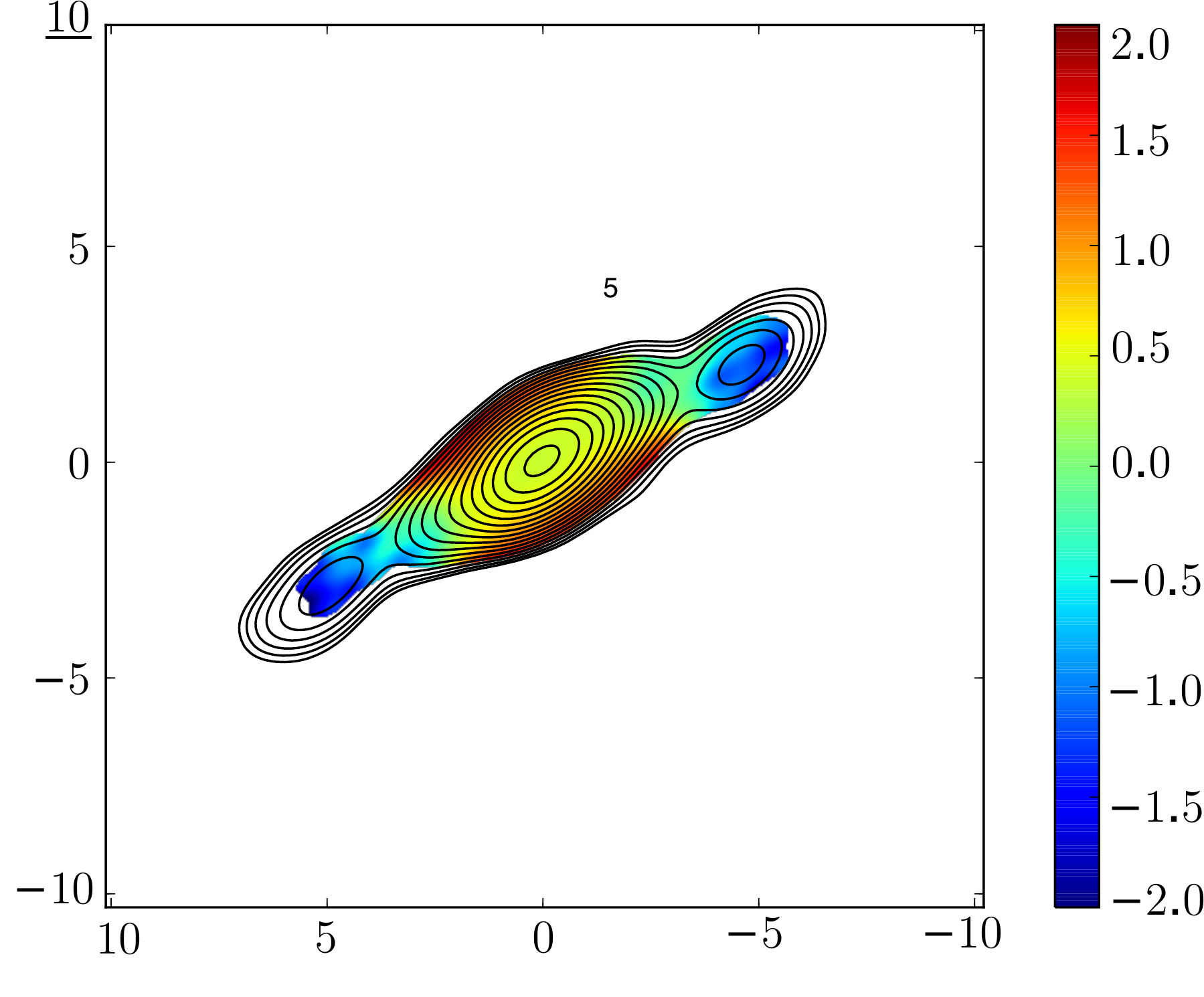}
 \includegraphics[height=3.8cm]{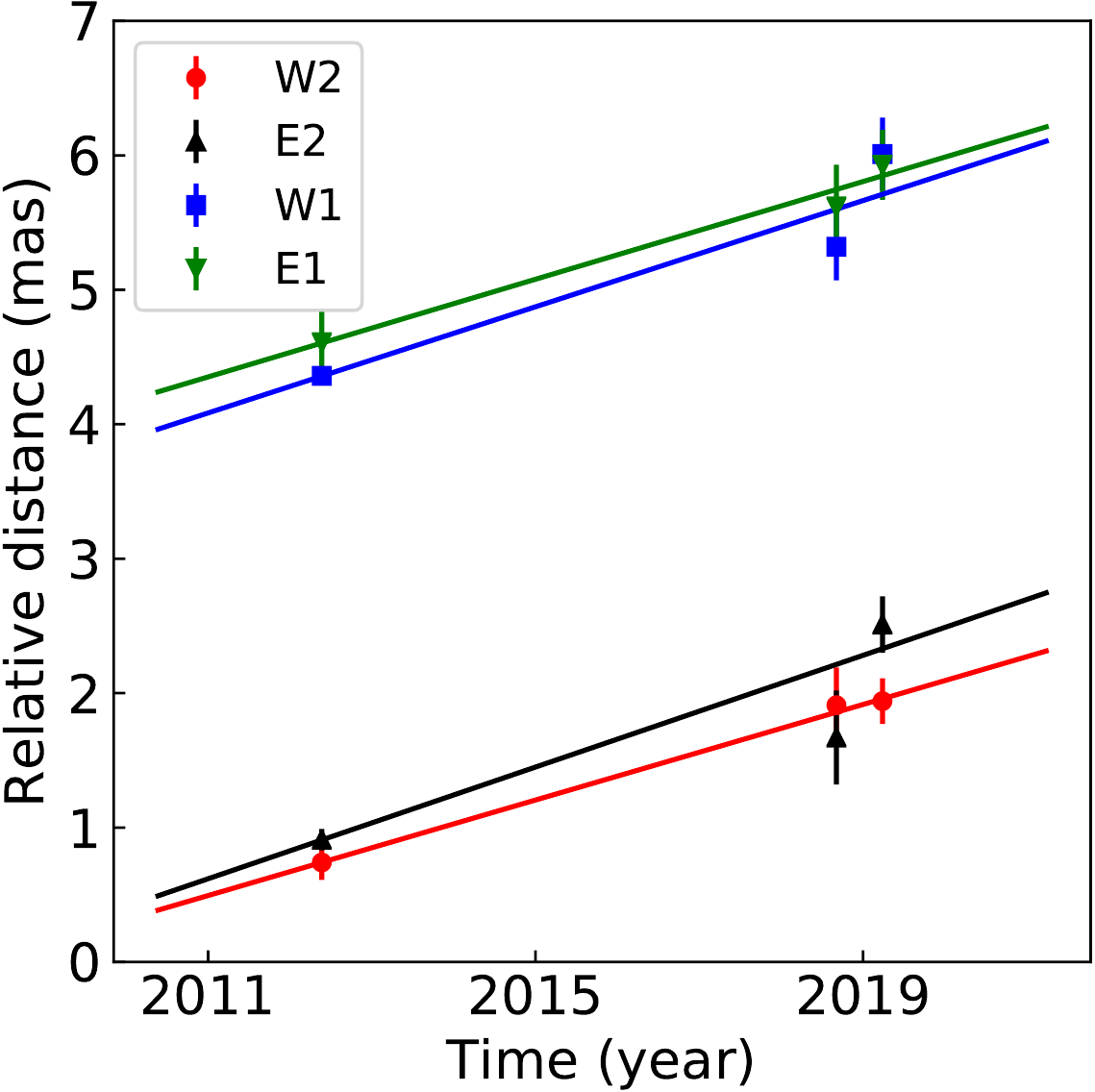}
\\
 \includegraphics[height=3.8cm]{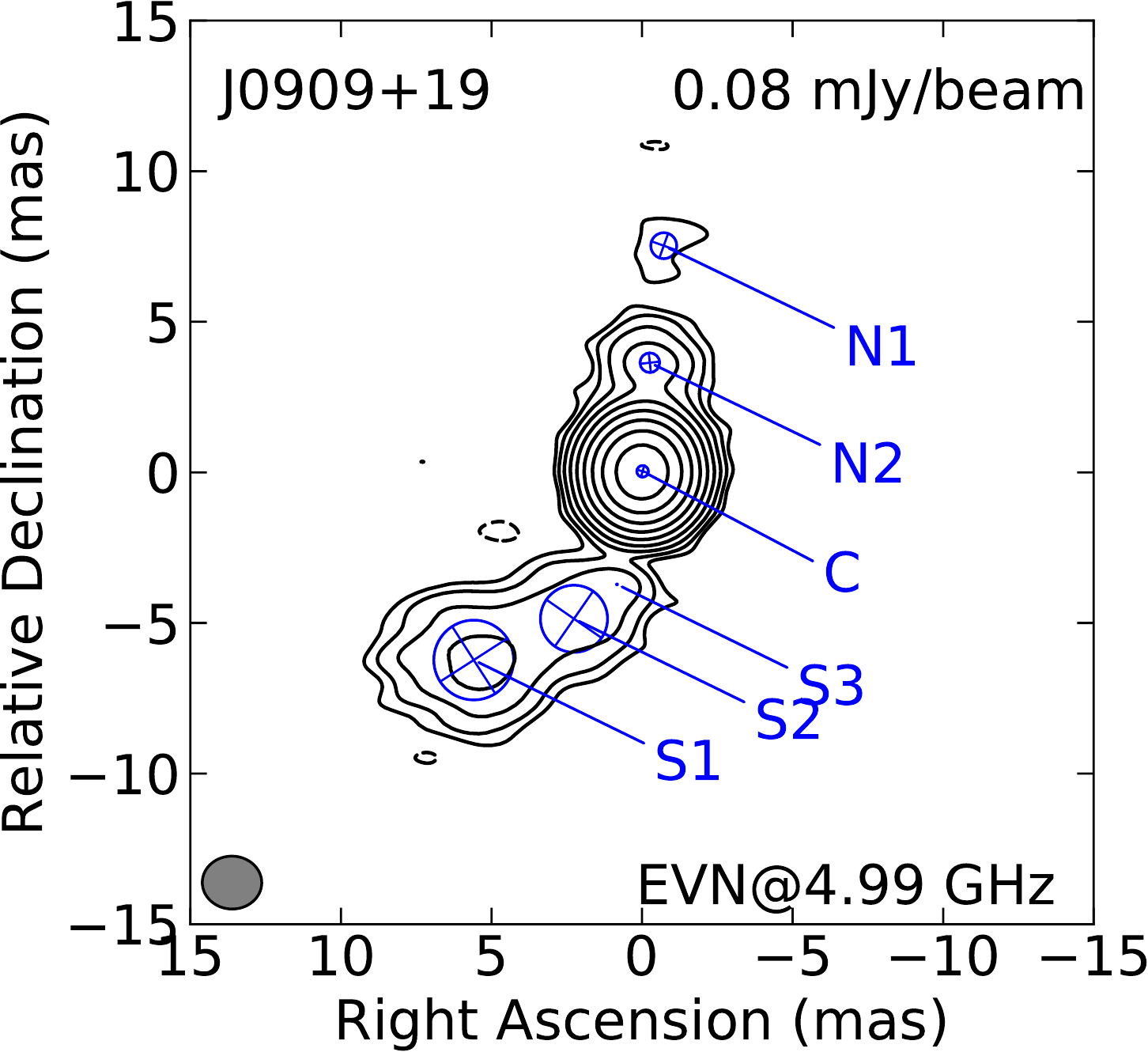}
 \includegraphics[height=3.8cm]{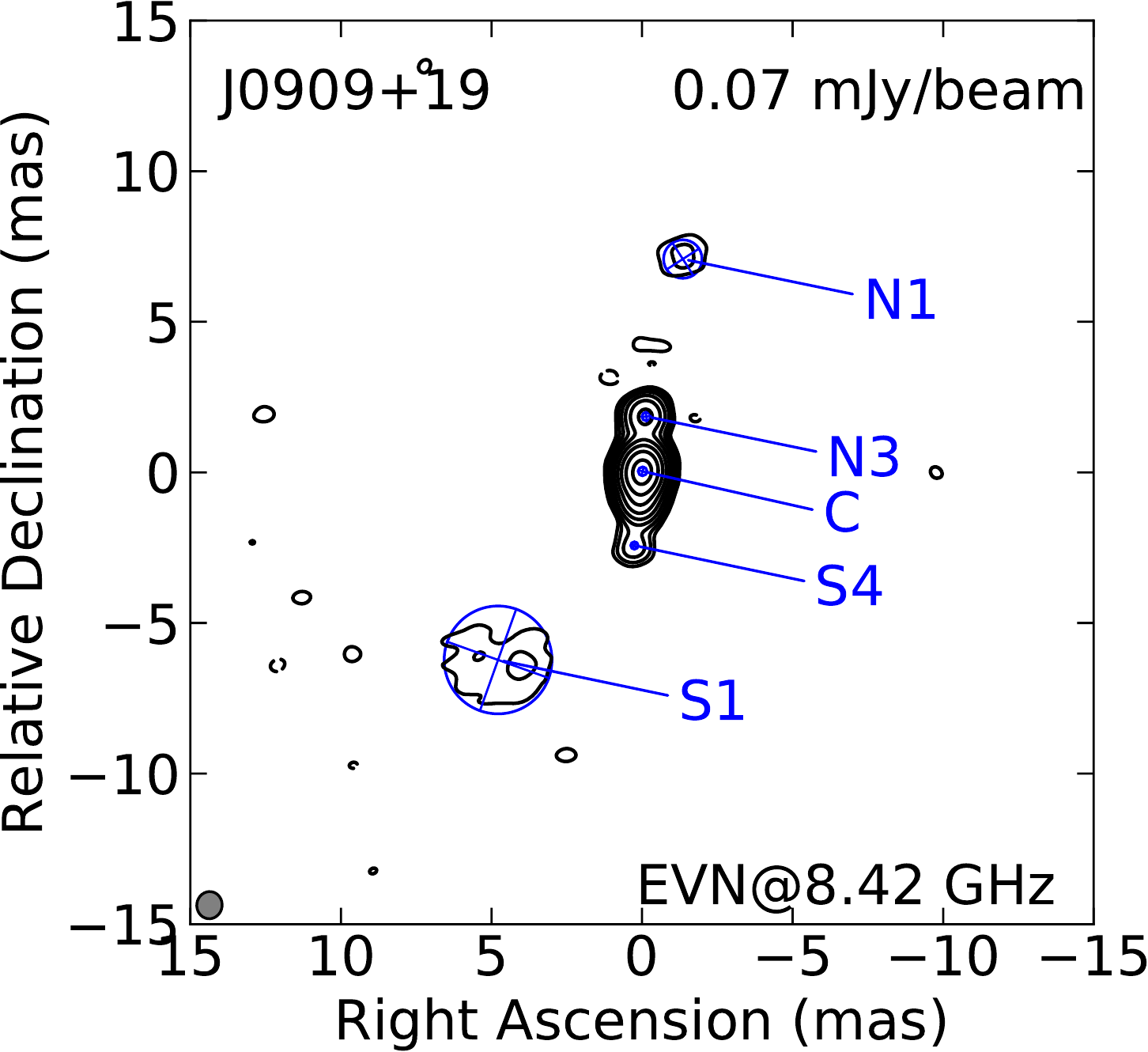}
 \includegraphics[height=3.8cm]{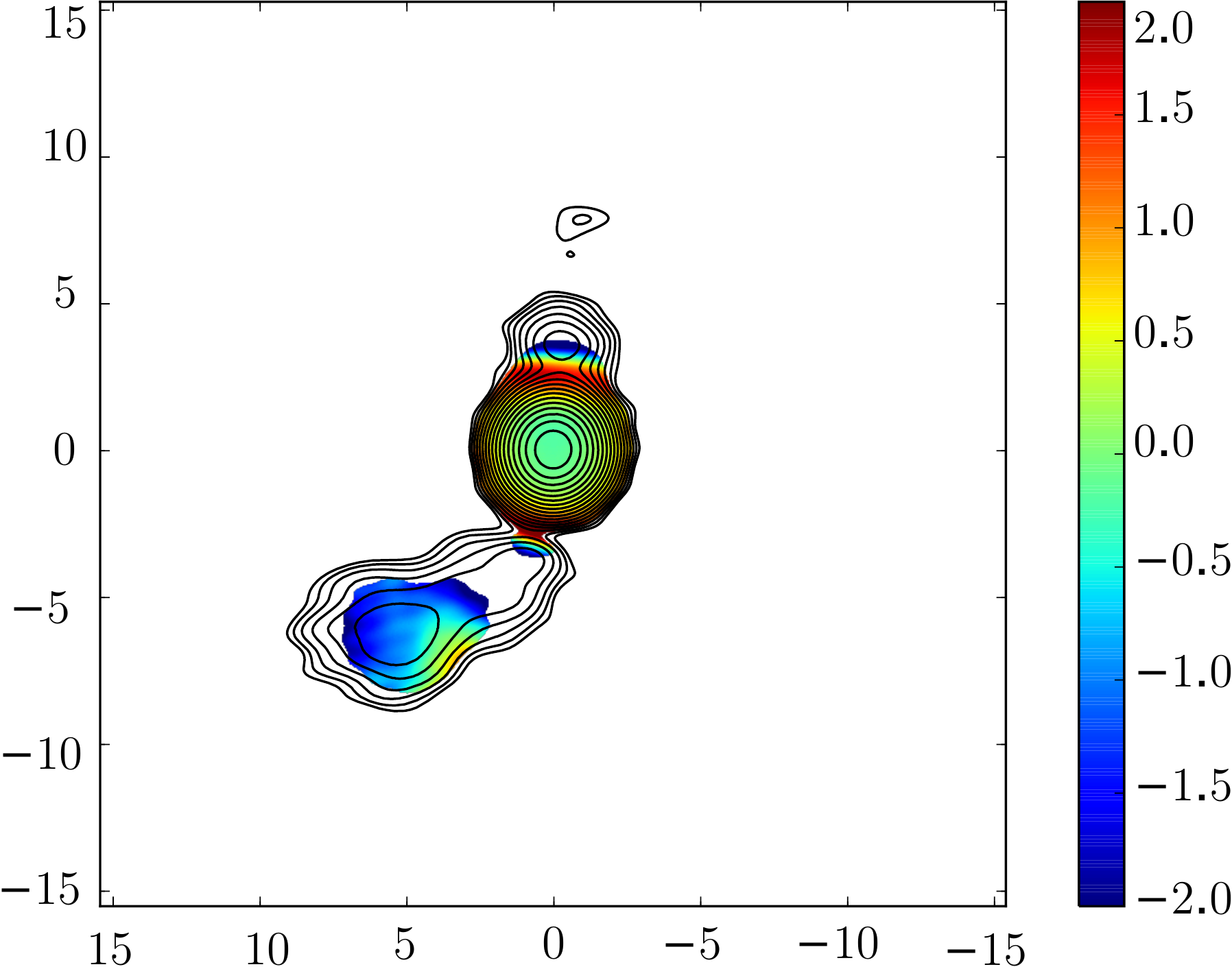}
 \includegraphics[height=3.8cm]{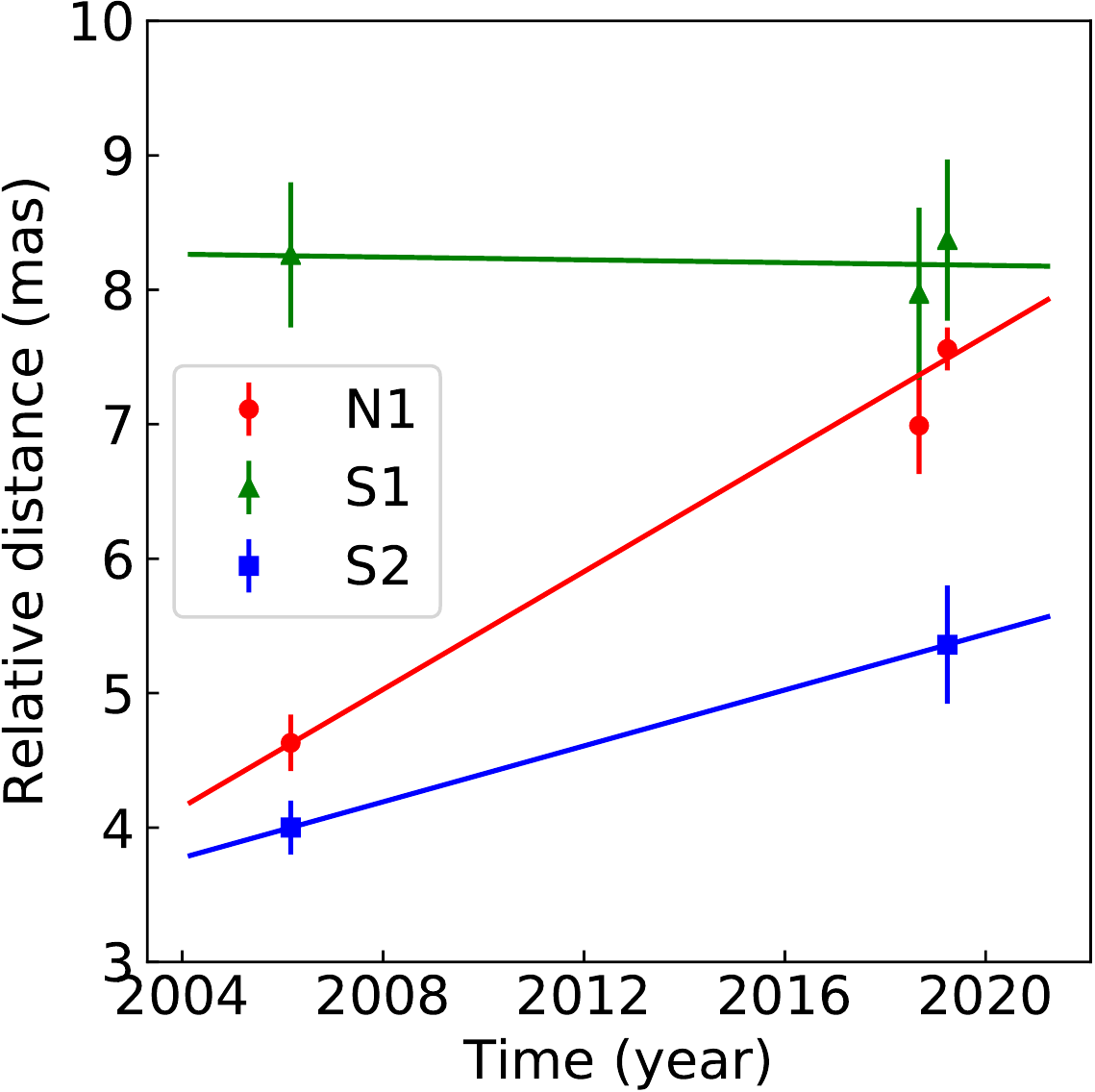}
\\
 \includegraphics[height=3.8cm]{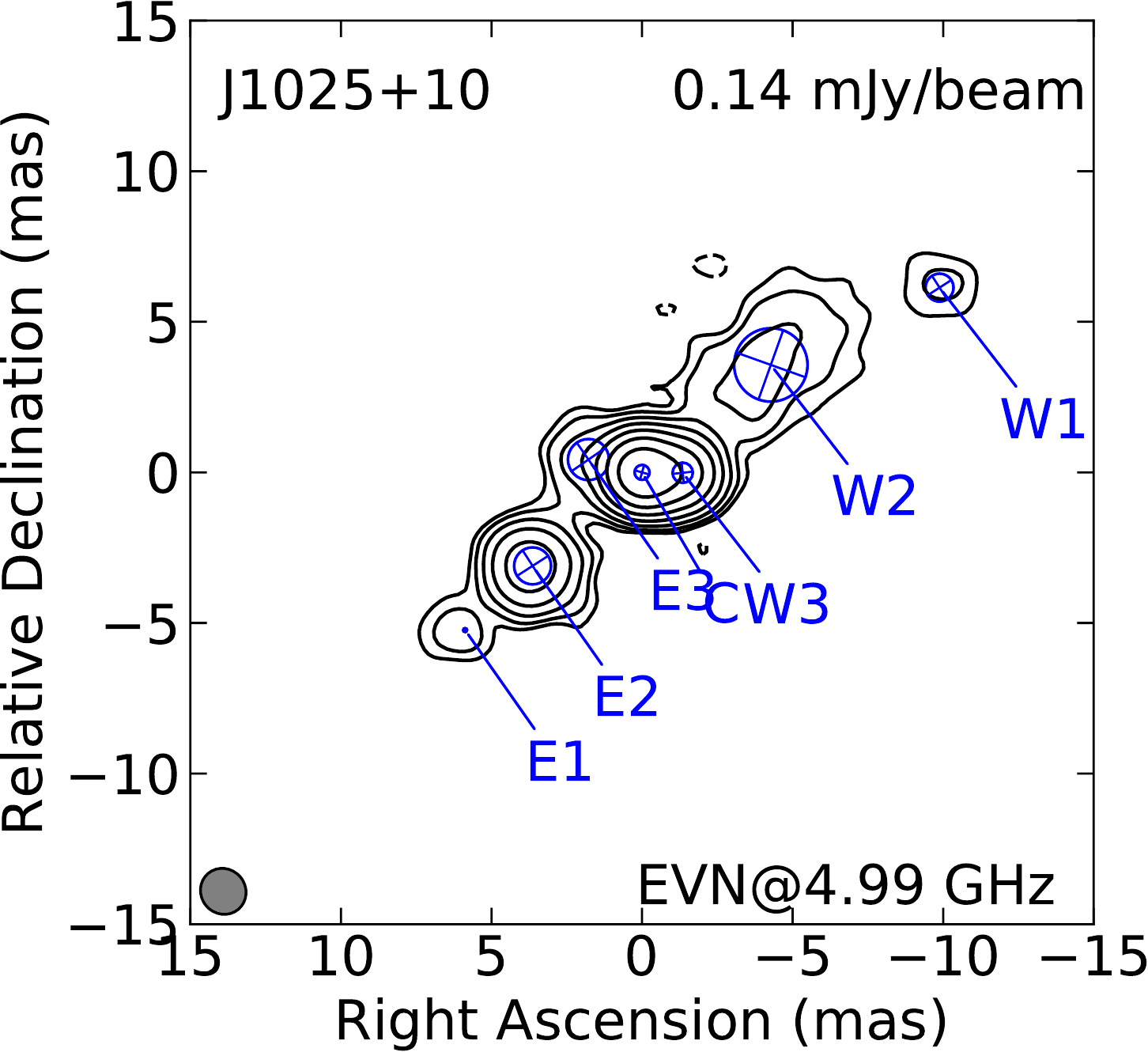}
 \includegraphics[height=3.8cm]{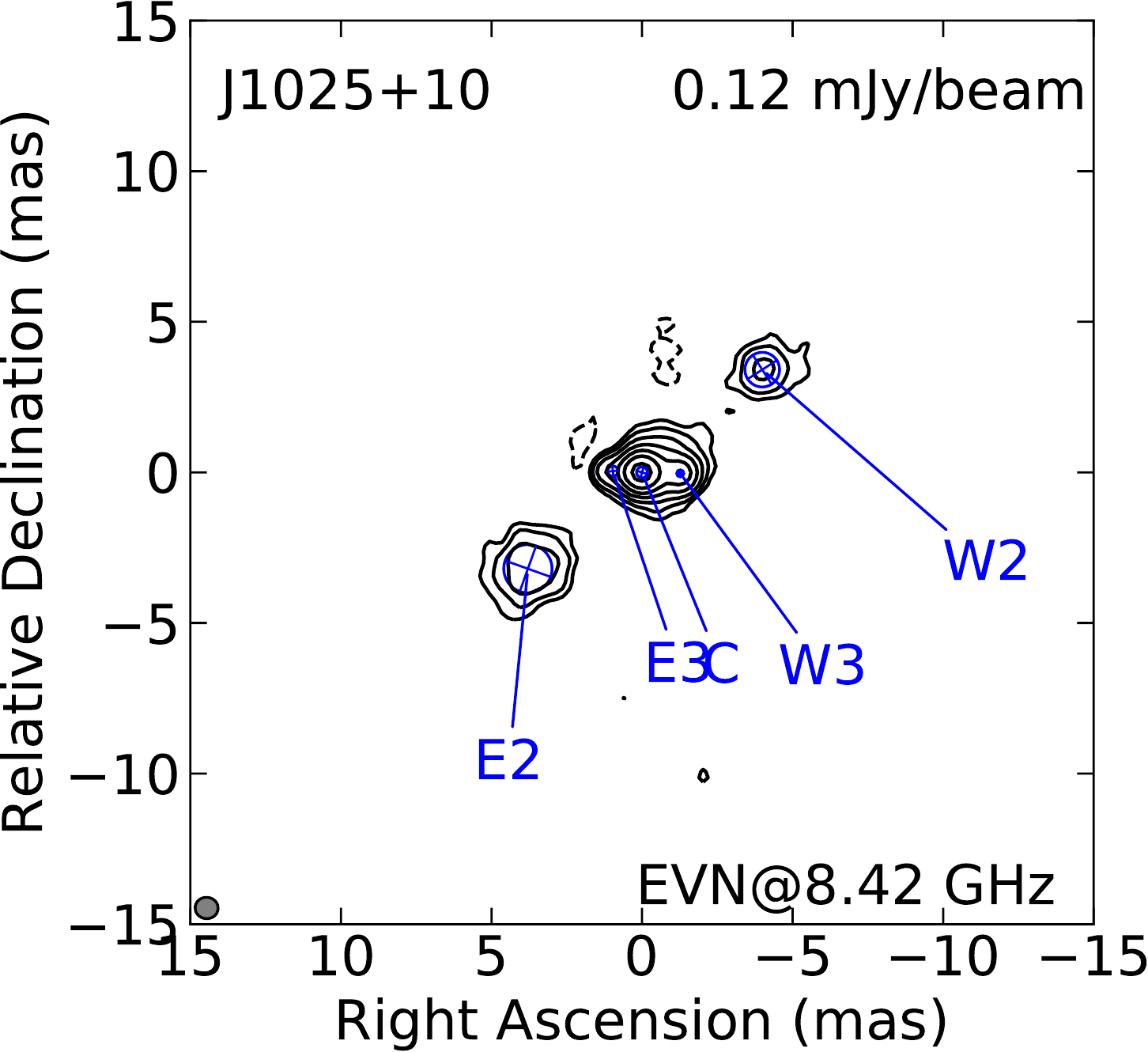}
 \includegraphics[height=3.8cm]{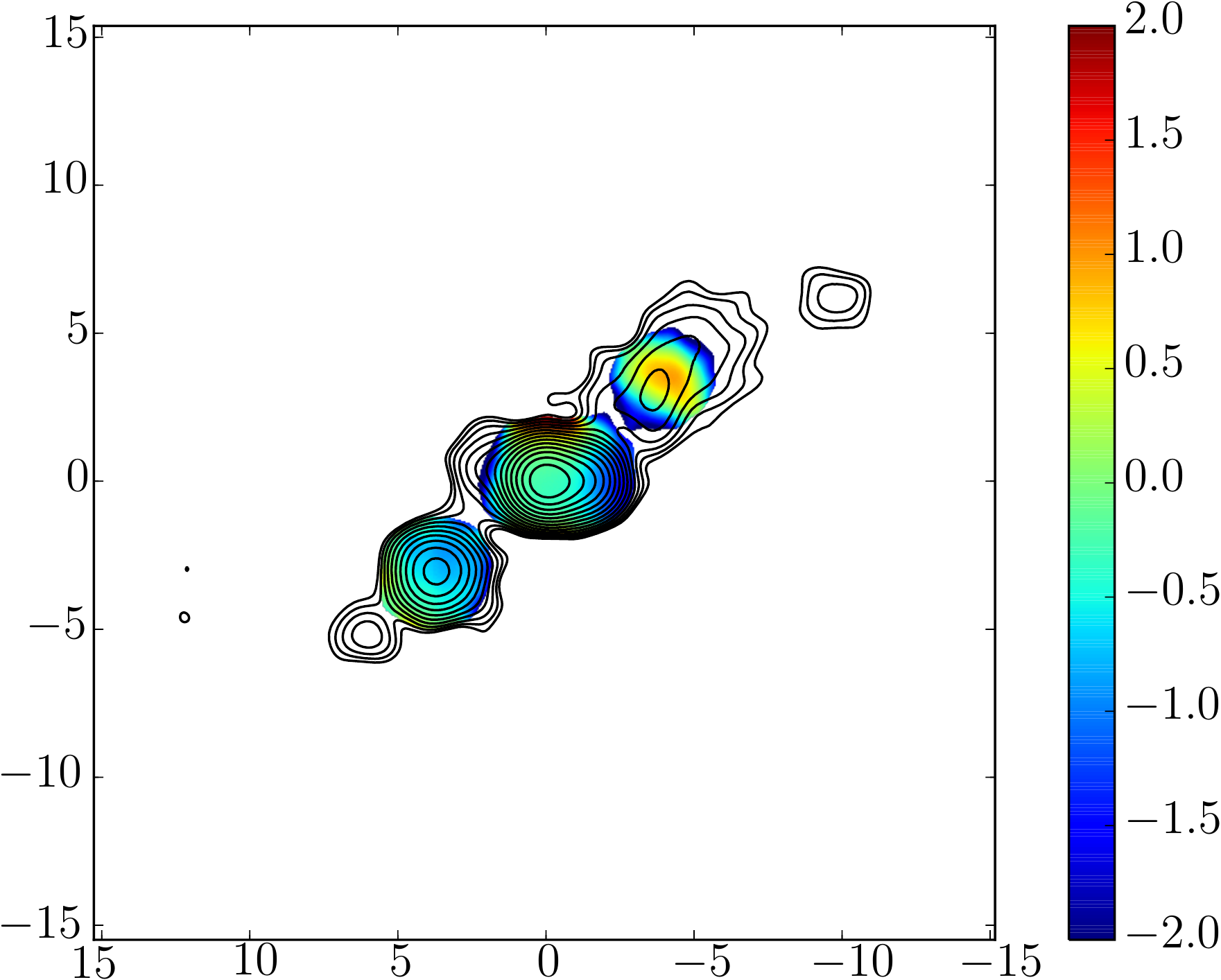}
 \includegraphics[height=3.8cm]{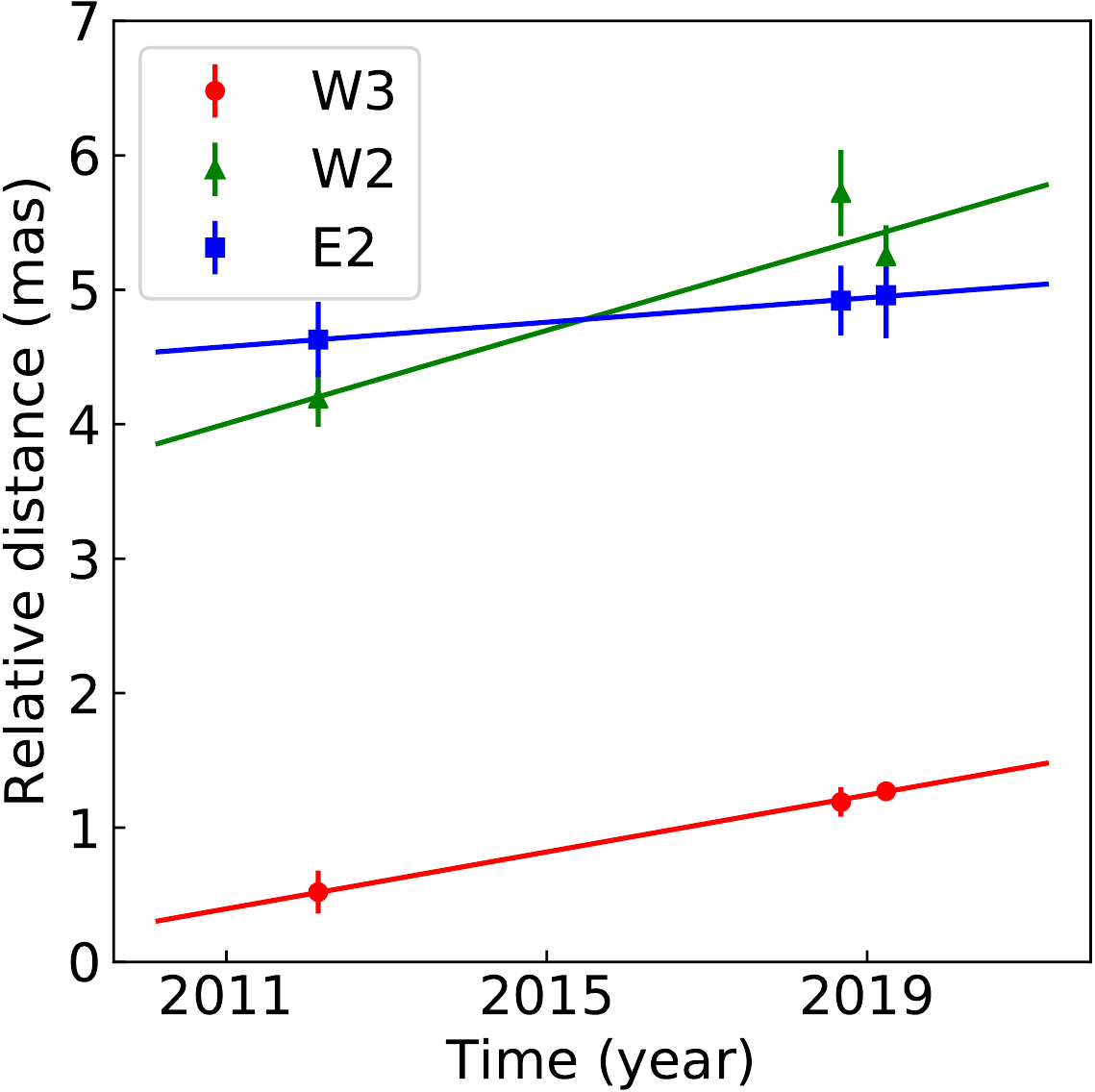} 
 \\
 \includegraphics[height=3.8cm]{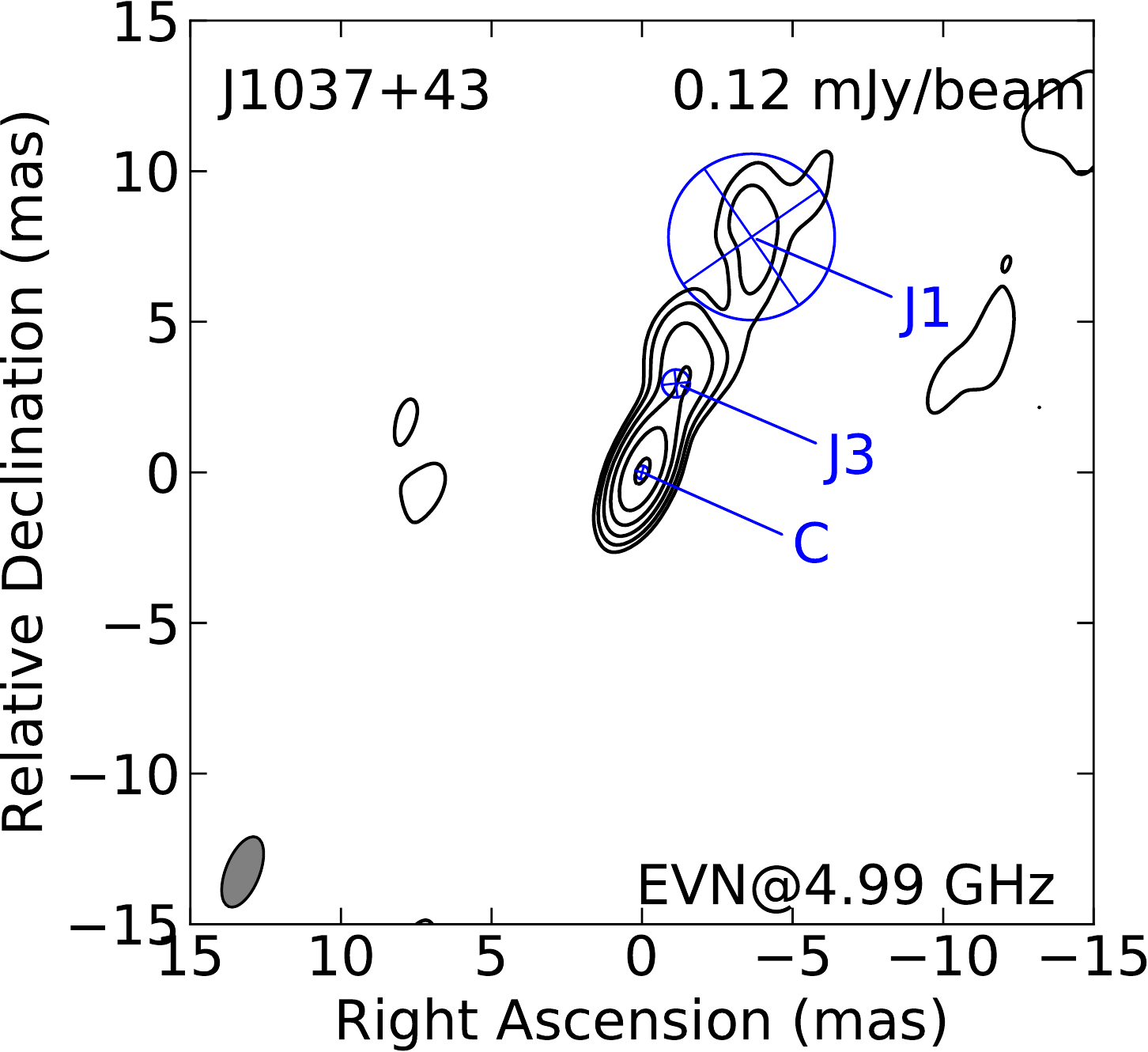}
 \includegraphics[height=3.8cm]{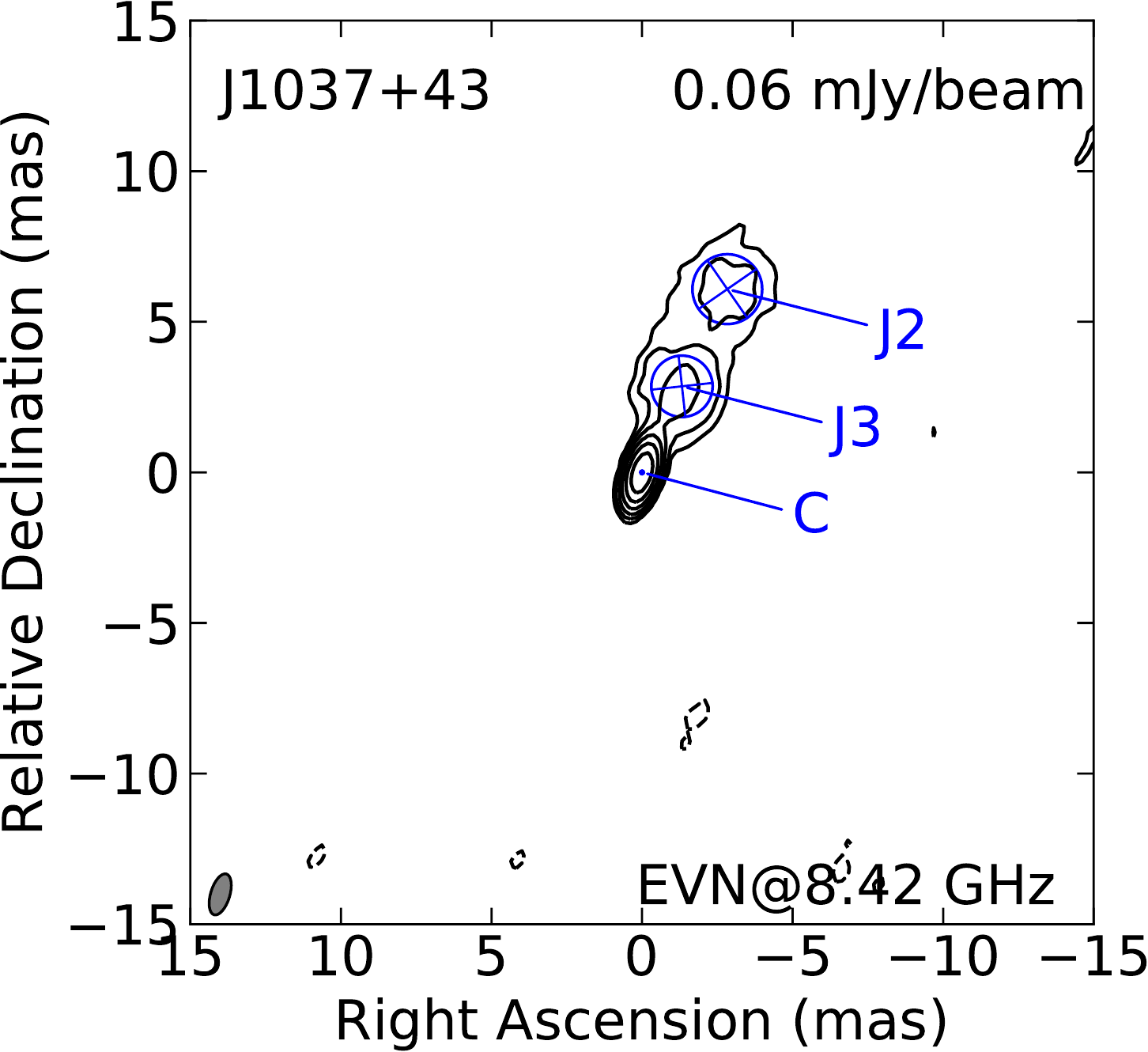}
 \includegraphics[height=3.8cm]{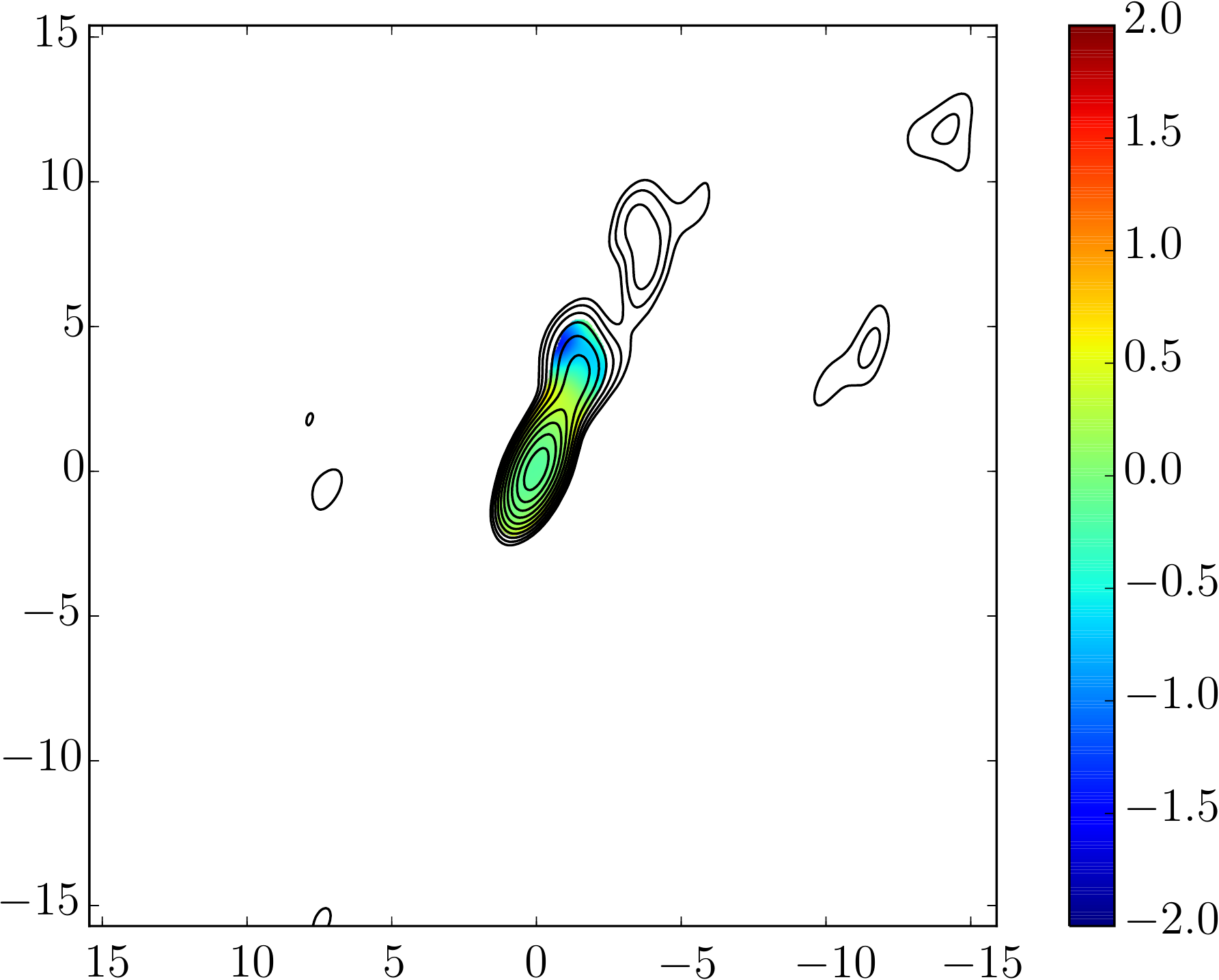}
 \includegraphics[height=3.8cm]{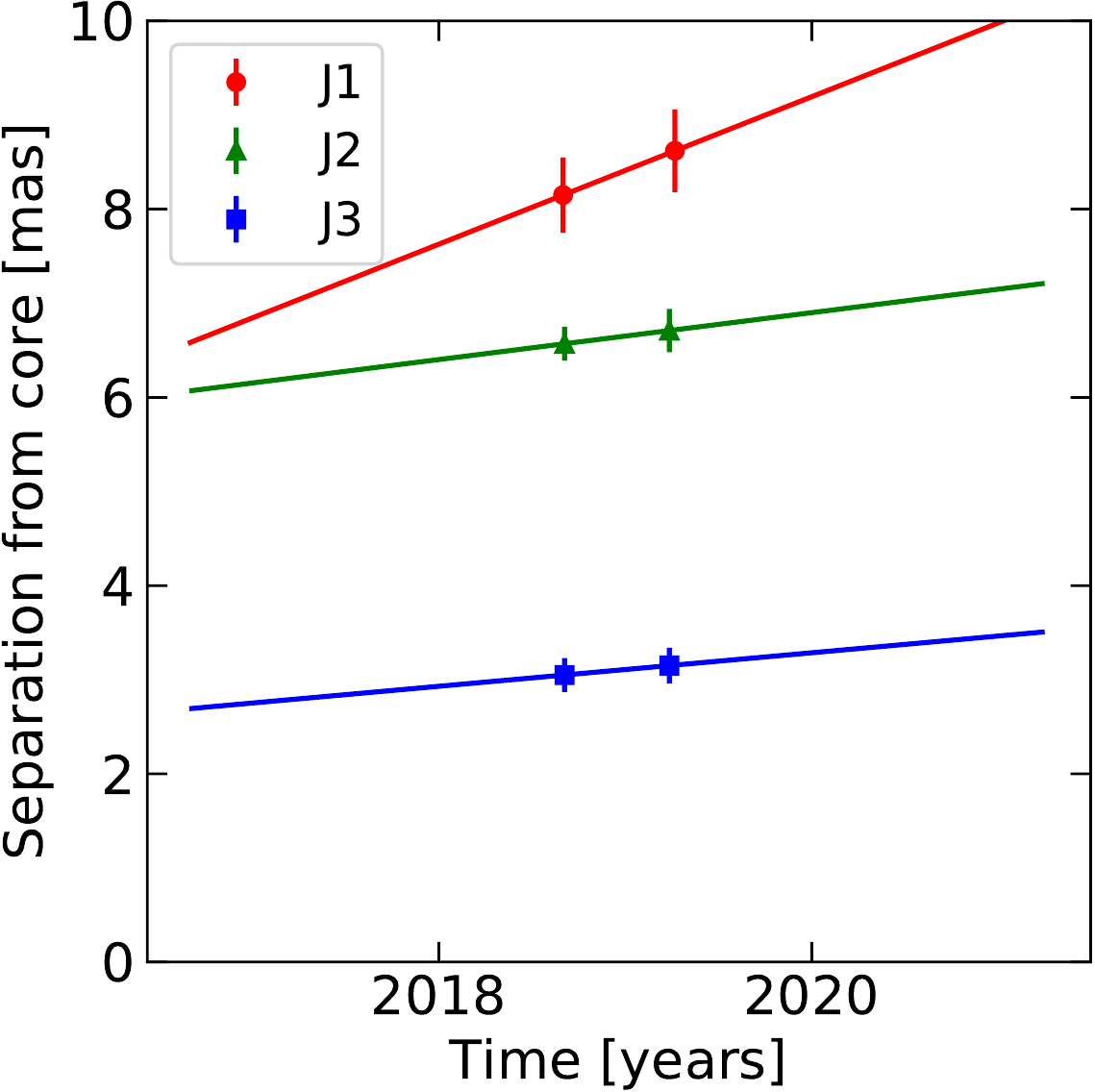}
\\
 \includegraphics[height=3.8cm]{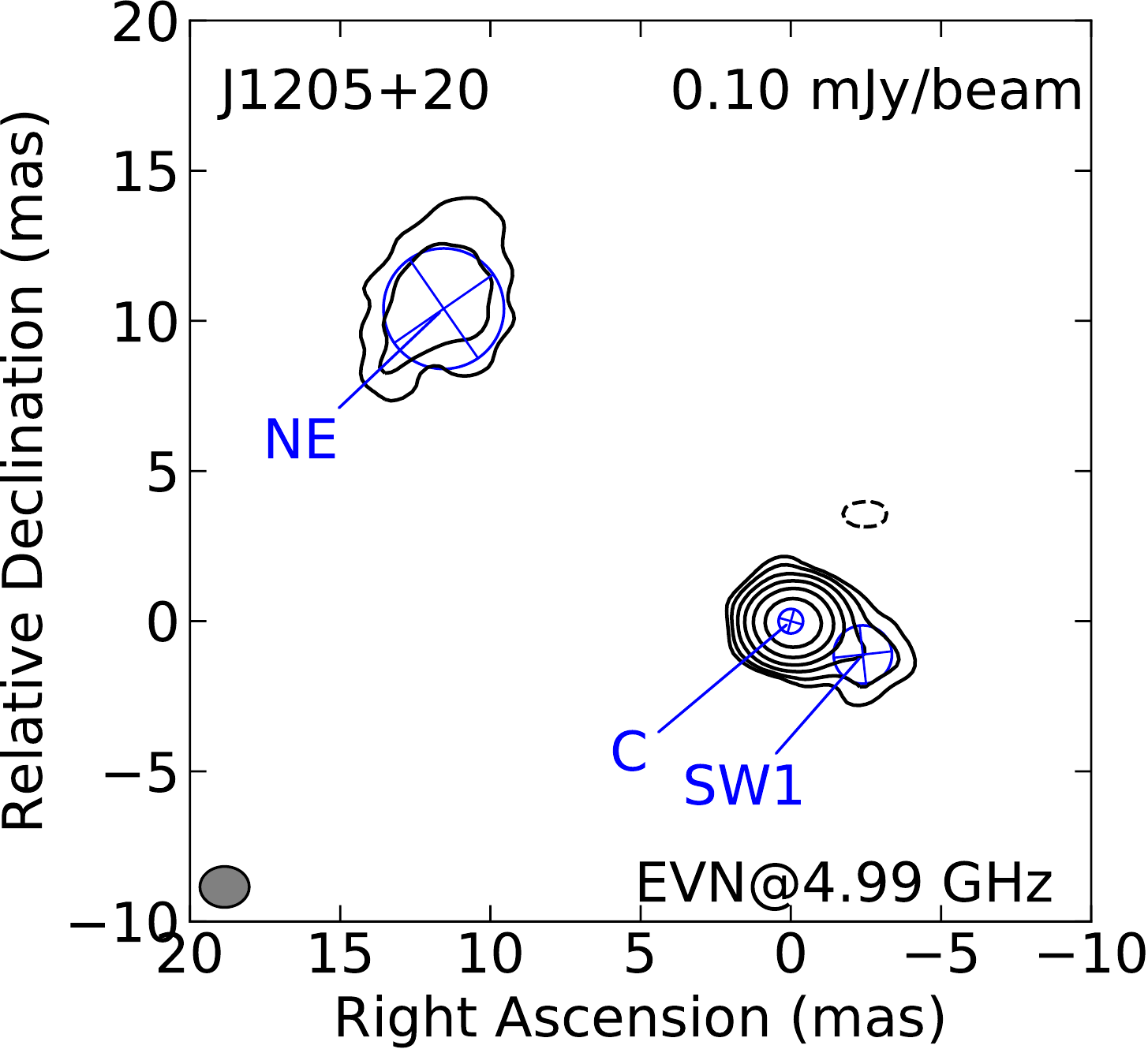}
 \includegraphics[height=3.8cm]{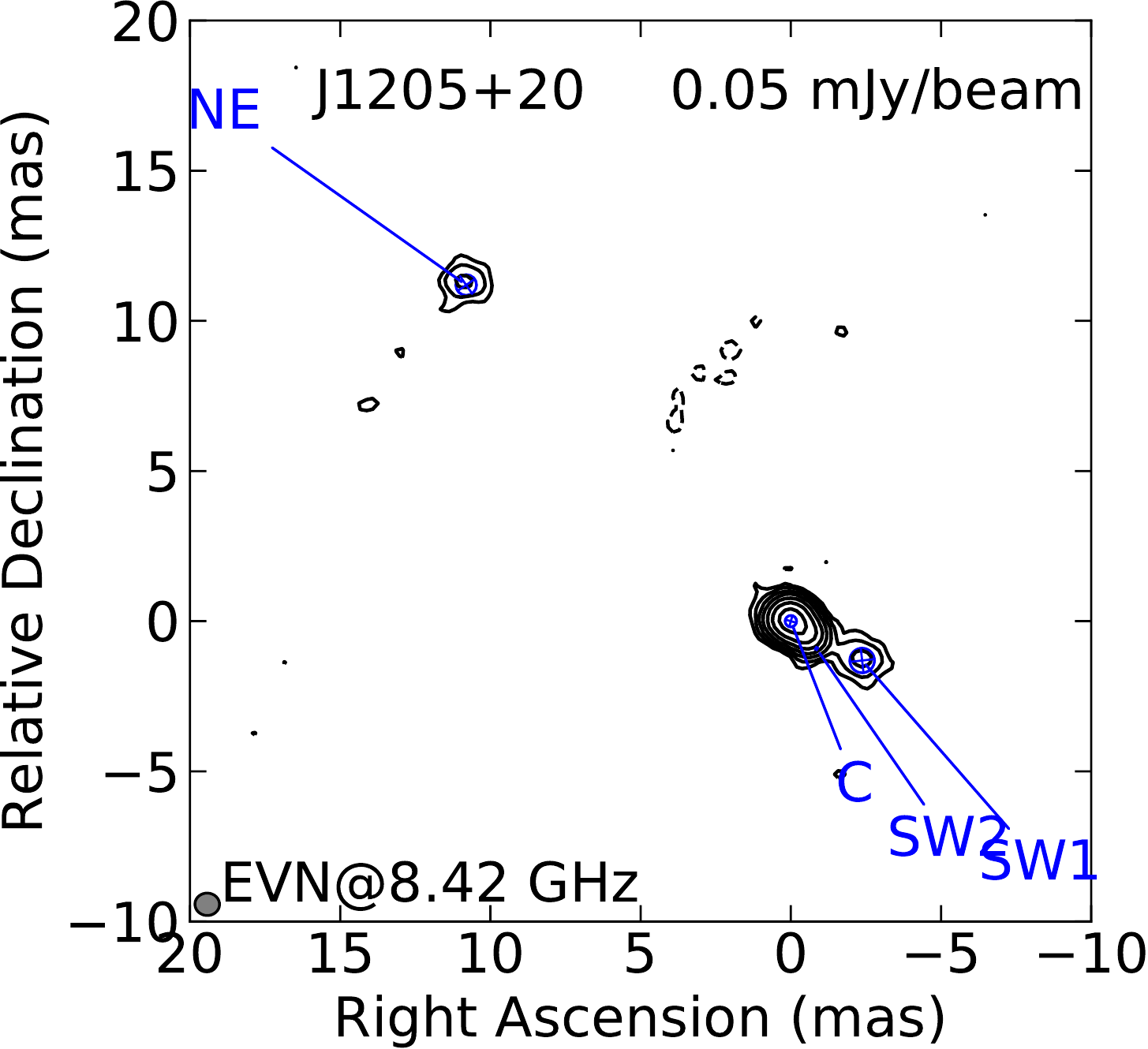}
 \includegraphics[height=3.8cm]{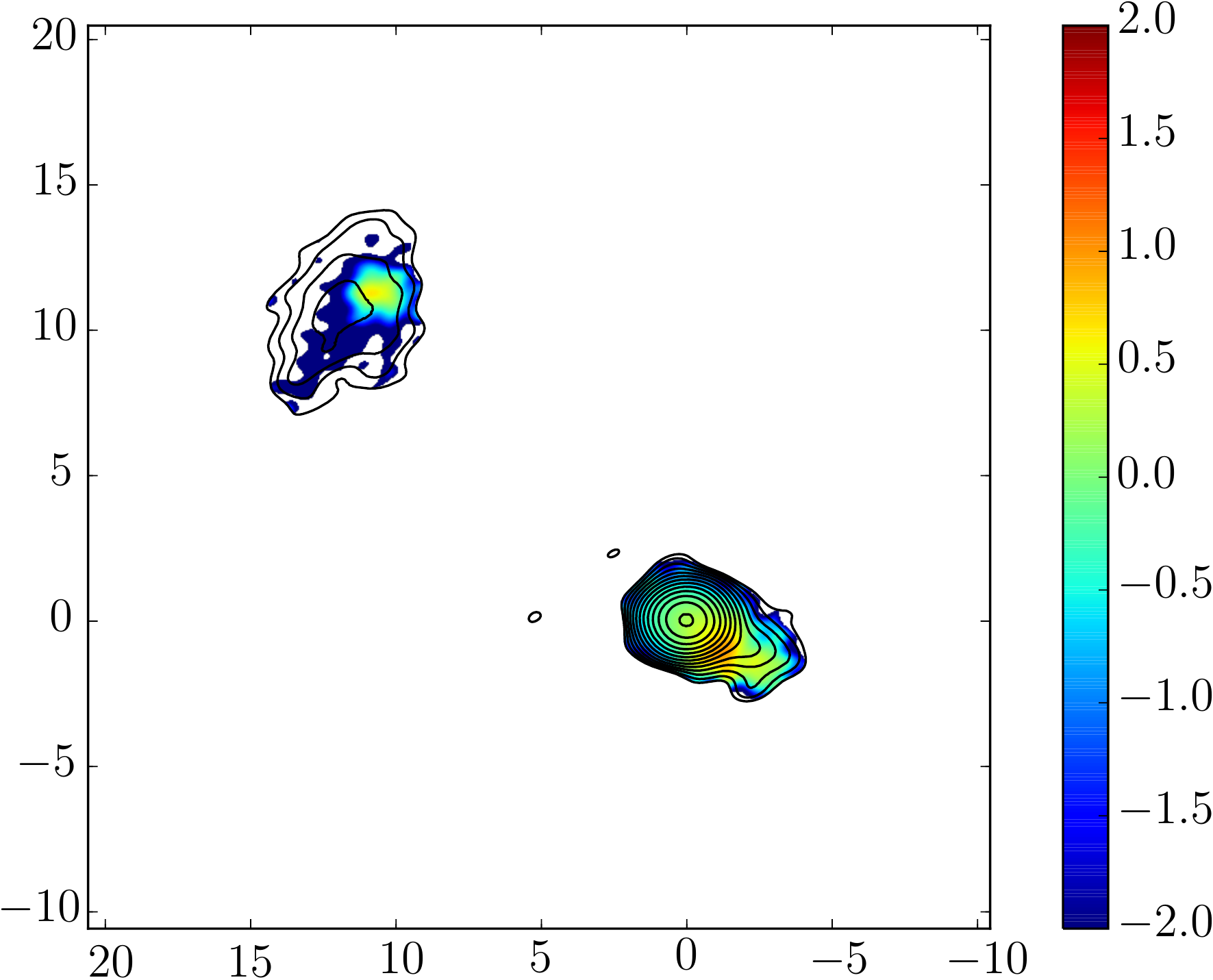}
 \includegraphics[height=3.8cm]{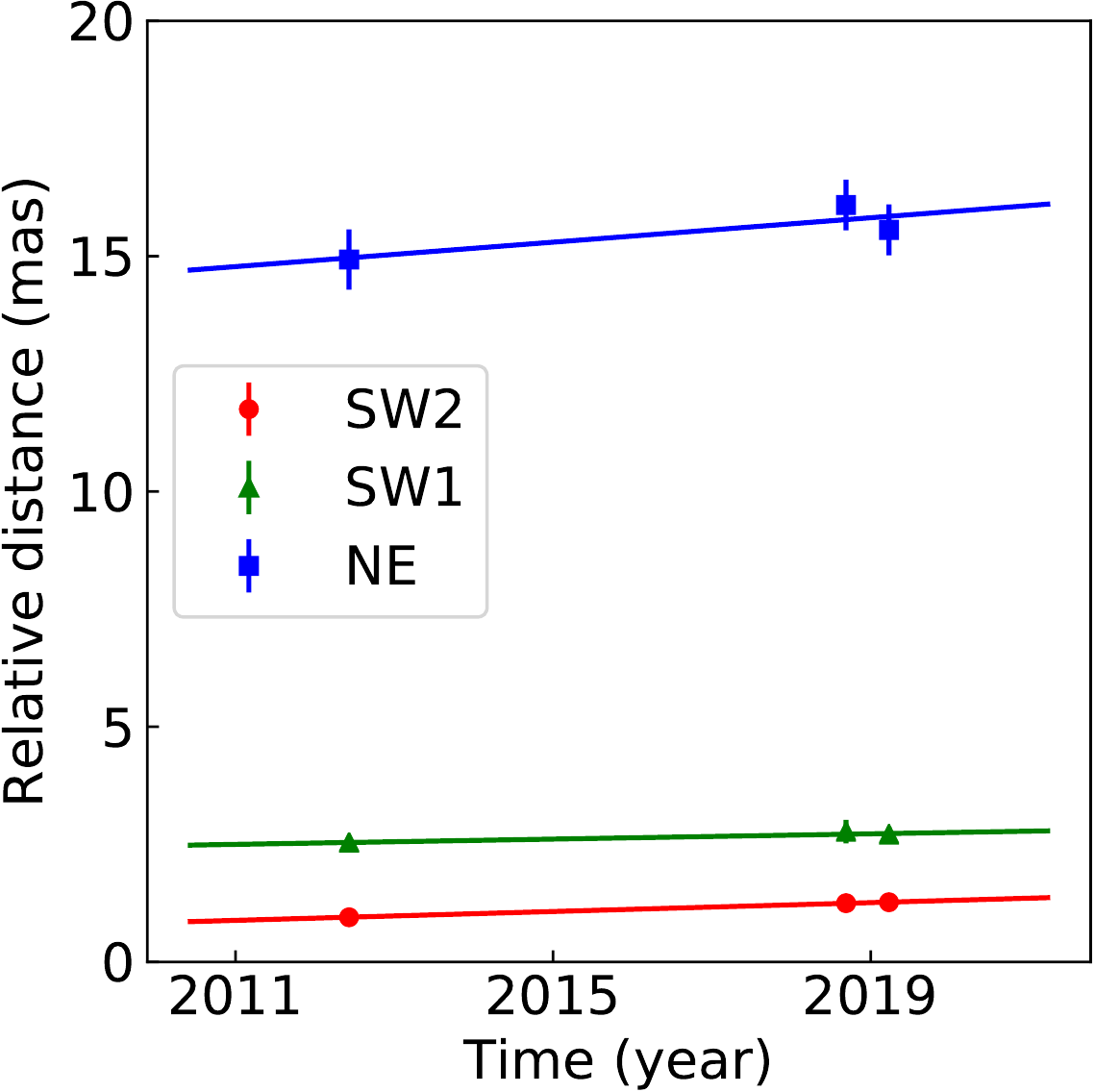}
\\
\caption{EVN images at 5 GHz, 8 GHz, spectral index maps using EVN data and jet proper motions of the eight FR 0s. The 5- and 8-GHz images are obtained from the new epoch of EVN observation. The peak intensity and lowest contour level ($\sim 3\sigma$) are listed in Table \ref{image result}. The contours increase in steps of 2. The grey-colored ellipse in the bottom-left corner of each panel denotes the restoring beam. The spectral index maps are obtained by using the EVN 5 and 8 GHz data. The proper motions of jet components are determined by the least-squares linear fit to the component positions as a function of time.
} \label{fig:VLBI}
\end{figure*}
\addtocounter{figure}{-1}
\begin{figure*}
\centering
 \includegraphics[height=3.8cm]{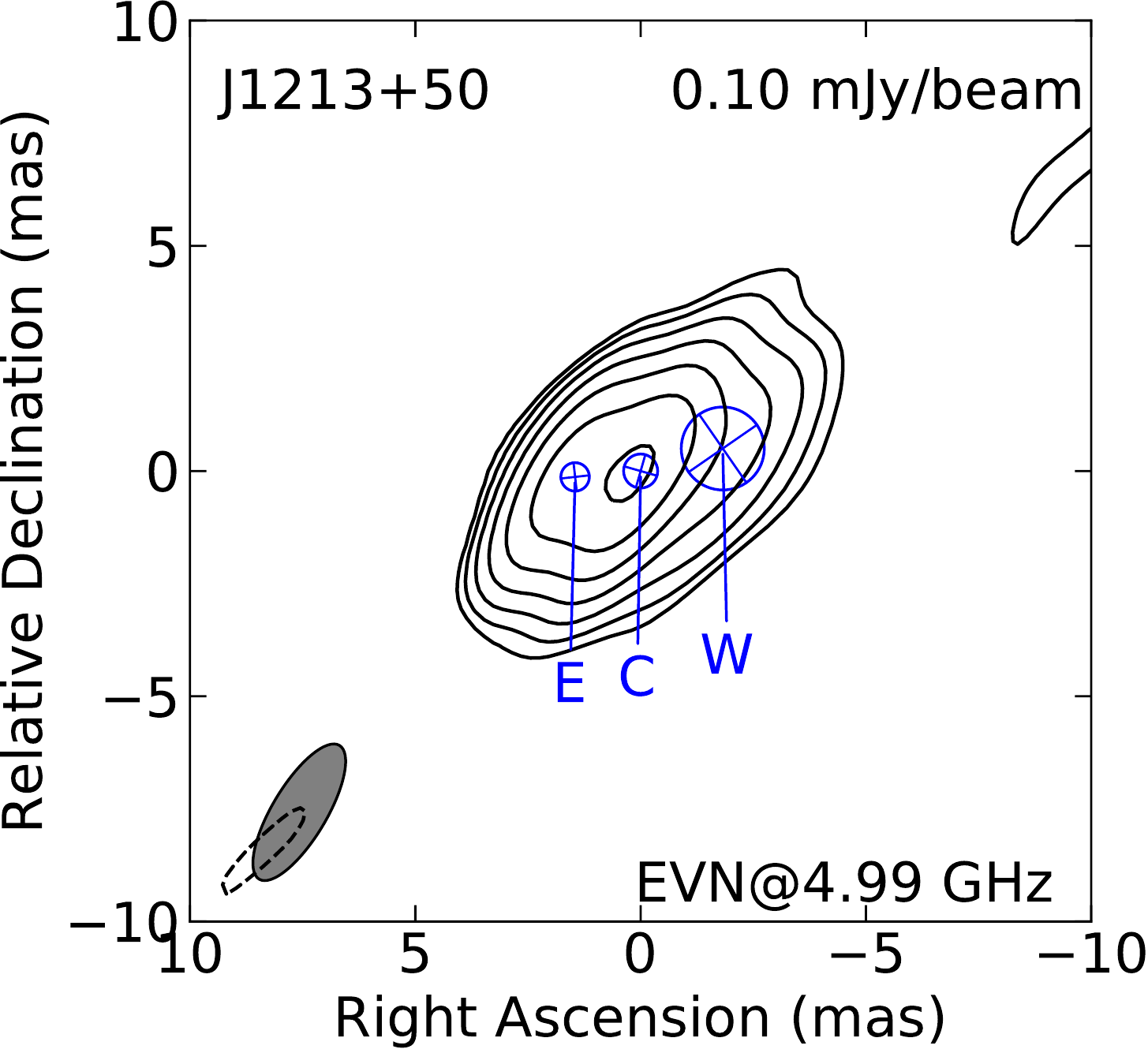}
 \includegraphics[height=3.8cm]{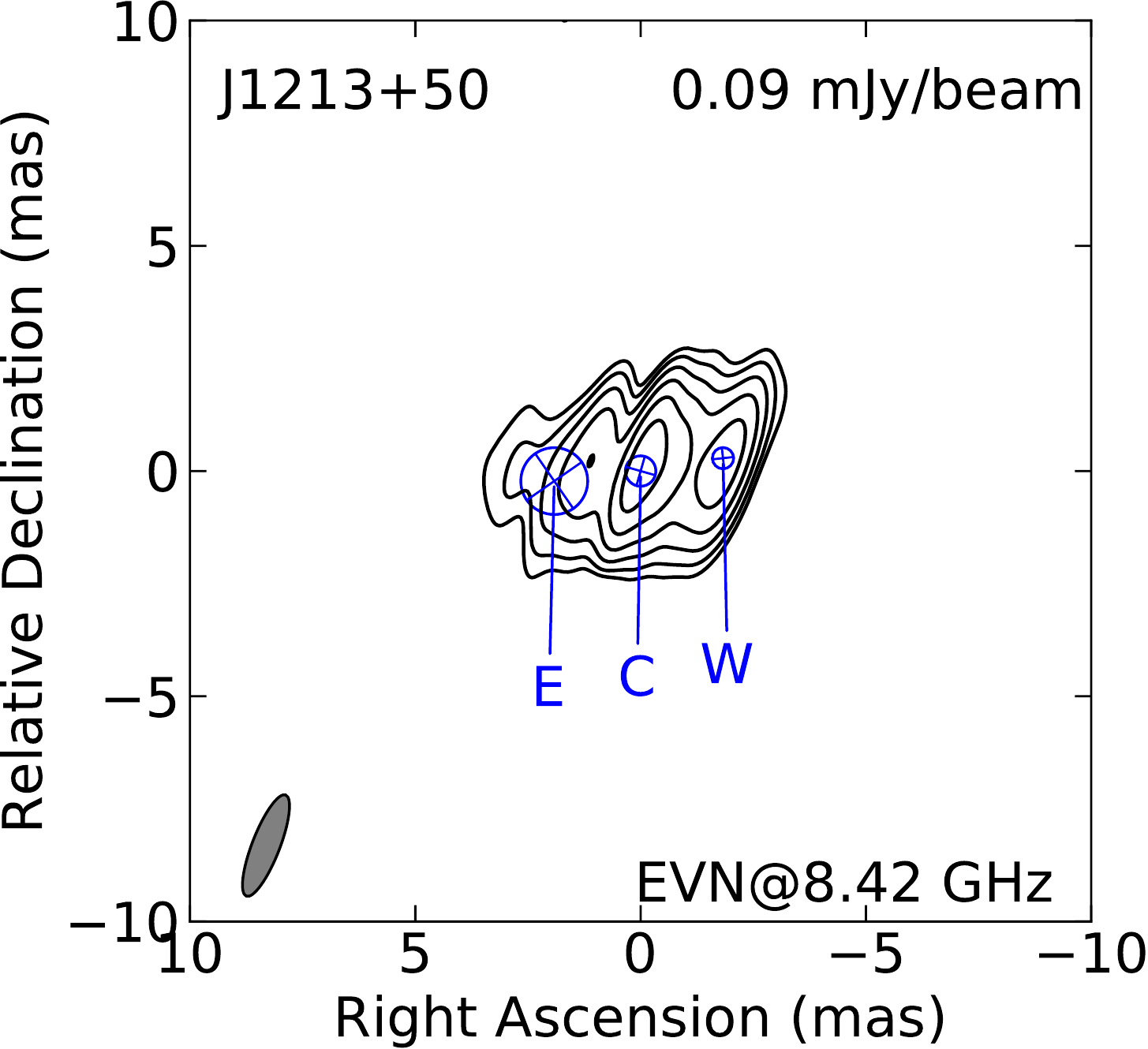}
 \includegraphics[height=3.8cm]{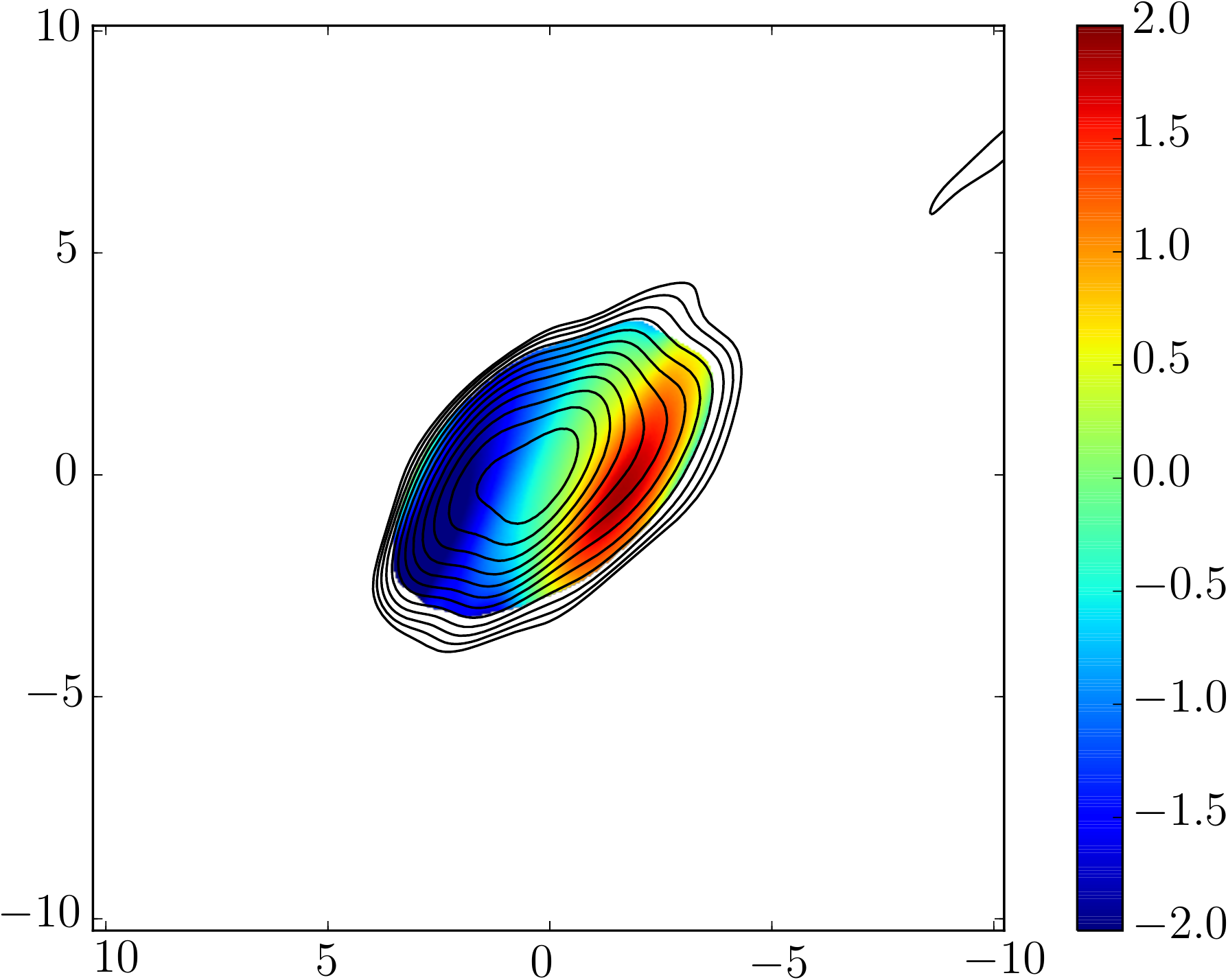}
 \includegraphics[height=3.8cm]{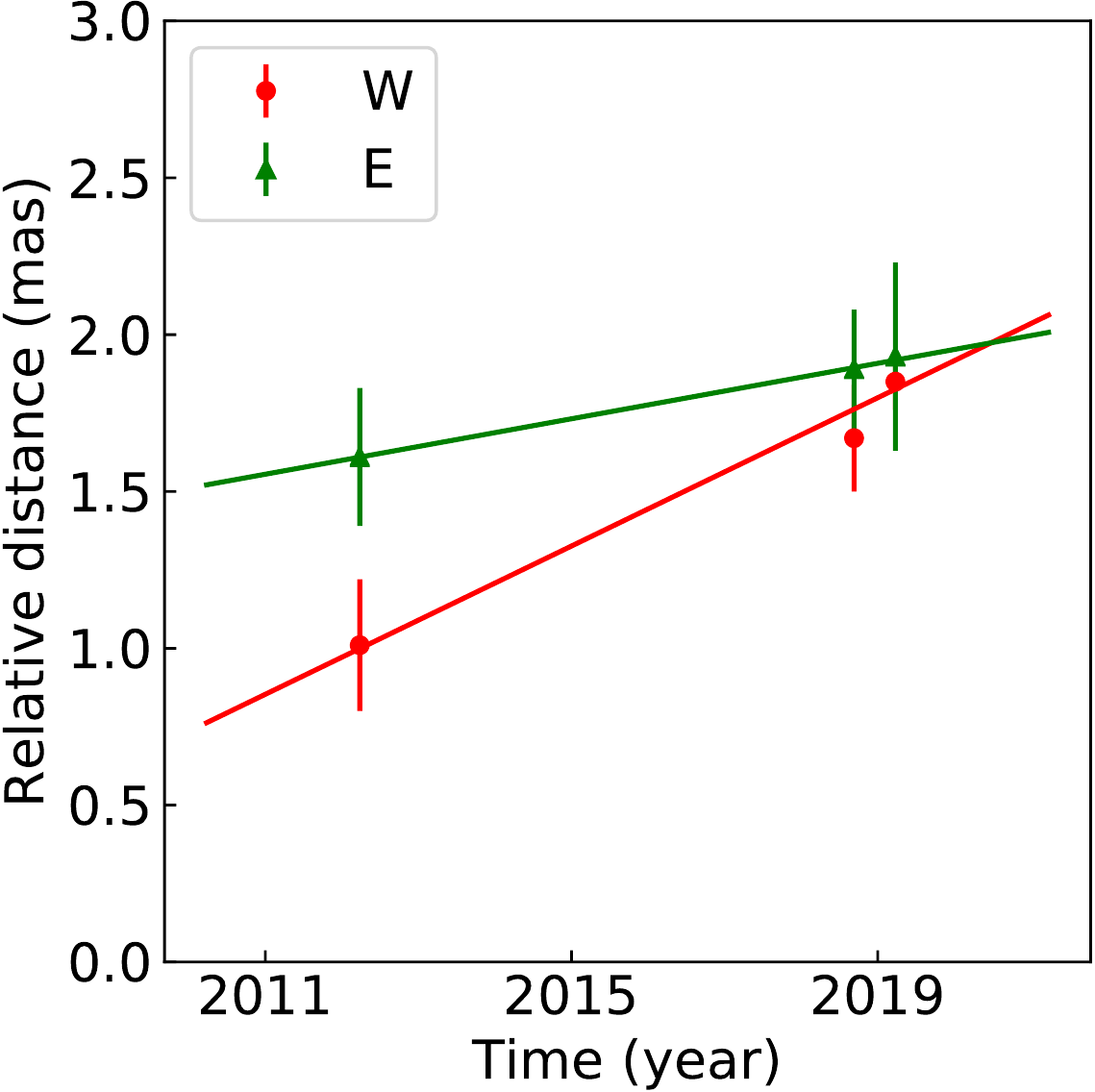}
\\
 \includegraphics[height=3.8cm]{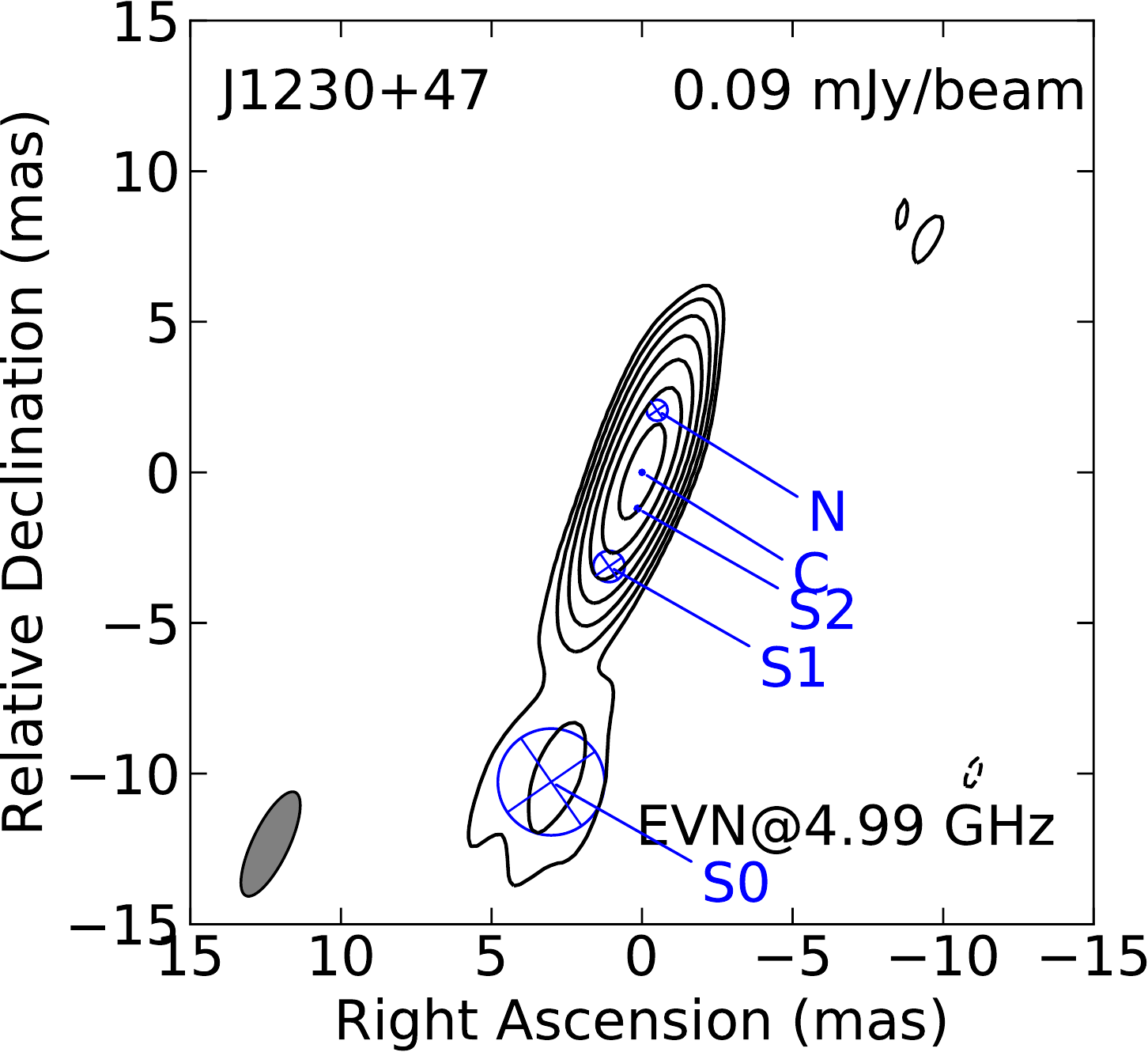}
 \includegraphics[height=3.8cm]{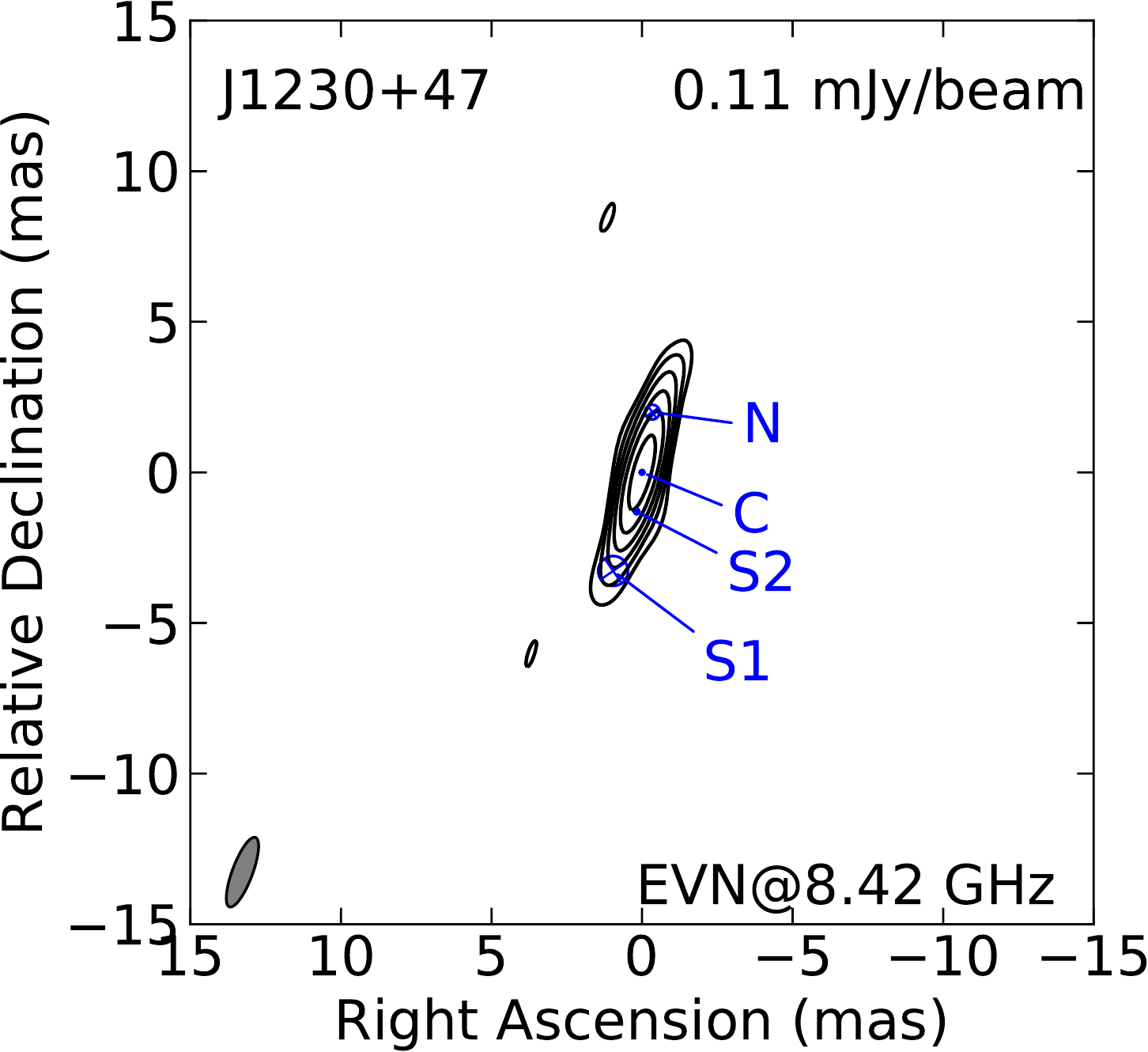} 
 \includegraphics[height=3.8cm]{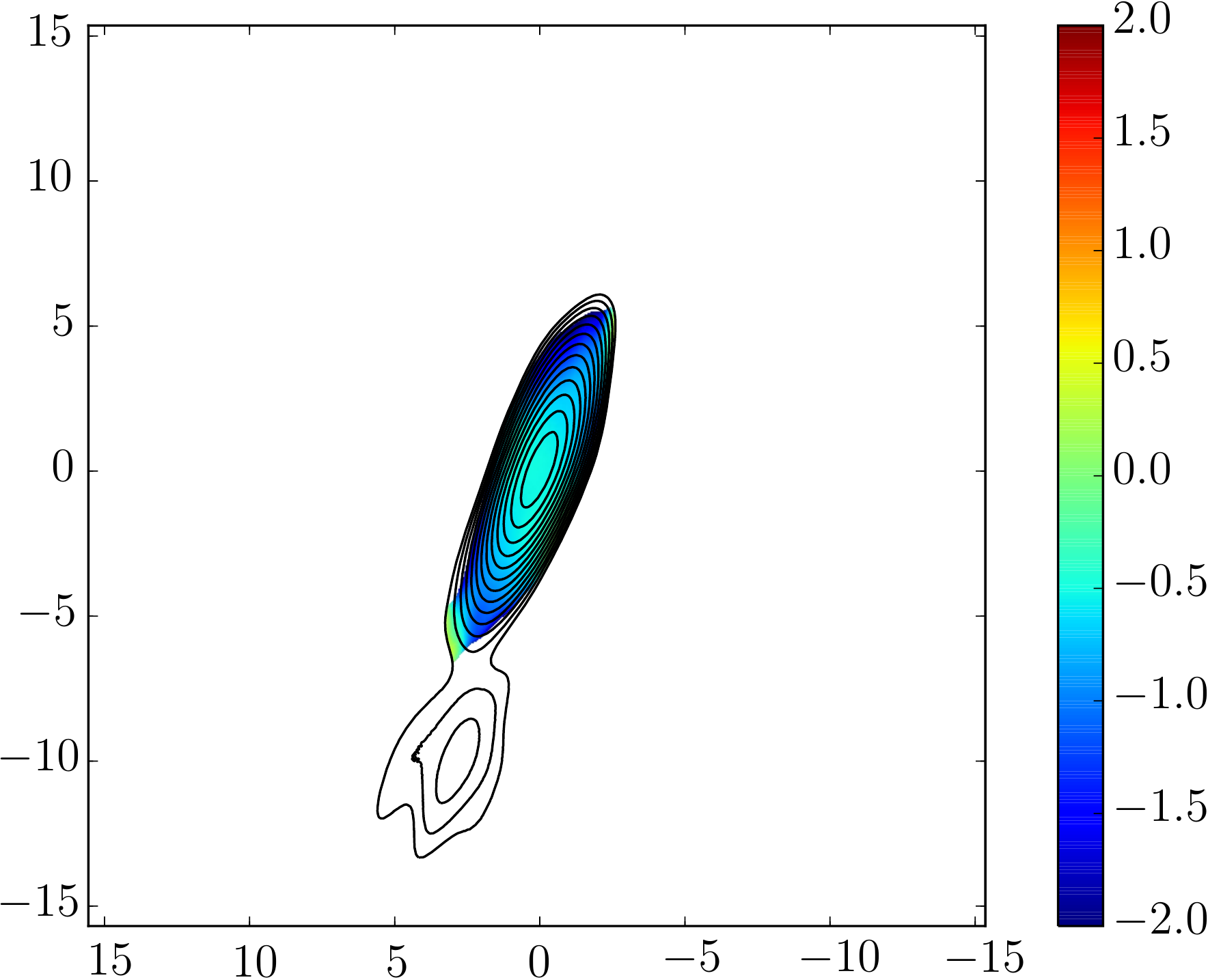}
 \includegraphics[height=3.8cm]{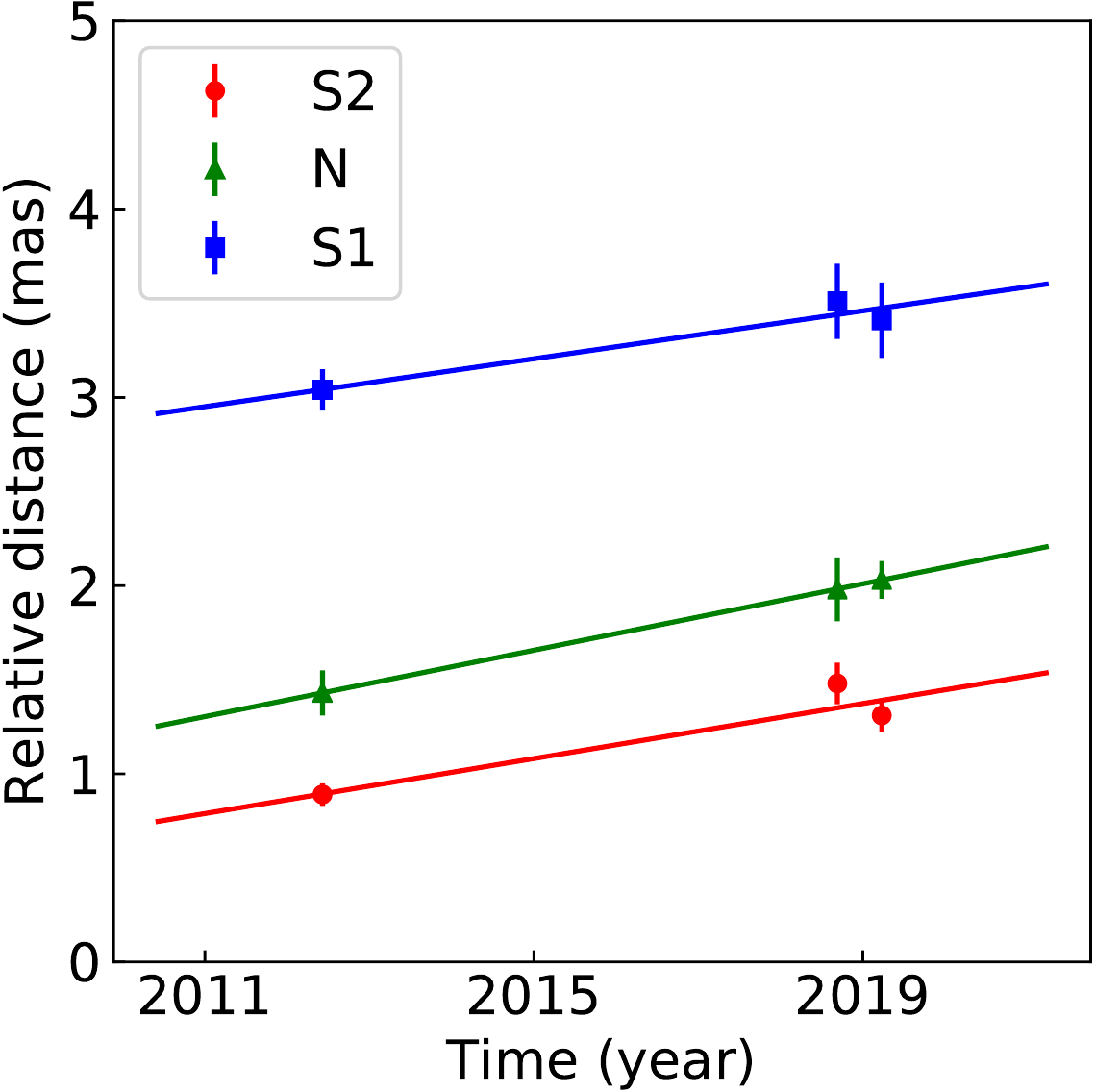} 
\\
 \includegraphics[height=3.8cm]{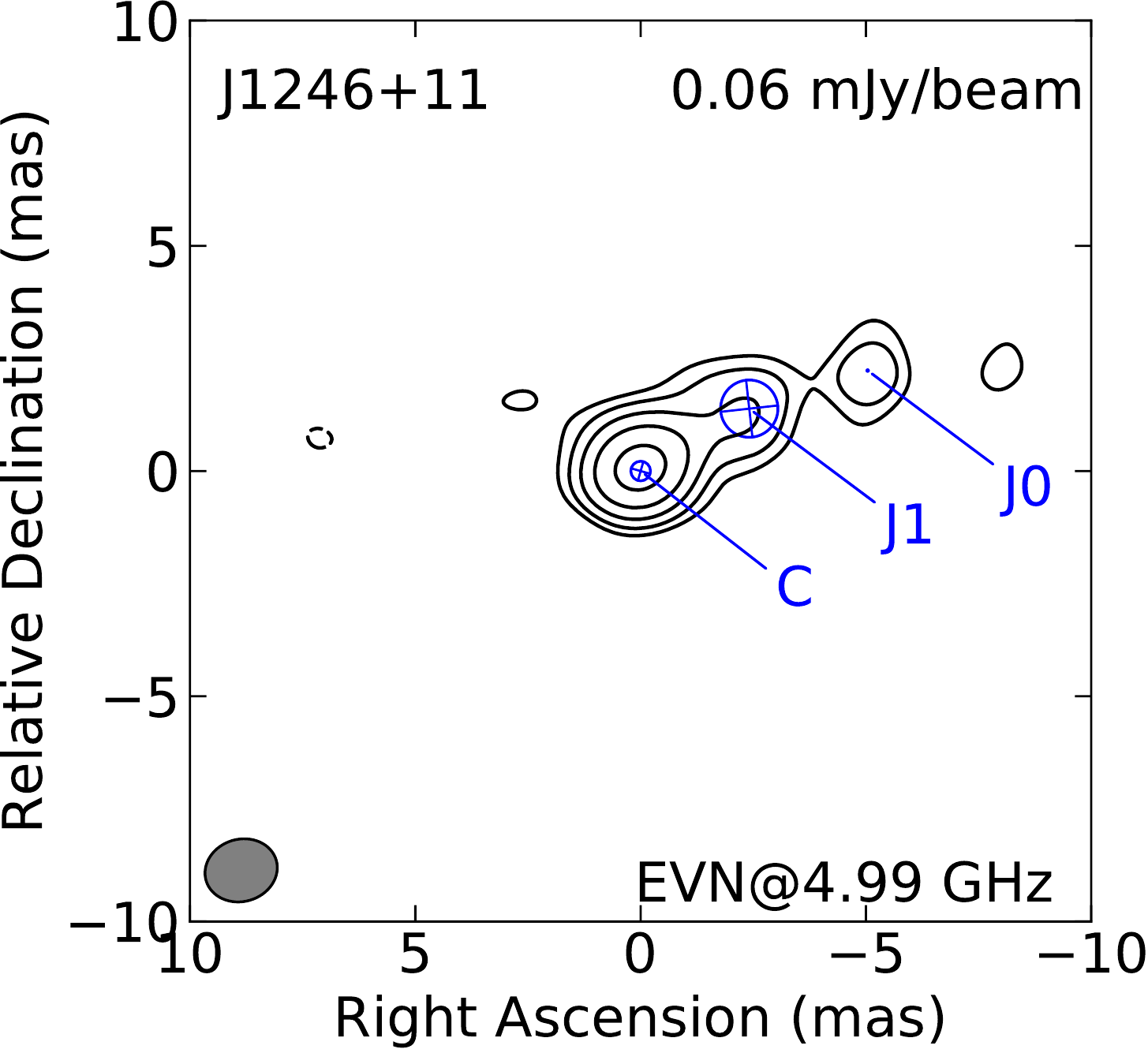}
 \includegraphics[height=3.8cm]{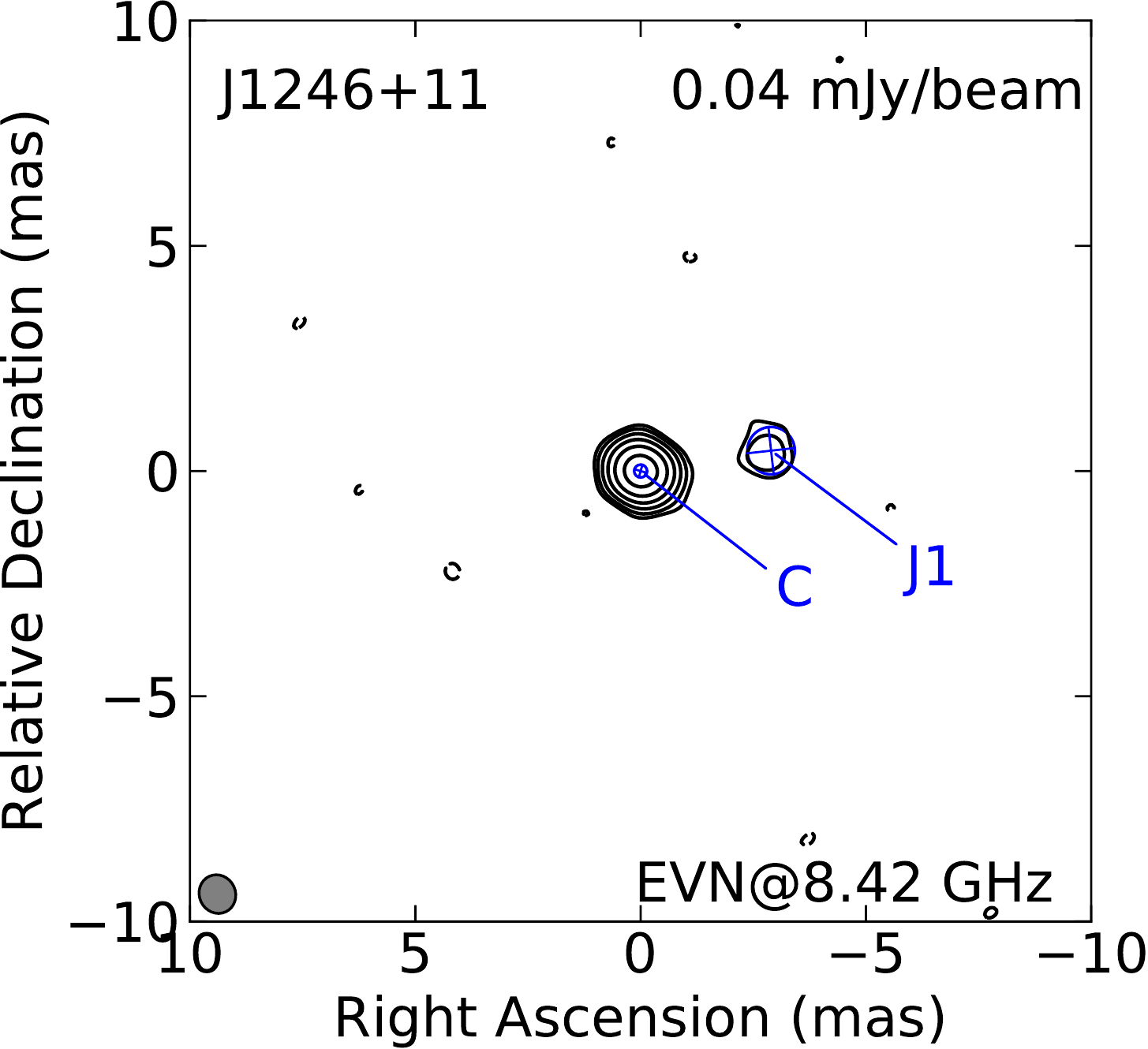} 
 \includegraphics[height=3.8cm]{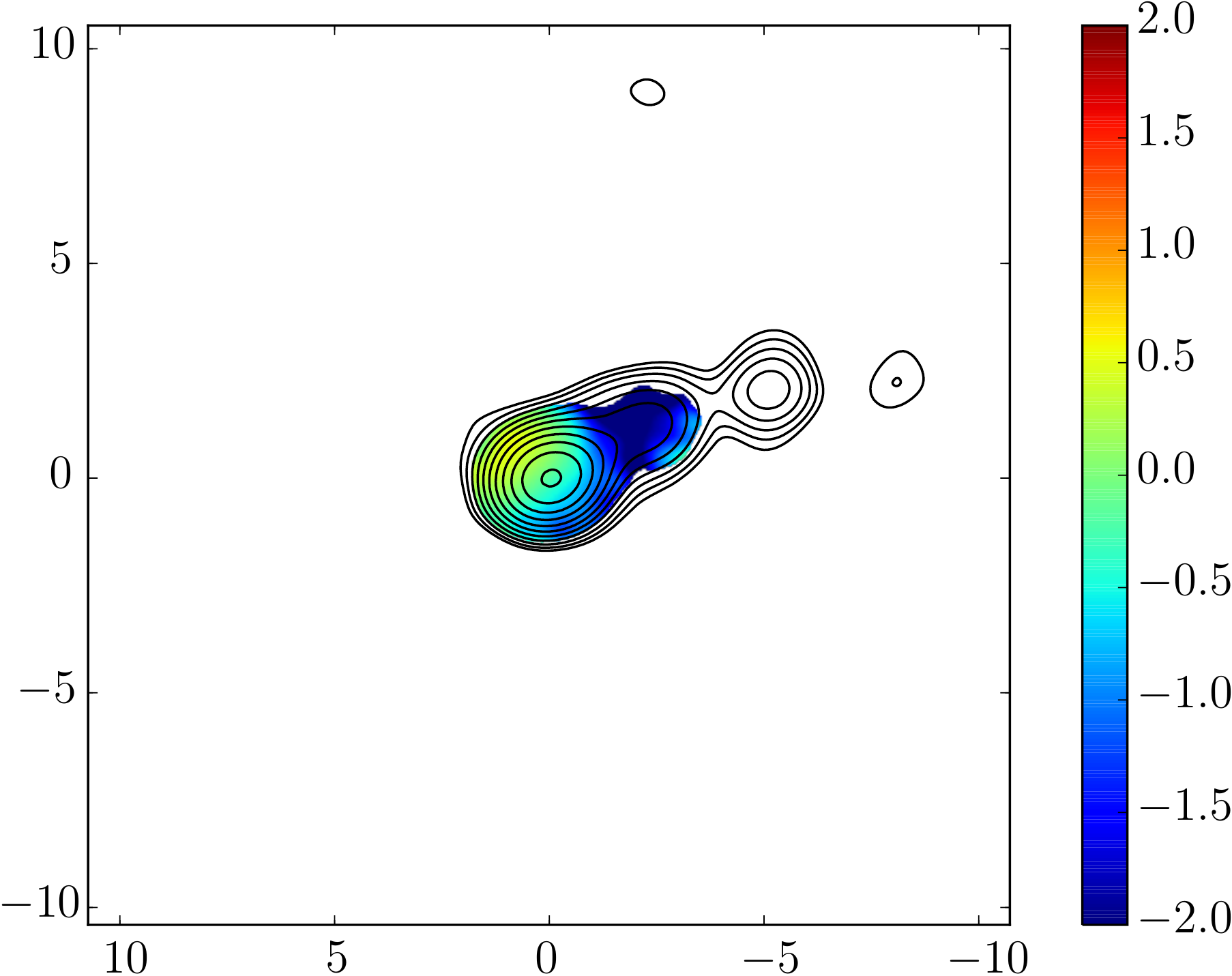}
 \includegraphics[height=3.8cm]{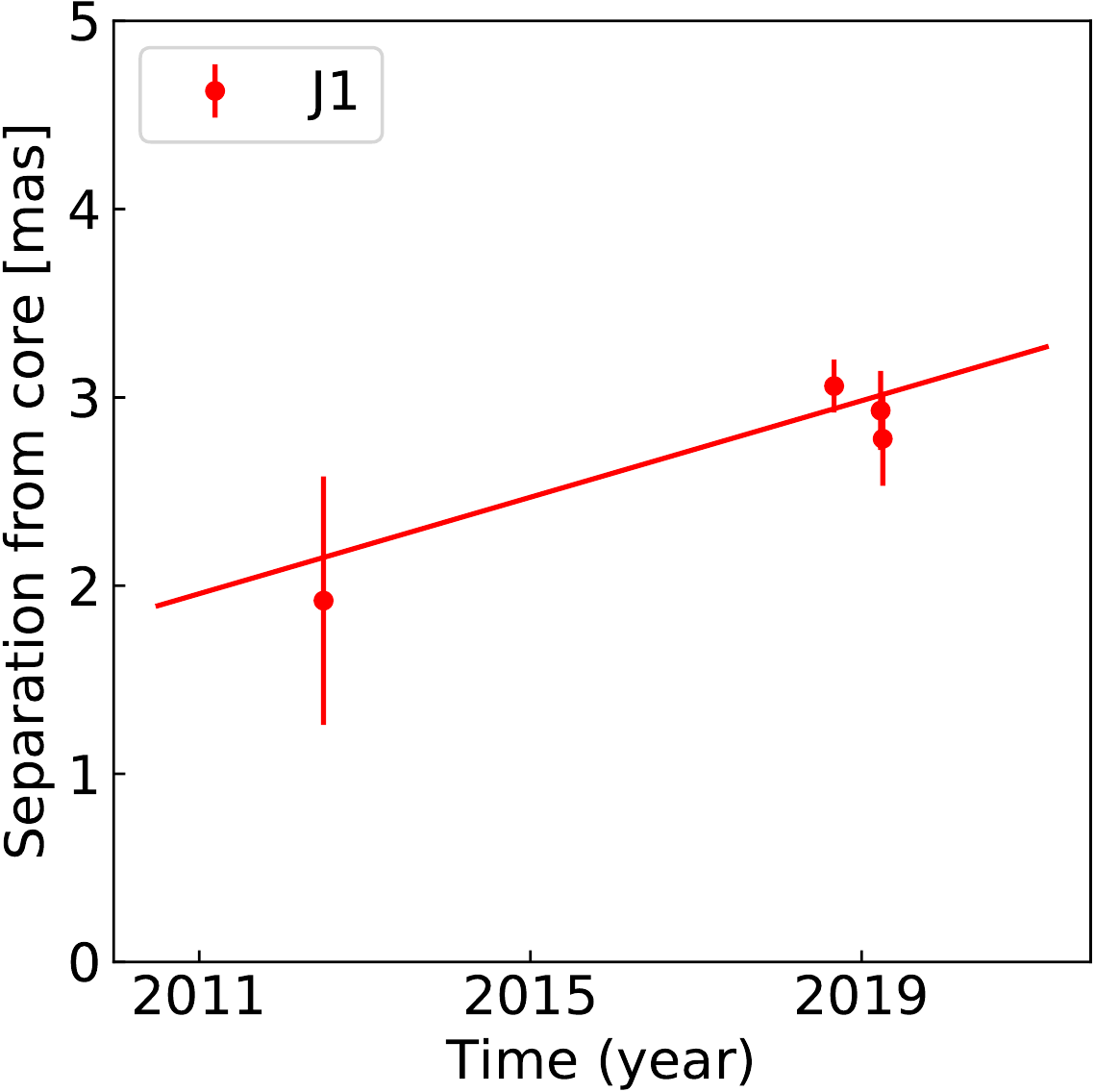}
\\
\caption{Continued}
\label{fig:continued}
\end{figure*}

\begin{table*}
\centering
    \caption{Radio properties of eight FR 0s studied in this paper}
	\begin{tabular}{lccccccr}
	\hline \hline
Name & Other name &    $z$      &    S  & $\rm L_{NVSS}$ & $P_{\rm 1.4 GHz}$ & $D_{\rm VLA}$ & Morphology   \\
     &            &                 & (mJy) &              \\
\hline
J0906+4124 & MCG +07-19-027 & 0.027 & 51.8  &  39.12 & 22.97 & 2.084  & two-sided Jet   \\
J0909+1928 & MRK 1226       & 0.028 & 69.1  &  39.26 & 23.11 & 2.156  & two-sided Jet   \\
J1025+1022 & ASK 378291.0   & 0.046 & 76.6  &  39.75 & 23.60 & 3.468  & two-sided Jet$^*$  \\
J1037+4335 & UGC 05771      & 0.025 & 132.2 &  39.44 & 23.29 & 1.932  & One-sided Jet   \\
J1205+2031 & NGC 4093       & 0.024 & 89.9  &  39.24 & 23.09 & 1.856  & two-sided Jet   \\
J1213+5044 & NGC 4187       & 0.031 & 96.5  &  39.50 & 23.35 & 2.380  & two-sided Jet   \\
J1230+4700 & CGCG 244-025   & 0.039 & 93.8  &  39.70 & 23.55 & 2.964  & two-sided Jet   \\
J1246+1153 & VCC 2041       & 0.047 & 61.2  &  39.68 & 23.53 & 3.540  & One-sided Jet   \\
\hline
	\end{tabular}
\\[0.1cm]
\label{general properties}
Notes. Column description: (1) J2000 name; (2) other name; (3) redshift; (4) NVSS 1.4 GHz flux density from \cite{1998AJ....115.1693C}; (5) logarithm of the NVSS radio luminosity \citet{1998AJ....115.1693C}; (6) the source radio power; (7) upper limit of the source size, assuming all the source is 4 arc min;(5) morphology of the source at parsec-scales.
$^*$: The inner 2-mas jet of  J1025+1022 shows a large misalignment with the outer jet, indicating an X-shaped jet.
\end{table*}

\begin{table*}
    \centering
    \caption{Details of the VLBA and EVN observations.}
	\begin{tabular}{cccccc}
	\hline \hline
Project code & Date          & $\nu$ &    Target sources  & Calibrators  & Participating stations$^{a}$ \\
             & (yyyy-mm-dd)    &  (GHz)     &                    &              &                        \\
\hline
BC241A & 2018-08-03 & 4.87 & J0906+4124, J0909+1928, J1025+1022, J1037+4335 & 4C39.25         & VLBA($-$LA, PT)   \\
BC241B & 2018-08-06 & 8.37 & J0906+4124, J0909+1928, J1025+1022, J1037+4335 & 4C39.25         & VLBA($-$PT)       \\
BC241C & 2018-08-04 & 4.87 & J1205+2031, J1213+5044, J1230+4700, J1246+1153 & 4C39.25, 3C345  & VLBA              \\
BC241D & 2018-08-12 & 8.37 & J1205+2031, J1213+5044, J1230+4700, J1246+1153 & 4C39.25, 3C345  & VLBA($-$PT, SC)   \\
EC063A & 2019-02-24 & 8.42 & J1205+2031, J1213+5044, J1230+4700, J1246+1153 & 4C39.25, 3C345  & EfWbO6HhYsMcIbT6SvBdZc   \\
EC063B & 2019-02-26 & 8.42 & J0906+4124, J0909+1928, J1025+1022, J1037+4335 & 4C39.25         & EfWbO6HhYsMcIbUrT6SvBdZc   \\
EC063C & 2019-03-03 & 4.99 & J1205+2031, J1213+5044, J1230+4700, J1246+1153 & 4C39.25, 3C345  & EfWbMcNtO8T6UrYsTrHhIbJbSvBdZc   \\
EC063D & 2019-03-07 & 4.99 & J0906+4124, J0909+1928, J1025+1022, J1037+4335 & 4C39.25         & EfWbMcNtO8T6UrYsTrHhIbJbSvBdZc    \\
\hline
	\end{tabular}
\\[0.1cm]
\label{observations log}
Notes. Column description: (1)observing code; (2) observation date; (3) observing frequency; (4) target sources; (5) calibrators; (6) participating stations. VLBA stations that were not used in individual observations are shown in brackets. EVN stations: Ef: Effelsberg, Wb: Westerbork, O6: Onsala-60, Hh: Hartebeesthoek, Ys: Yebes, Mc: Medicina, Ib: Irbene, Ur: Urumqi, T6: Tianma, Sv: Svetloe, Bd: Badary, Zc: Zelenchukskaya, Nt: Noto, O8: Onsala-85, Tr: Torun and Jb: Jodrell Bank. 
\end{table*}

\begin{table*}
\centering
\caption{Imaging results of 8 FR 0s}
\begin{tabular}{ccccccccccc}
\hline \hline
Name &   Obs. date  & Array & $\rm \nu$ &  $\rm S_{tot}$ & $\rm S_{peak}$   &  $\rm \sigma$ & $\rm B_{maj}$  & $\rm B_{min}$  & $\rm \theta$ & $\rm \alpha$ \\
     &  (yyyy-mm-dd)  &    &  (GHz)    &      (mJy)      & (mJy beam$^{-1}$) &  (mJy beam$^{-1}$) &      (mas)     &        (mas)   &     ($\degr$)  &  \\
\hline
J0906+4124 & 2018-08-03  & VLBA & 4.87 & 75.93  & 37.72  & 0.07 & 3.34 & 1.22 & 1.64     & 0.22     \\
           & 2019-03-07  & EVN  & 4.99 & 74.77  & 54.71  & 0.06 & 2.08 & 1.32 & $-$45.87 & 0.26     \\
           & 2018-08-06  & VLBA & 8.37 & 85.95  & 40.01  & 0.09 & 2.09 & 0.78 & 0.74     &          \\
           & 2019-02-26  & EVN  & 8.42 & 85.87  & 46.08  & 0.11 & 1.05 & 0.69 & $-$30.47 &          \\
J0909+1928 & 2018-08-03  & VLBA & 4.87 & 133.71 & 110.49 & 0.09 & 3.11 & 1.21 & $-$1.39  & 0.29     \\
           & 2019-03-07  & EVN  & 4.99 & 135.13 & 103.11 & 0.08 & 1.97 & 1.77 & 89.47    & $-$0.18  \\
           & 2018-08-06  & VLBA & 8.37 & 156.65 & 126.09 & 0.08 & 1.98 & 0.78 & $-$1.26  &          \\
           & 2019-02-26  & EVN  & 8.42 & 123.11 & 76.66  & 0.07 & 0.90 & 0.84 & $-$7.88  &          \\
J1025+1022 & 2018-08-03  & VLBA & 4.87 & 114.45 & 56.37  & 0.11 & 3.09 & 1.28 & 1.96     & $-$0.04  \\
           & 2019-03-07  & EVN  & 4.99 & 114.60 & 50.67  & 0.12 & 1.46 & 1.39 & 40.19    & $-$0.13  \\
           & 2018-08-06  & VLBA & 8.37 & 112.63 & 71.89  & 0.11 & 1.97 & 0.82 & 1.56     &          \\
           & 2019-02-26  & EVN  & 8.42 & 106.65  & 31.20  & 0.12 & 0.71 & 0.67 & 79.54    &         \\
J1037+4335 & 2018-08-03  & VLBA & 4.87 & 25.82  & 11.12  & 0.12 & 3.71 & 1.53 & 8.48     & $-$0.58  \\
           & 2019-03-07  & EVN  & 4.99 & 25.57  & 12.48  & 0.12 & 2.48 & 1.13 & $-$21.57 & $-$0.72  \\
           & 2018-08-06  & VLBA & 8.37 & 18.82  & 13.24  & 0.18 & 2.29 & 0.84 & 8.71     &          \\
           & 2019-02-26  & EVN  & 8.42 & 17.54  & 10.85  & 0.06 & 1.25 & 0.59 & $-$16.00 &          \\
J1205+2031 & 2018-08-04  & VLBA & 4.87 & 37.35  & 20.74  & 0.10 & 3.43 & 1.57 &  19.89   & 0.15     \\
           & 2019-03-03  & EVN  & 4.99 & 35.80  & 18.41  & 0.10 & 1.54 & 1.27 & $-$89.8  & $-$0.04  \\
           & 2018-08-12  & VLBA & 8.37 & 40.49  & 32.58  & 0.07 & 2.87 & 1.02 & 10.00    &          \\
           & 2019-02-24  & EVN  & 8.42 & 32.28  & 18.07  & 0.05 & 0.76 & 0.70 & 87.79    &          \\
J1213+5044 & 2018-08-04  & VLBA & 4.87 & 47.90  & 27.67  & 0.08 & 3.54 & 1.92 & 12.15    & 0.19     \\
           & 2019-03-03  & EVN  & 4.99 & 45.58  & 20.48  & 0.11 & 3.24 & 1.12 & $-$30.96 & $-$0.61  \\
           & 2018-08-12  & VLBA & 8.37 & 56.85  & 24.91  & 0.19 & 2.45 & 0.99 & 2.81     &          \\
           & 2019-02-24  & EVN  & 8.42 & 33.18  & 14.56  & 0.09 & 2.25 & 0.61 & $-$21.4  &          \\
J1230+4700 & 2018-08-04  & VLBA & 4.87 & 41.51  & 26.18  & 0.11 & 3.47 & 1.63 & 19.31    & $-$0.06  \\
           & 2019-03-03  & EVN  & 4.99 & 39.31  & 24.49  & 0.09 & 3.55 & 1.13 & $-$25.59 & $-$0.21  \\
           & 2018-08-12  & VLBA & 8.37 & 40.38  & 28.16  & 0.06 & 2.51 & 0.98 & 1.80     &          \\
           & 2019-02-24  & EVN  & 8.42 & 35.35  & 15.76  & 0.10 & 2.36 & 0.61 & $-$18.82 &          \\
J1246+1153 & 2018-08-04  & VLBA & 4.87 & 8.03   & 5.71   & 0.08 & 3.77 & 1.75 & 19.99    & 0.08     \\
           & 2019-03-03  & EVN  & 4.99 & 6.83   & 3.88   & 0.07 & 1.60 & 1.37 & $-$69.81 & 0.12     \\
           & 2018-08-12  & VLBA & 8.37 & 8.38   & 8.05   & 0.08 & 3.10 & 1.09 & $-$13.61 &          \\
           & 2019-02-24  & EVN  & 8.42 & 7.29   & 6.10   & 0.03 & 0.86 & 0.80 &  25.12   &          \\
\hline
\end{tabular}
\\[0.1cm]
\label{image result}
Notes. Column description: (1) J2000 name; (2) observing date; (3) VLBI array; (4) observing frequency; (5) integrated flux density; (6) peak specific intensity; (7) off-source rms noise in the clean image; (8) major axis of the restoring beam (FWHM); (9) minor axis of the restoring beam (FWHM); (10) position angle of the major axis, measured from north through east; (11) spectral index.
\end{table*}





\appendix

\section{VLBA images} \label{app1}
\label{A}

\begin{figure*}
\centering
 \includegraphics[height=3.8cm]{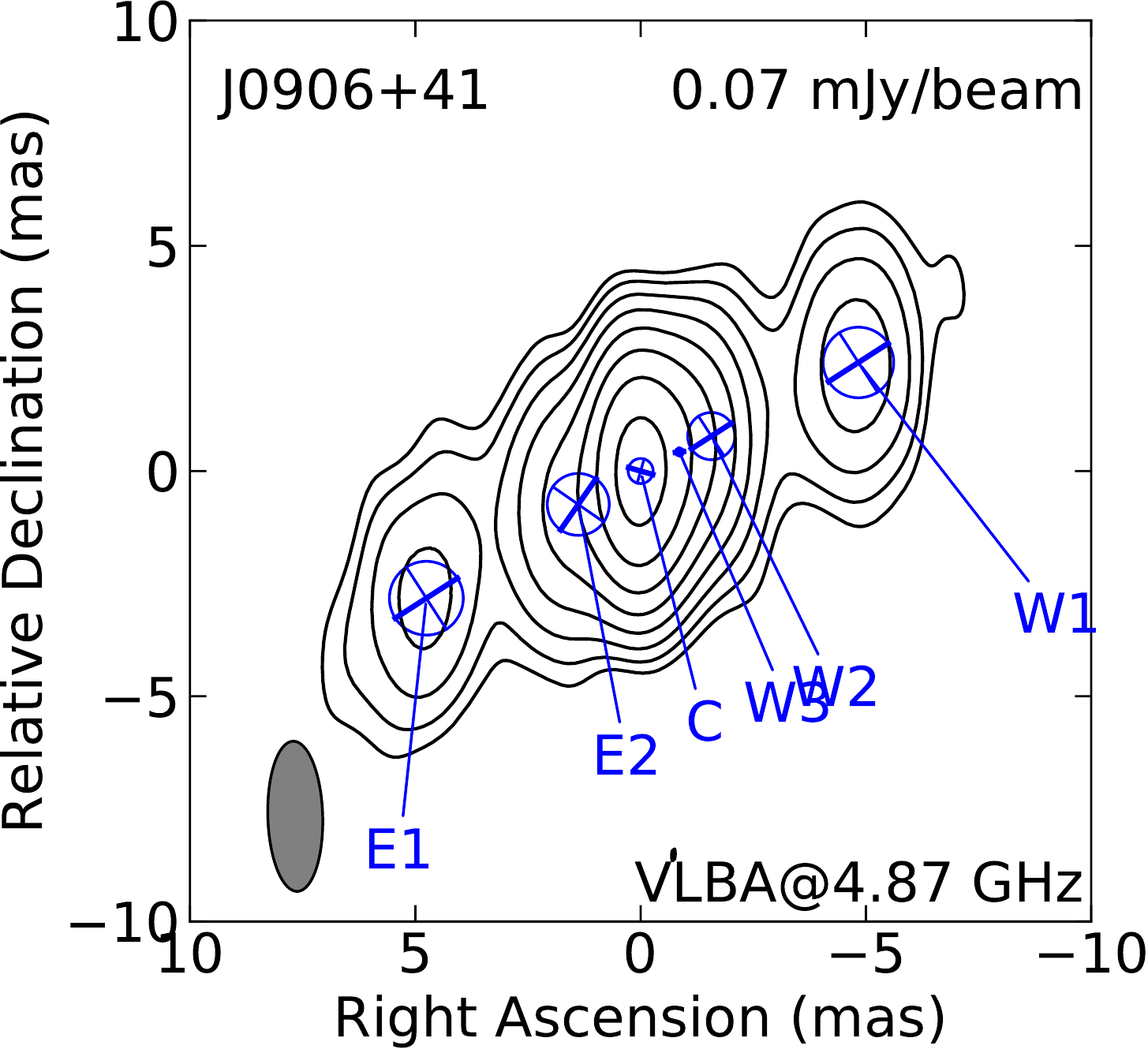}
 \includegraphics[height=3.8cm]{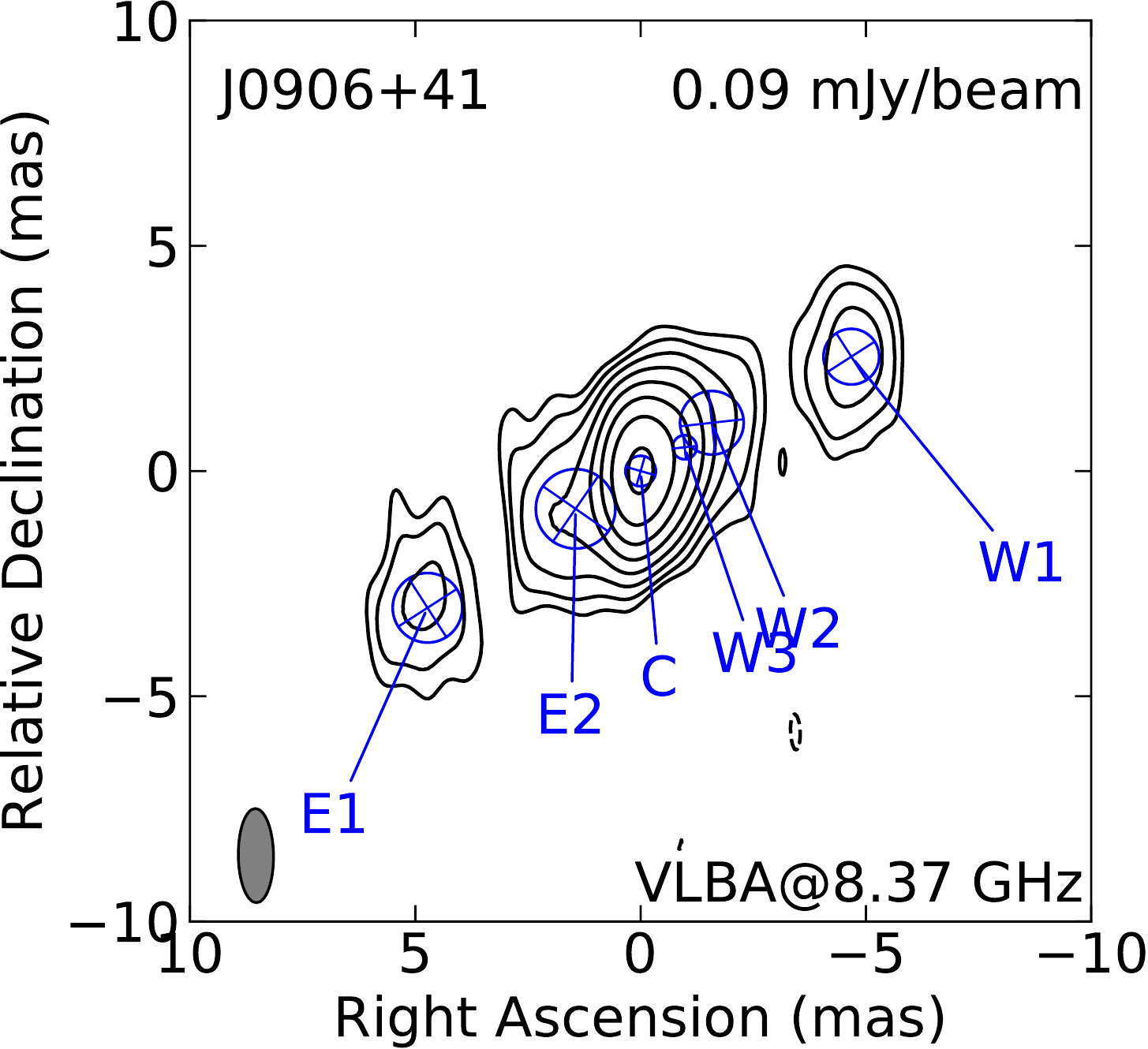}
  \includegraphics[height=3.8cm]{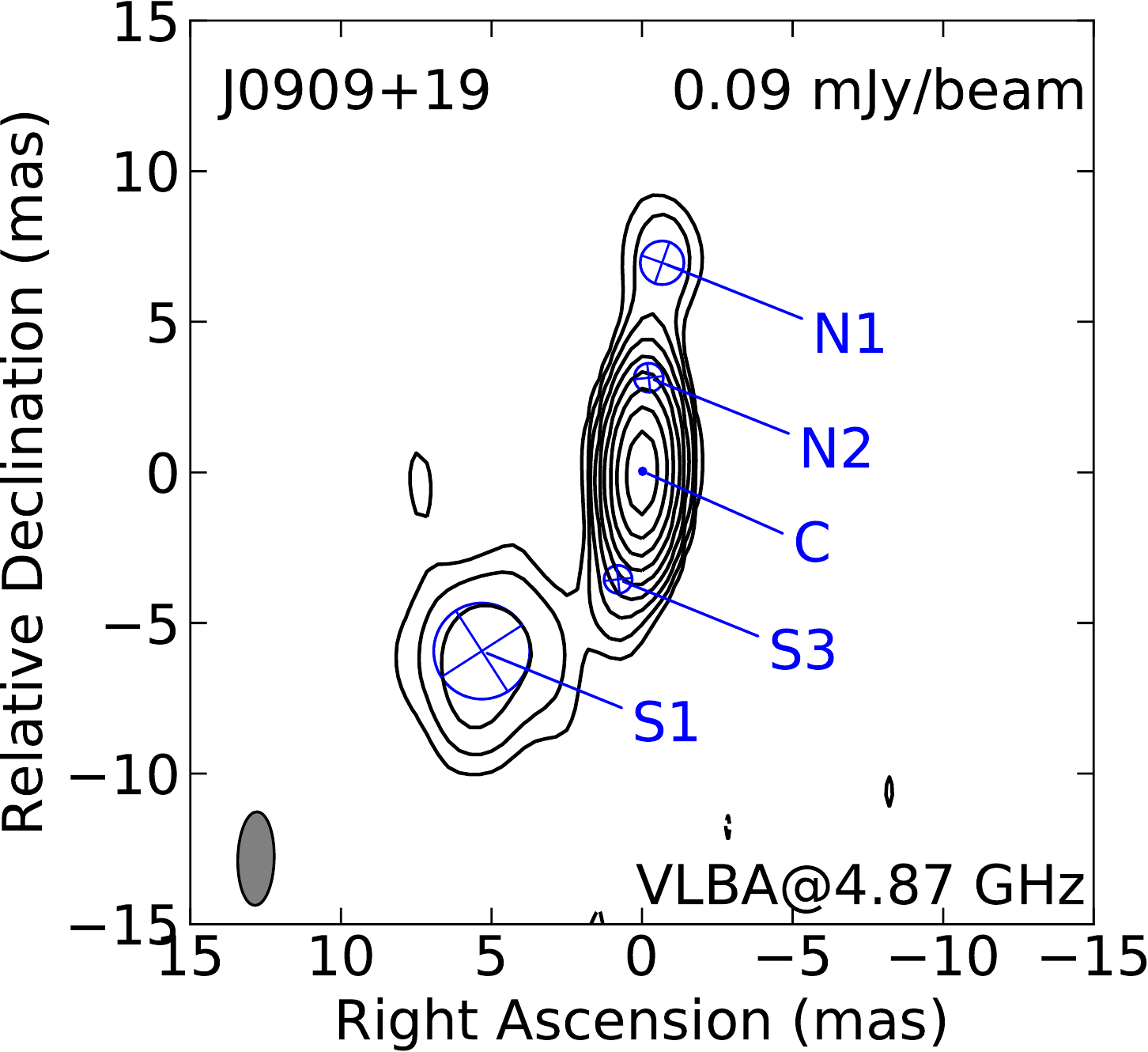}
 \includegraphics[height=3.8cm]{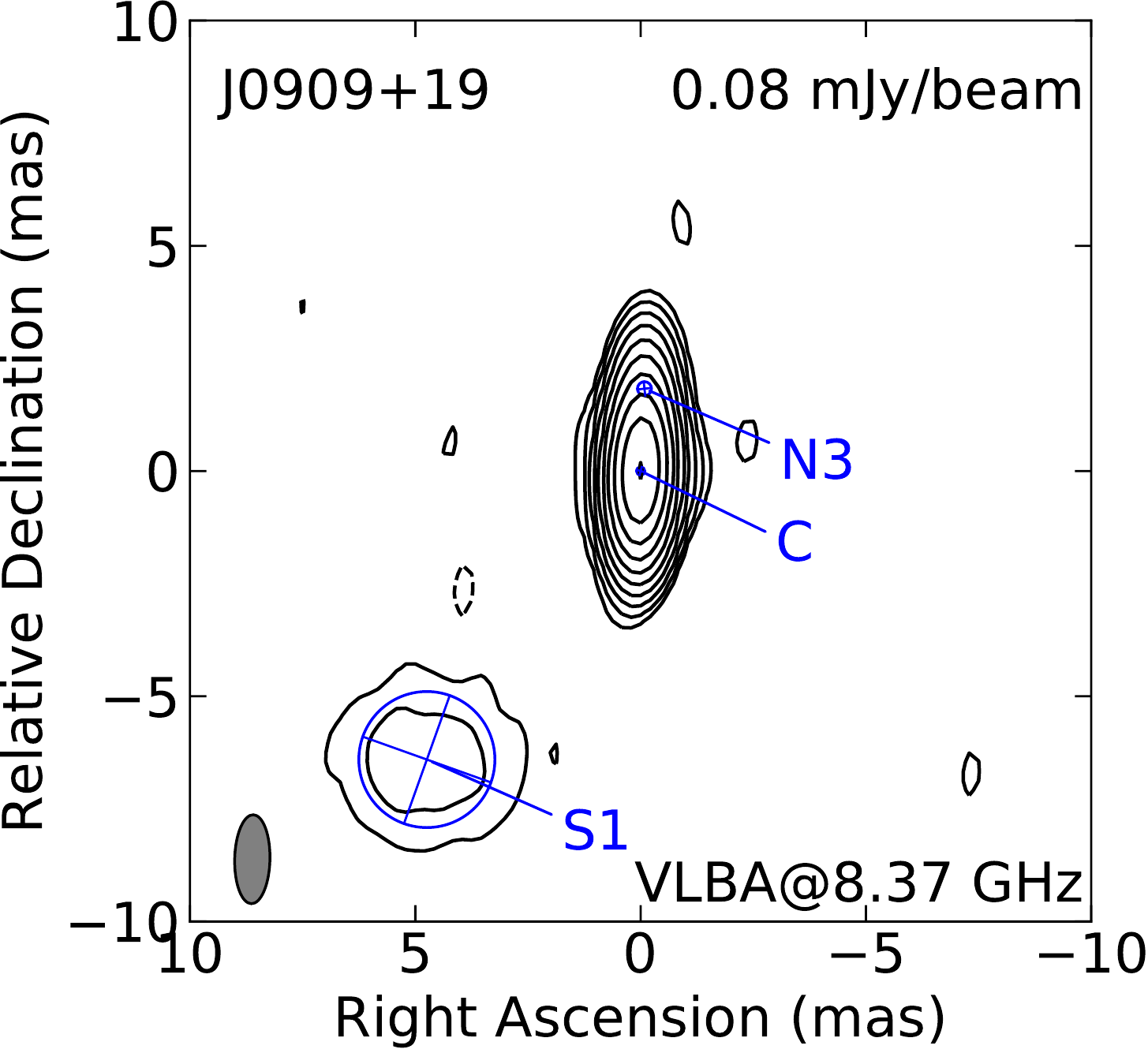}
\\
 \includegraphics[height=3.8cm]{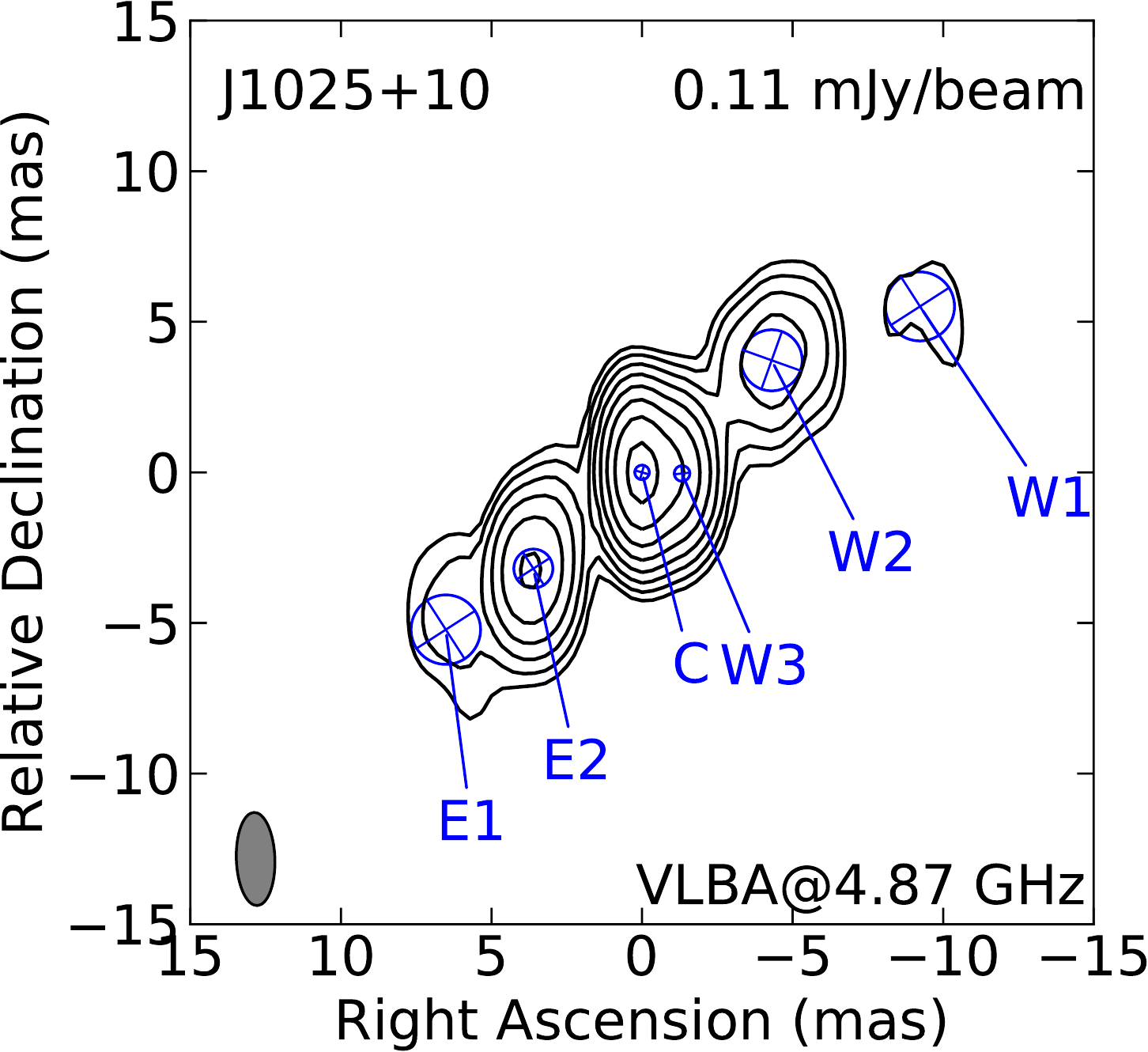}
 \includegraphics[height=3.8cm]{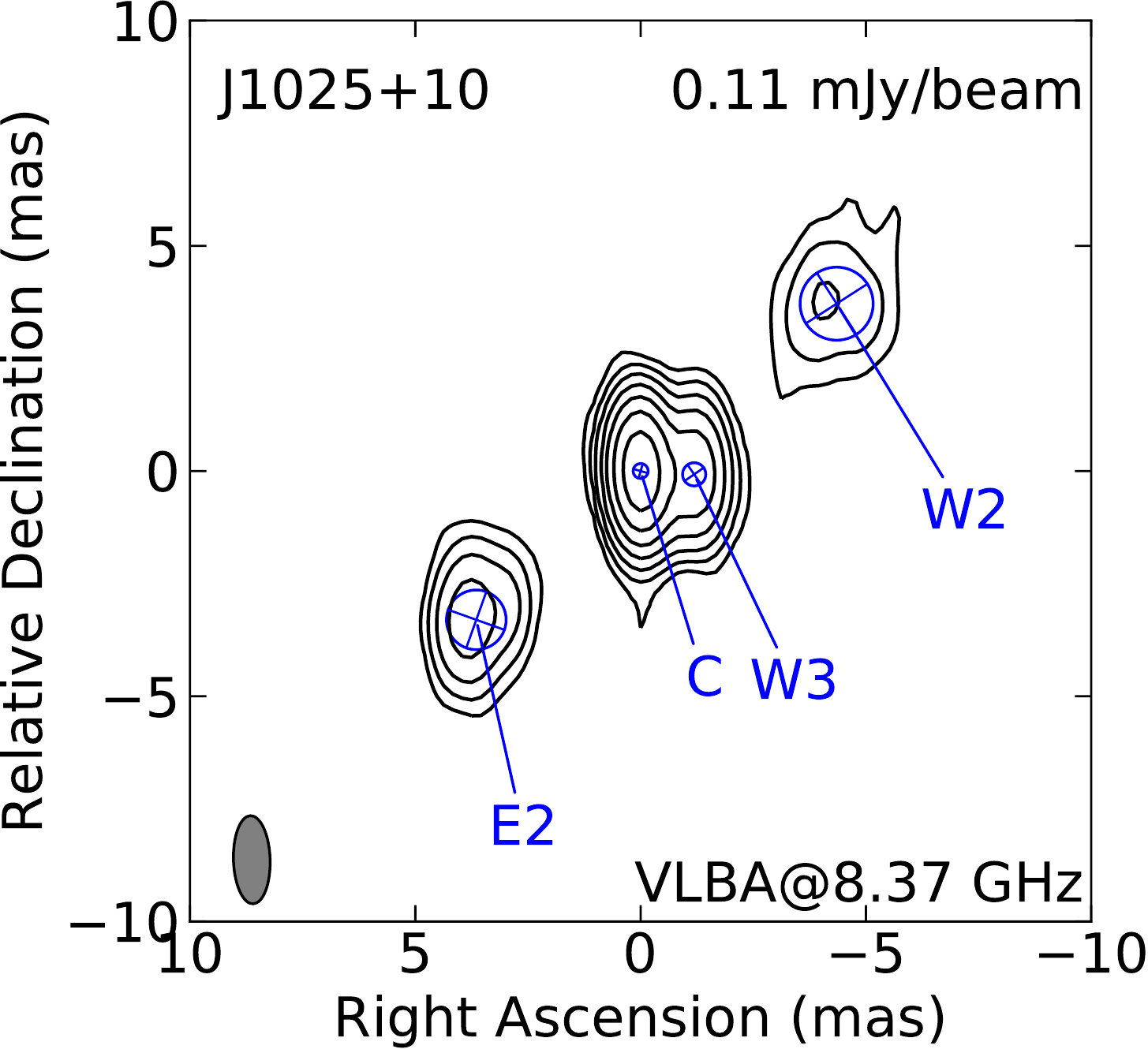}
 \includegraphics[height=3.8cm]{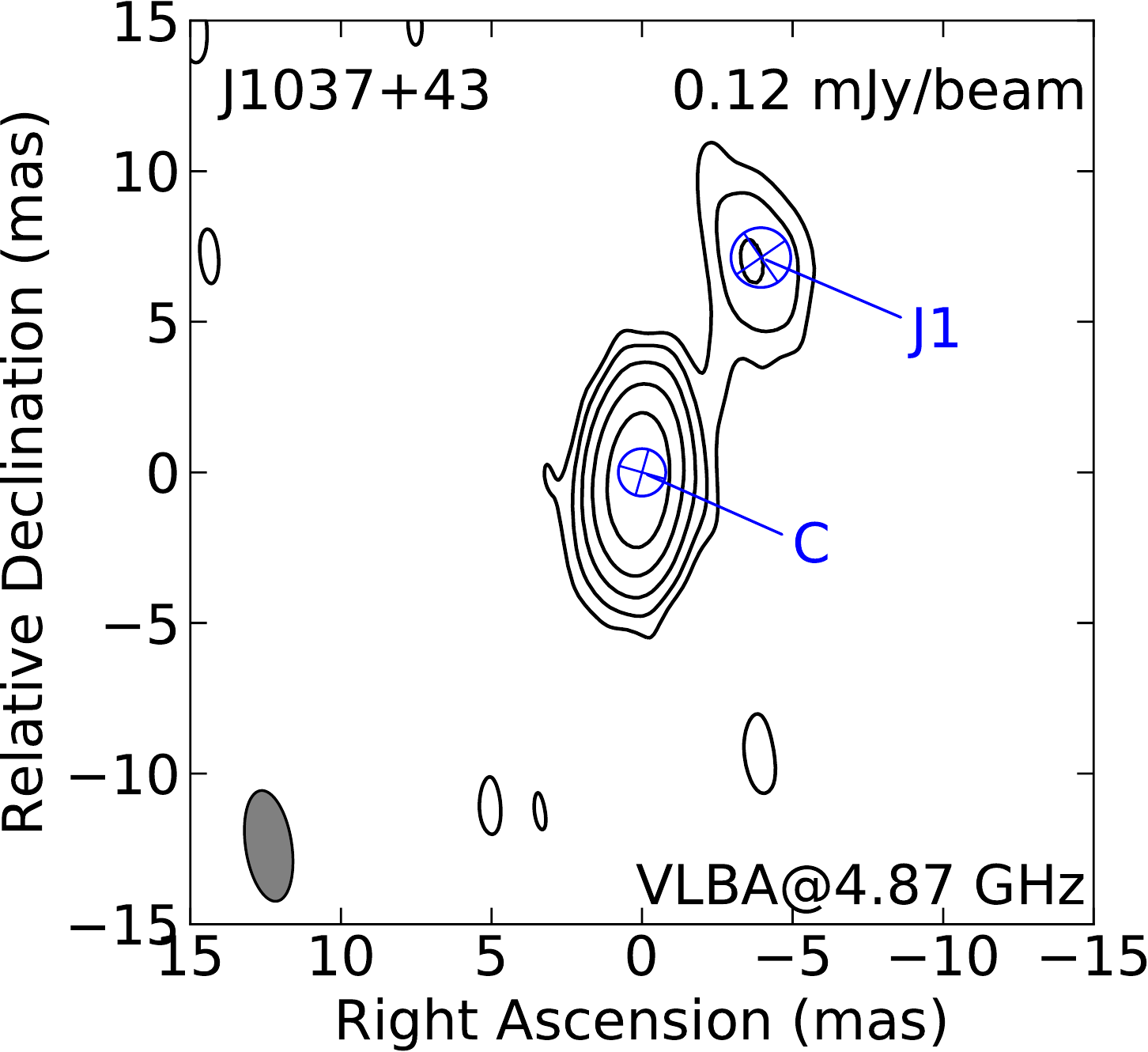}
 \includegraphics[height=3.8cm]{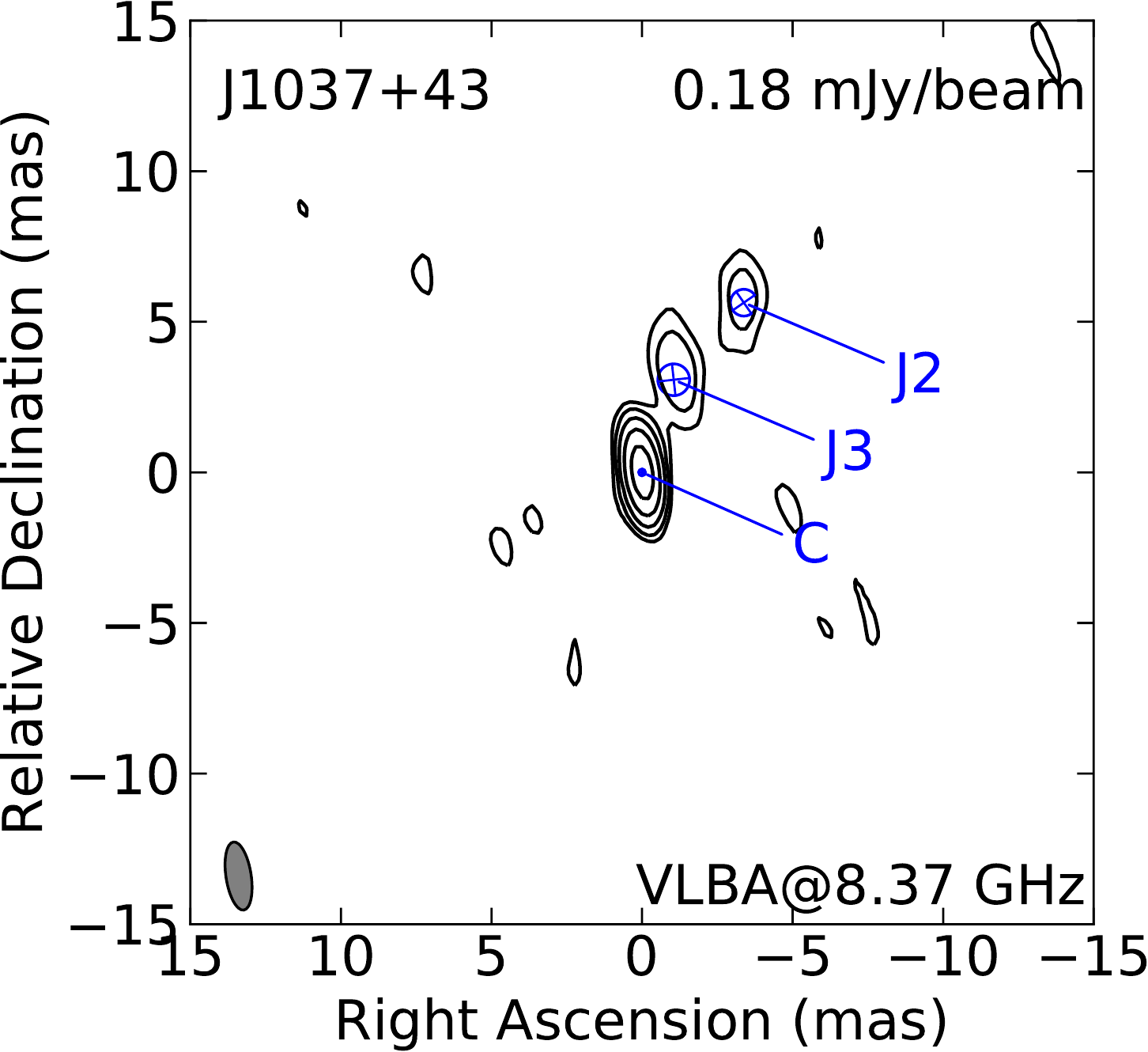}
\\
 \includegraphics[height=3.8cm]{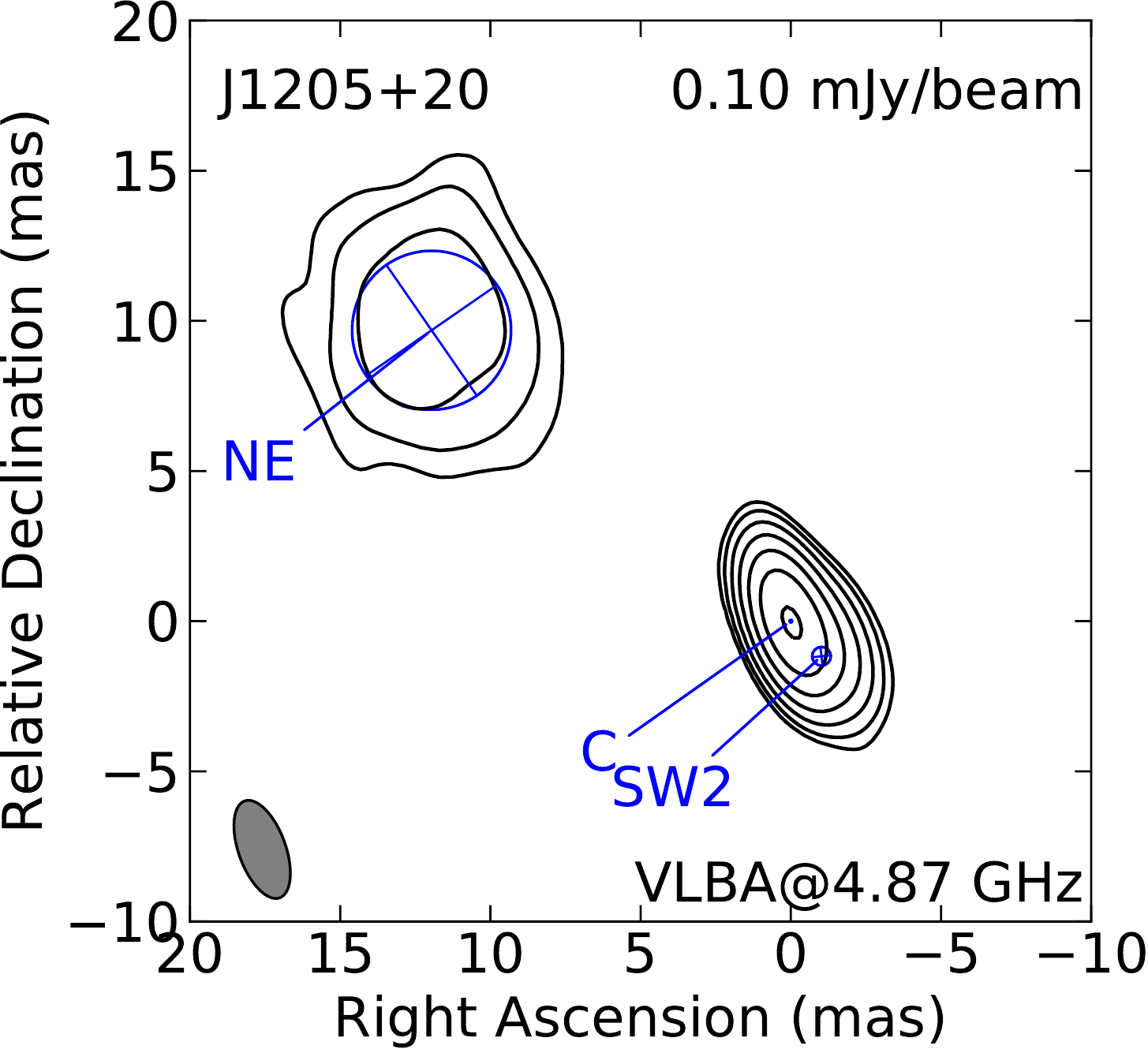}
 \includegraphics[height=3.8cm]{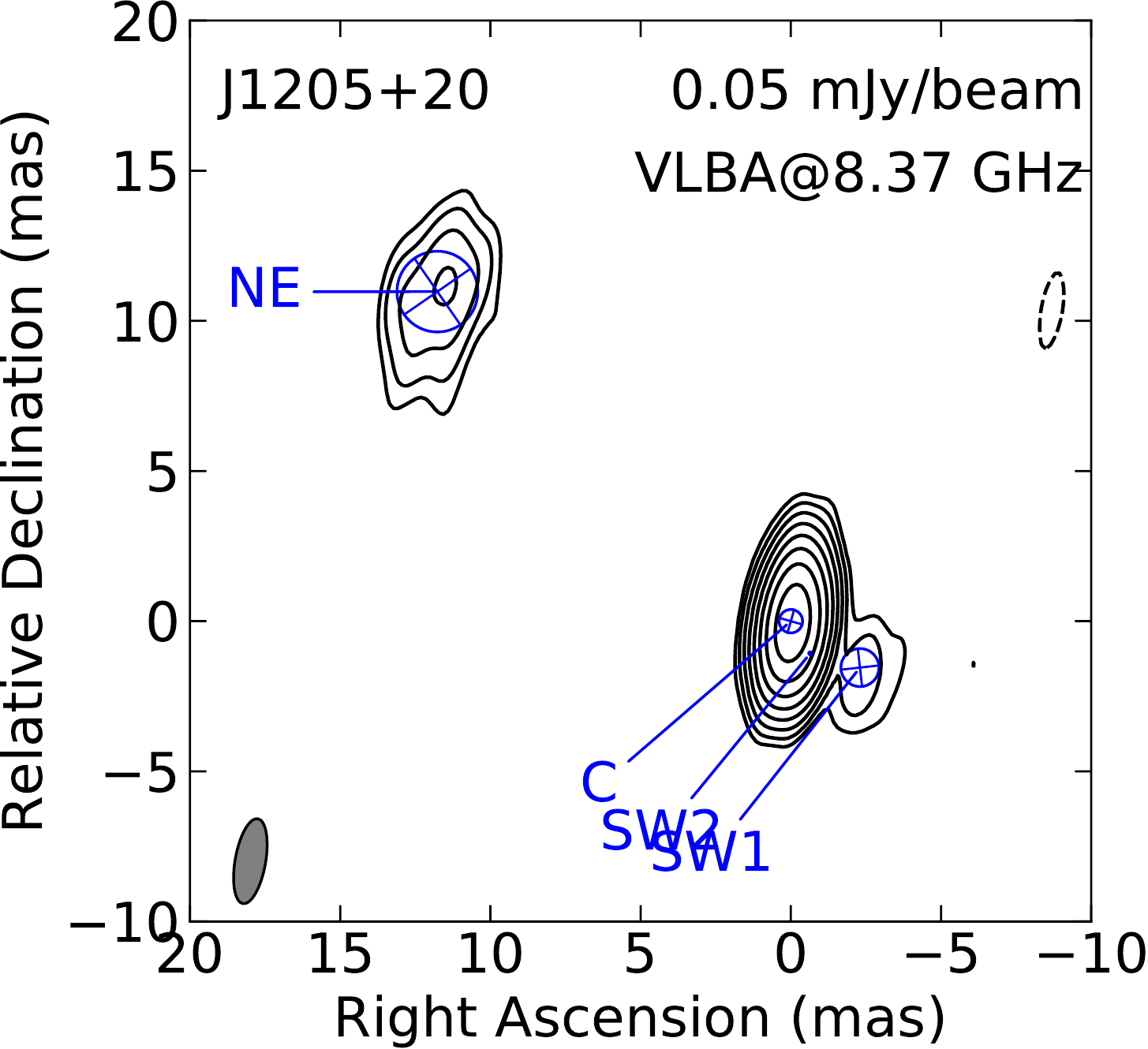}
 \includegraphics[height=3.8cm]{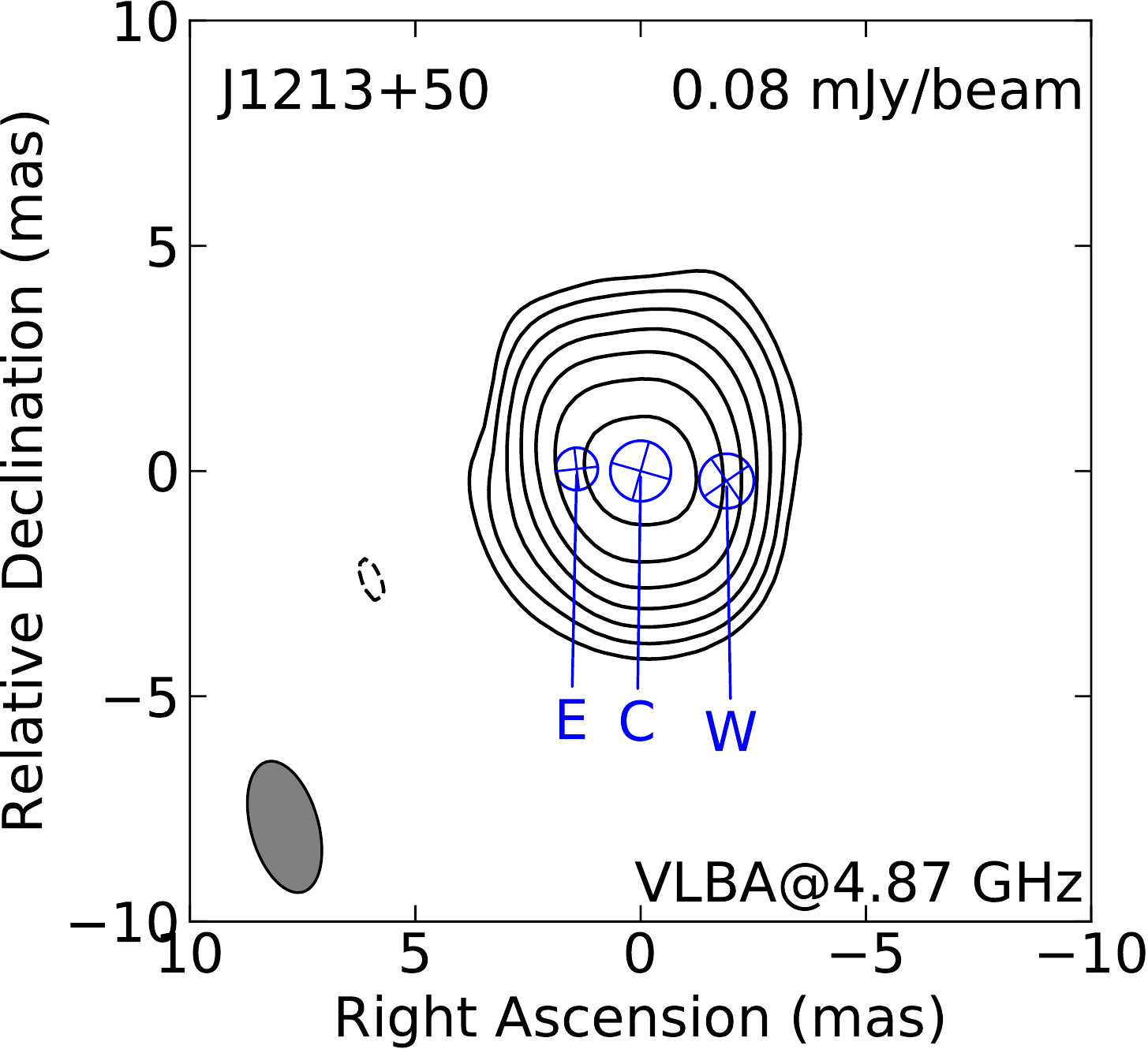}
 \includegraphics[height=3.8cm]{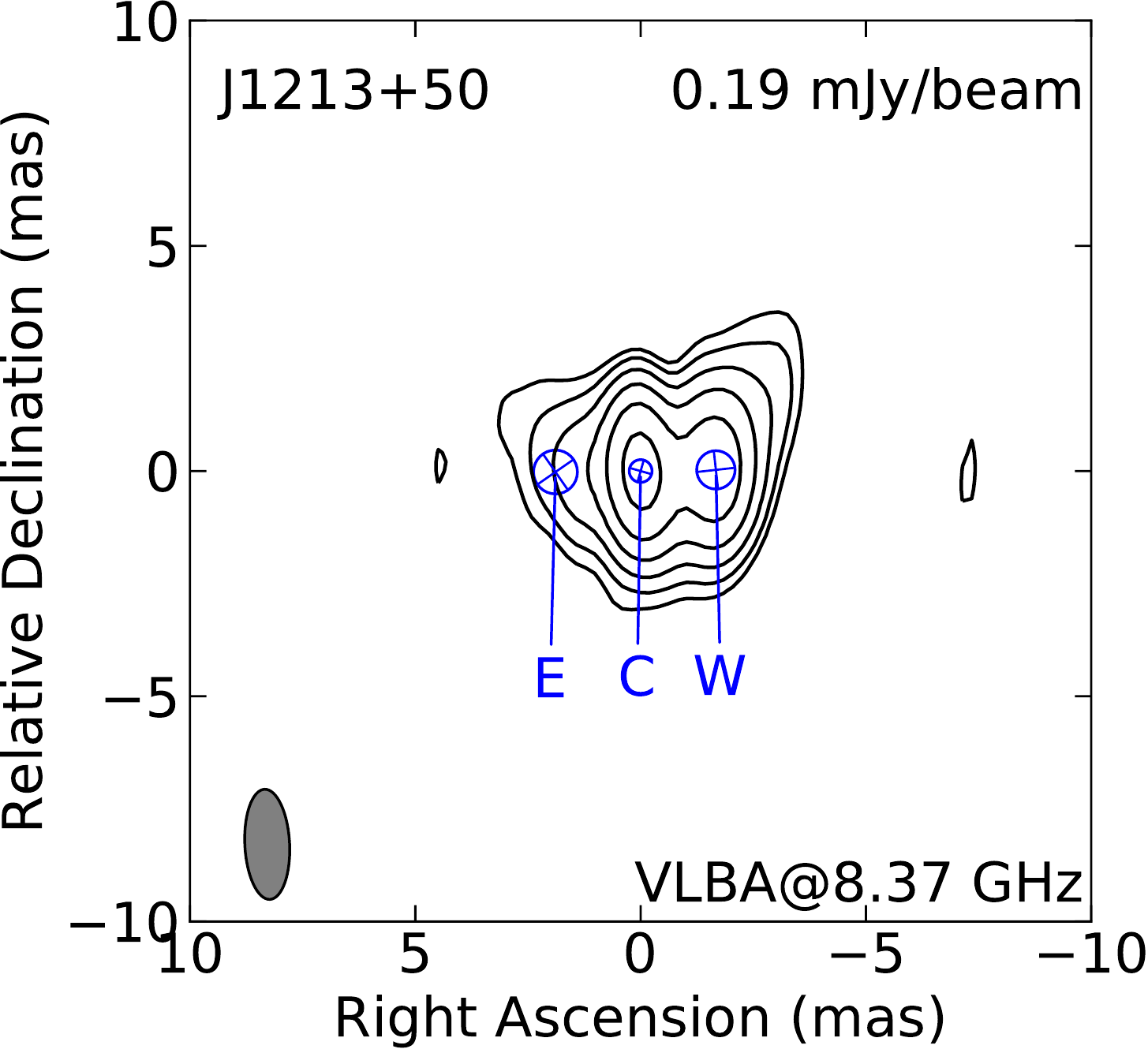}
 \\
 \includegraphics[height=3.8cm]{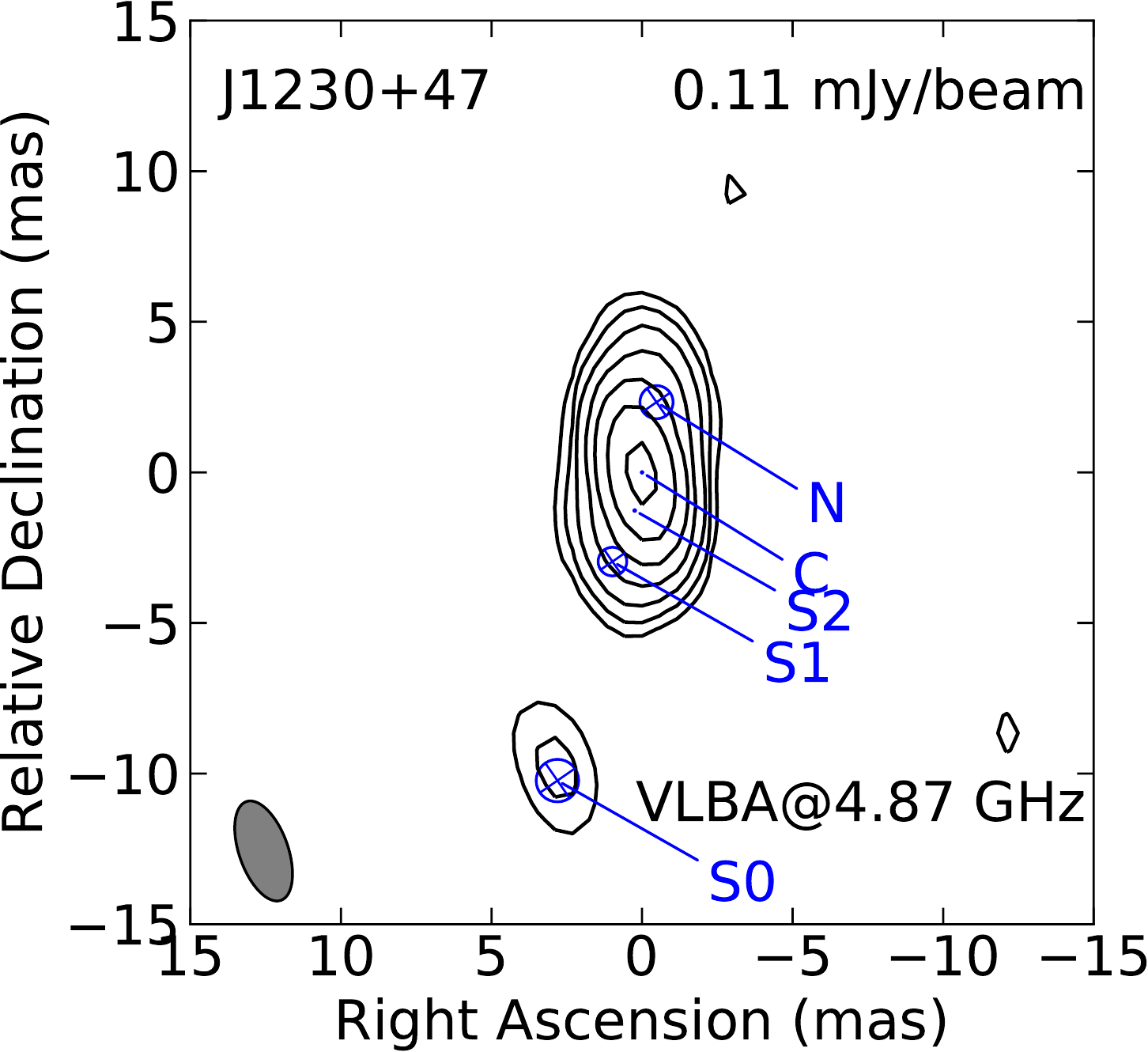}
 \includegraphics[height=3.8cm]{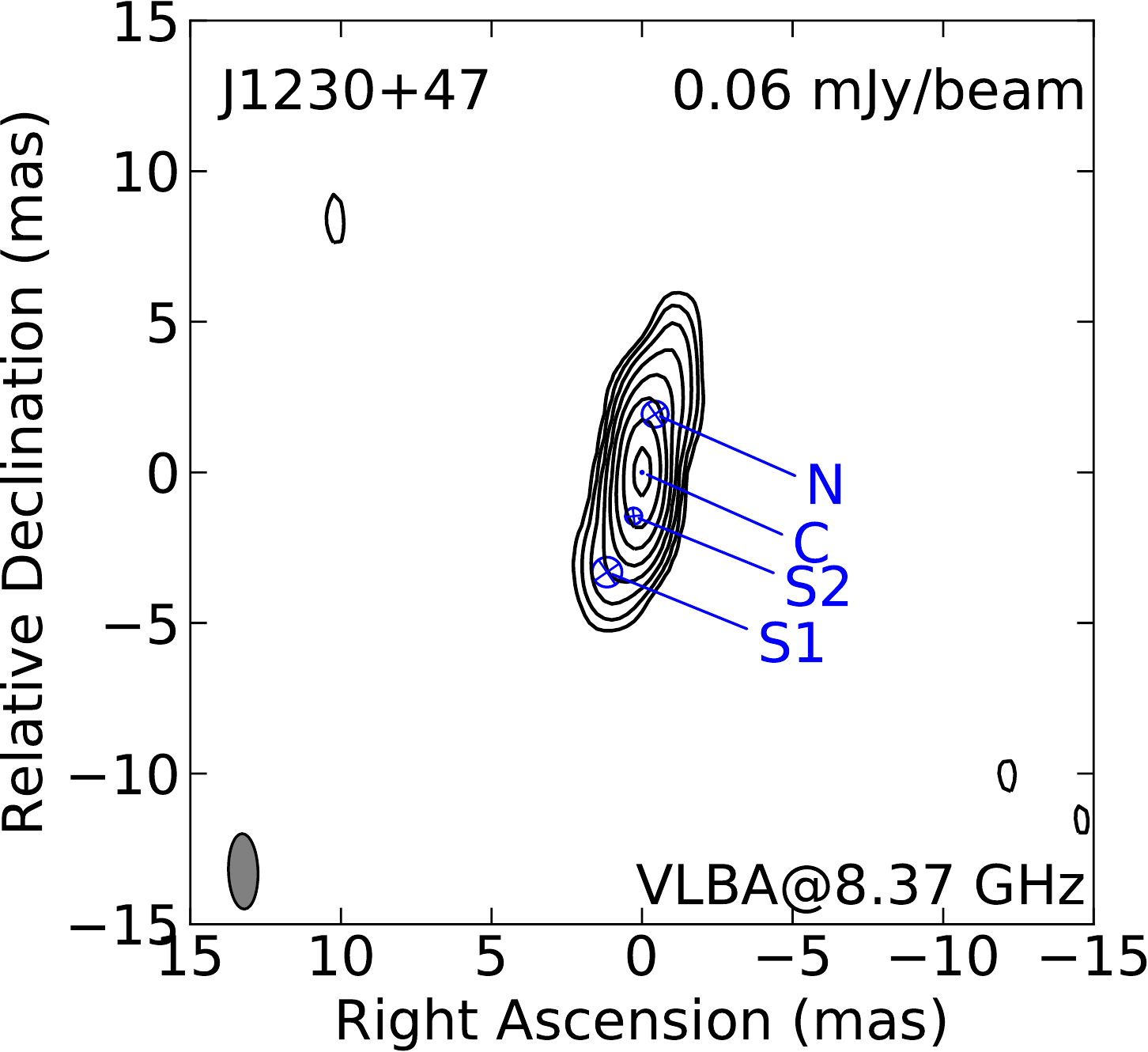}
 \includegraphics[height=3.8cm]{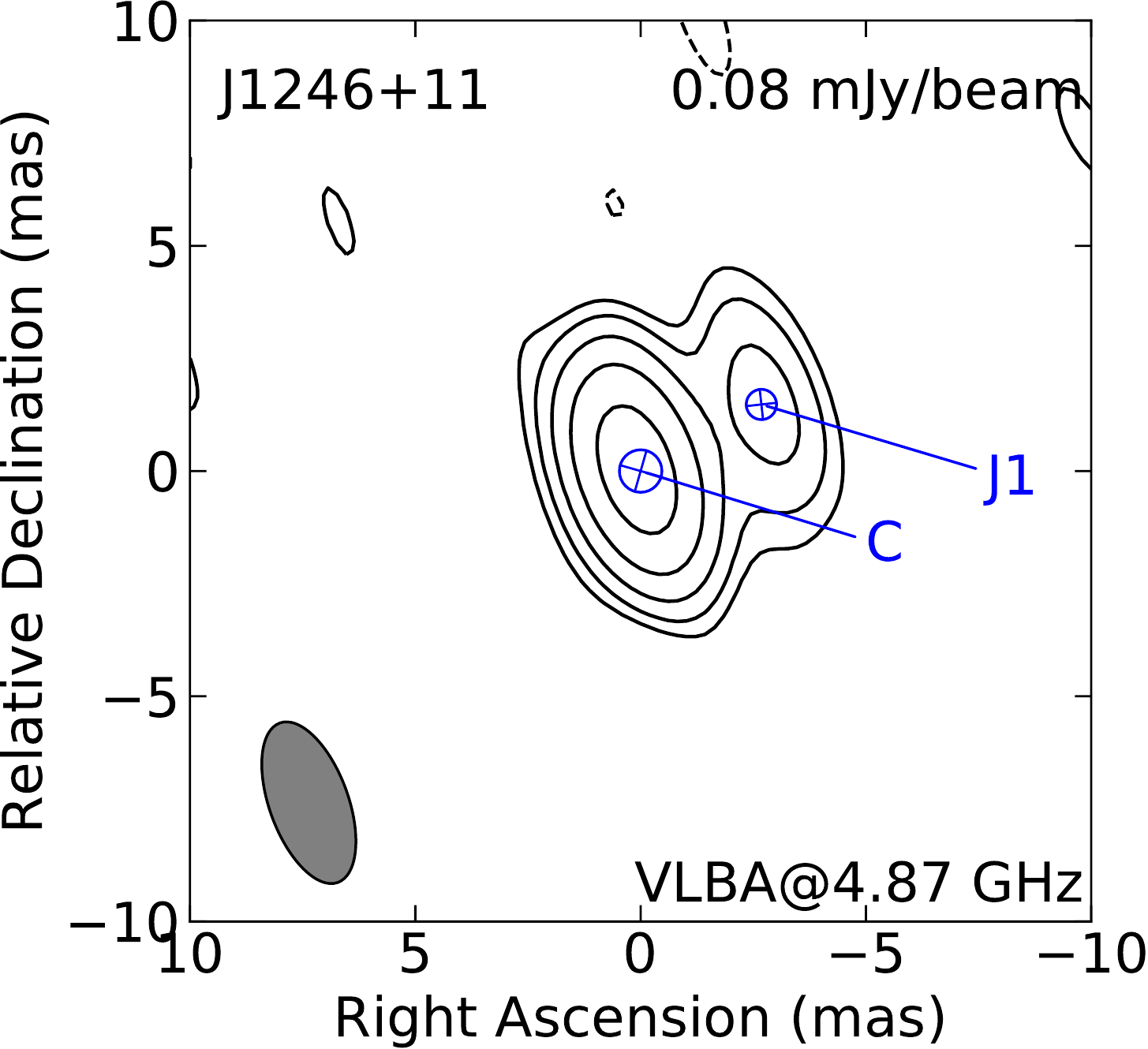}
 \includegraphics[height=3.8cm]{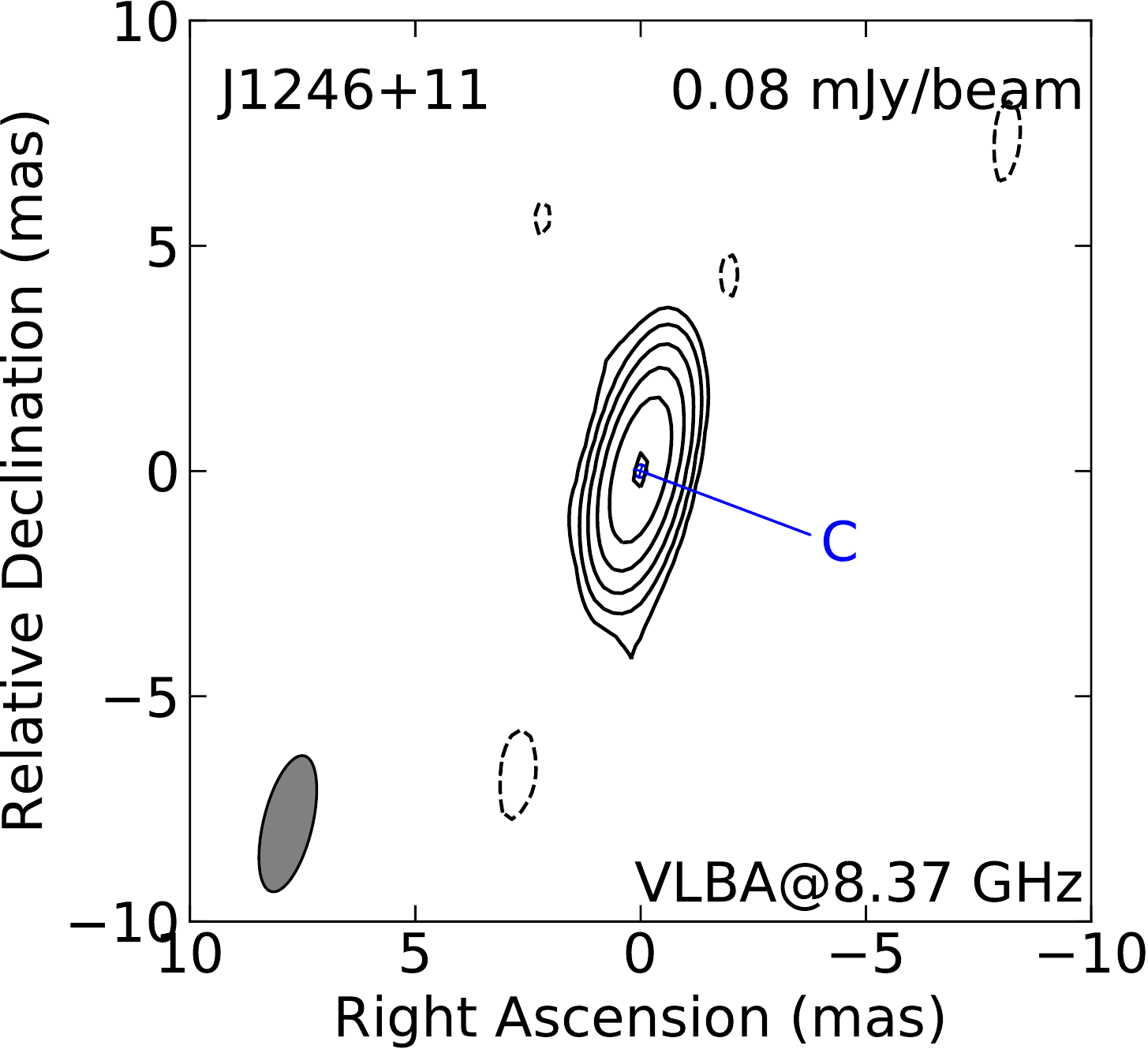}
\\
\caption{VLBA images at 5 GHz and 8 GHz of the eight FR 0s. The peak intensity and lowest contour level ($\sim 3\sigma$) are listed in Table \ref{image result}. The contours increase in steps of 2. The grey-colored ellipse in the bottom-left corner of each panel denotes the restoring beam.
} \label{image:VLBA}
\end{figure*}

The VLBA images of eight FR 0 sources are presented in Figure~\ref{image:VLBA}. The images parameters are referred to Tables~\ref{image result}.

\section{Description on Individual Sources}
\label{3.4}

Here, we report the detailed image structure and component identification based on the properties of morphology, spectral index and kinematics obtained from the previous results and our observations.

\subsection{J0906+4124}

The source J0906+4124 is associated with a host galaxy at redshift $z = 0.027$ \citep{1999PASP..111..438F}. 
The 150-MHz LOFAR image and 1.4-GHz NVSS image show a compact unresolved core with flux densities of $\sim$ 17.0 mJy and $\sim$ 51.8 mJy \citep{1998AJ....115.1693C,2020A&A...642A.107C}. 
Its total flux density spectrum is inverted  at low frequencies ($\rm \alpha^{150 MHz-1.4 GHz}$ = 0.55) \citep{2020A&A...642A.107C}.
The total flux density spectrum is flat ($\rm \alpha^{4.87-8.37}_{VLBA}$ = 0.22, $\rm \alpha^{4.99-8.42}_{EVN}$ = 0.26) between 5 and 8 GHz, suggesting an emission peak at $\nu > 5$ GHz.
The EVN images at 5 and 8 GHz  (first row of Figure \ref{fig:VLBI}) display a two-sided jet structure extending in the northwest-southeast direction on parsec scales, in agreement with the previous results \citep{2018ApJ...863..155C}.
Two new components (WN,EN) are detected at 5 GHz due to the high sensitivity of the EVN observation.
The eastern and western components show an almost aligned structure. 
The two hotspots show steep spectra: E1 ($\rm \alpha^{4.99-8.42}_{EVN}$ = $-$1.61) and W1 ($\rm \alpha^{4.99-8.42}_{EVN}$ = $-$1.25). 
The central component C is most compact, and has the highest flux density. It has an inverted spectrum ($\rm \alpha^{4.87-8.37}_{VLBA}$ = 0.59, $\rm \alpha^{4.99-8.42}_{EVN}$ = 0.54), which is identified as the core  (\textit{i.e.}, the jet base).

Using the core (component C) as a reference, the linear regression fitting gives the proper motion velocities of the four components along the jet axis: W2, 0.178$\pm$0.029 mas $\rm yr^{-1}$; E2, 0.208$\pm$0.029 mas $\rm yr^{-1}$; W1, 0.198$\pm$0.030 mas $\rm yr^{-1}$; and E1, 0.182$\pm$0.046 mas $\rm yr^{-1}$.
These angular velocities correspond to $0.32\,c$, $0.37\,c$, $0.35\,c$, $0.32\,c$ for W2, E2, W1, E1, respectively.
The kinematic ages of the jet components derived from the proper motions are 11$\pm$2 years (W2), 12$\pm$2 years (E2), 31$\pm$4 years (W1) and 33$\pm$7 years (E1), respectively.

\subsection{J0909+1928}

The source J0909+1928 is associated with a galaxy at redshift $z = 0.028$\footnote{SDSS Data Release 6 as obtained 2007 September 17 from \url{http://www.sdss.org/dr6/products/catalogs/index.html}}.
The source is unresolved at 150-MHz LOFAR observation and 1.4-GHz NVSS observation \citep{1998AJ....115.1693C,2020A&A...642A.107C}.
Our 5- and 8- GHz EVN images (second row of Figure \ref{fig:VLBI}) exhibit an asymmetric two-sided structure at mas resolution, with one jet extending to the north and the counter jet apparently bending eastward at $\sim$5 mas (2.5 pc), suggesting a local interaction between the jet and dense interstellar medium (ISM) in the narrow-line region of the host galaxy.
Such jet–ISM interaction could deflect the jet flow and change its direction \citep{2018ApJ...864..118K}. 
The source has a complex radio spectrum: a flat spectrum with a spectral index of $\rm \alpha$ = $-$0.42 between 150 MHz and 1.4 GHz \citep{2020A&A...642A.107C}, an inverted  spectrum with $\rm \alpha$ = 0.34 between 1.4 GHz and 5 GHz \citep{2019A&A...631A.176C}, and a flat spectrum with $\rm \alpha^{4.99-8.42}_{EVN}$ = $-$0.18.
The central component C shows the highest compactness, the highest flux density and the flat spectrum ($\rm \alpha^{4.87-8.37}_{VLBA}$ = 0.28, $\rm \alpha^{4.99-8.42}_{EVN}$ = $-$0.32) and is identified as the core. 
S2, S1 and N1 are detected at 5 GHz, in agreement with the previous result \citep{2018ApJ...863..155C}. 
Four new components are detected in our new images: N3 at ~2 mas away from the core in a position angle of ~$\rm -4\degr$, N2 at ~3.6 mas in P.A. = $\rm -6\degr$, S4 at ~1.9 mas in P.A. = 175$\degr$ and S3 at ~3.6 mas in P.A. = 165$\degr$.

Using the core as a reference, a linear regression fit gives the proper motion velocities of three components: N1 (along the jet axis) 0.219$\pm$0.017 mas $\rm yr^{-1}$, S1 (almost no motion and position angle changing) $-$0.005$\pm$0.054 mas $\rm yr^{-1}$, and  S2 ( non-radial motion) 0.104$\pm$0.037 mas $\rm yr^{-1}$.
These angular velocities correspond to $0.39\,c$, $-0.08\,c$, $0.18\,c$ for N1, S1, S2 respectively.

\subsection{J1025+1022}

The source J1025+1022 is 
at a redshift of 0.046 \citep{2012ApJS..199...26H}.
The source is unresolved at 150-MHz LOFAR observation and 1.4-GHz NVSS observation \citep{1998AJ....115.1693C,2020A&A...642A.107C}.
Our observation shows a rich two-sided structure with the jet extending roughly in the northwest-southeast direction on parsec scales. 
The inner jet follows a straight trajectory along the east-west direction (~$\rm \pm90\degr$) until 3 mas, but the outer jet has a position angle P.A.= $\rm \sim -50 \degr$ (or 130$\degr$) and extends from northwest to southeast.
The high-resolution VLA observations reveal a slightly convex radio spectrum: the flux density at 4.5 GHz is typical $\sim$20 per cent higher than that at 1.5 GHz and 15 per cent higher than that at 7.5 GHz \citep{2019MNRAS.482.2294B}.
We calculated the total flux density spectrum $\alpha^{4.99-8.42}_{EVN}$ = $-$0.13, with a flat radio spectrum.
The central component C shows the most compact size, the highest flux density and a flat spectrum ($\rm \alpha^{4.87-8.37}_{VLBA}$ = 0.49, $\rm \alpha^{4.99-8.42}_{EVN}$ = 0.17) and is identified as the core. 
Five components (W1 -- W3, E1, E2) are detected and denoted in a manner consistent with the previous results, identifying the components with the same position relative to the central core as corresponding components \citep{2018ApJ...863..155C}. 
A new component E3 is identified due to the high EVN resolution in the east-west direction. 
J1025+1022 could be a restart of the central power source at a different jet axis direction, leading to an X-shaped source.
The possible adiabatic expansion of the components E1 and W1 result in the eventual dissipation and disappearance at 8 GHz \citep{2012ApJ...760...77A}. Alternatively, the apparent X-shaped morphology is a projection of a helical jet.

Using the core as a reference, a linear regression fit gives the proper motion velocities of three components: for W3 0.106$\pm$0.024 mas $\rm yr^{-1}$, for W2 0.173$\pm$0.041 mas $\rm yr^{-1}$, and  for E2 0.045$\pm$0.051 mas $\rm yr^{-1}$.
These angular velocities correspond to $0.31\,c$, $0.51\,c$, and $0.13\,c$ for W3, W2, and E2, respectively.

\subsection{J1037+4335}

The source J1037+4335 is associated with a quasar at redshift $z = 0.025$ \citep{1999PASP..111..438F}. 
The 150-MHz LOFAR image and 1.4-GHz NVSS image only detect a compact core with a flux density of $\sim$ 260.9 mJy and $\sim$ 132.2 mJy \citep{1998AJ....115.1693C,2020A&A...642A.107C}. 
J1037$+$4335 was resolved into two isolated components, separated by 19.7 pc in the previous study \citep{2018ApJ...863..155C}.
Our high resolution observations first show a faint one-sided jet extending to the north-west (see  Fig. \ref{fig:VLBI}). 
The core component shows the most compact size and the highest flux.
Three new components (J1 -- J3) are detected. 
The SW component is not detected in our EVN and VLBA observations at 5 and 8 GHz, although our sensitivity is higher (1$\sigma$=0.12 mJy beam$^{-1}$ at 8 GHz) compared to the previous image \citep{2018ApJ...863..155C}. 
Considering the large distance from the core and almost being perpendicular to the jet, we suggest that the SW component might be an artifact.
The total flux density spectrum is steep ($\rm \alpha^{0.15-1.4}$ = $-$0.30, $\rm \alpha^{1.4-5}$ = $-$0.56) at low frequencies \citep{2020A&A...642A.107C}.
The total flux density spectrum is also steep ($\rm \alpha^{4.87-8.37}_{VLBA}$ = $-$0.58, $\rm \alpha^{4.99-8.42}_{EVN}$ = $-$0.72) between 5 and 8 GHz with the quasi-simultaneous two frequency observations, suggesting an emission peak at $\nu$ < 1.4 GHz.
The flux density variation is less than 10 per cent from the three epochs of 8.4 GHz VLBI data.

The source is unresolved in the previous data, so we used the data from only two new epochs to determine the proper motion.
Using the core as a reference, a linear regression fit yields the proper motion velocities of the three components: for J3 0.178$\pm$0.465 mas $\rm yr^{-1}$, for J2 0.249$\pm$0.520 mas $\rm yr^{-1}$, and  for J1 0.783$\pm$0.991 mas $\rm yr^{-1}$.
These angular velocities correspond to $0.30\,c$, $0.42\,c$, and $1.32\,c$ for J3, J2, and J1 respectively.
We note that we measured the separation velocities between two epochs on a small time baseline, so the uncertainties in the estimated values are high.
A longer time span is needed to better constrain the proper motion.

\subsection{J1205+2031}

The source J1205+2031 is associated with the galaxy NGC~4093 at redshift $z = 0.024$ \citep{2004ApJ...607..202M}.
The source is unresolved in 150-MHz LOFAR observation and 1.4-GHz NVSS observation \citep{1998AJ....115.1693C,2020A&A...642A.107C}.
Our 5- and 8- GHz EVN images (last row of Figure \ref{fig:VLBI}) show a two-sided structure with jet extending in the northeast-southwest direction. 
The source has a flat spectrum with a spectral index of $\rm \alpha$ = $-$0.01 between 150 MHz and 1.4 GHz \citep{2020A&A...642A.107C}, a spectrum with $\rm \alpha$ = $-$0.44 between 1.4 GHz and 5 GHz \citep{2019A&A...631A.176C}, and a flat spectrum with $\rm \alpha^{4.99-8.42}_{EVN}$ = $-$0.13.
The central component C shows the most compact morphology, highest flux density and a flat spectrum ($\rm \alpha^{4.87-8.37}_{VLBA}$ = 0.49, $\rm \alpha^{4.99-8.42}_{EVN}$ = 0.22) and is identified as the core. 
Three components (NE, SW1, SW2) are detected and denoted in a manner consistent with the previous results \citep{2018ApJ...863..155C}. Components with similar positions  are identified as counterparts.

Using the core as a reference, a linear regression fit gives the proper motion velocities for three components: 0.047$\pm$0.009 mas $\rm yr^{-1}$ for SW2, $-$0.028$\pm$0.021 mas $\rm yr^{-1}$ for SW1, and 0.130$\pm$0.114 mas $\rm yr^{-1}$  for NE.
These angular velocities correspond to $0.07\,c$, $0.04\,c$, and $0.21\,c$ for SW2, SW1, and NE, respectively.
The low radial velocity obtained from proper motion measurements suggests an intrinsic slow jet.

\subsection{J1213+5044}

The source J1213+5044 is associated with the galaxy NGC~4187 at redshift $z = 0.031$ \citep{1999PASP..111..438F}.
The 150-MHz LOFAR image and 1.4-GHz NVSS image only detected a compact core with a flux density of $\sim$ 178.1 mJy and $\sim$ 96.5 mJy, respectively \citep{1998AJ....115.1693C,2020A&A...642A.107C}. 
J1213+5044 shows a symmetric two-sided structure with jet extending in the west-east direction on parsec scales, in agreement with the previous results \citep{2018ApJ...863..155C}. 
The total flux density spectrum is steep ($\rm \alpha^{0.15-1.4}$ = $-$0.27, $\rm \alpha^{1.4-5}$ = $-$1.35) at low frequencies \citep{2020A&A...642A.107C}.
The total EVN flux density spectrum is also steep ($\rm \alpha^{4.99-8.42}_{EVN}$ = $-$0.61) between 5 and 8 GHz, indicating an emission peak at $\nu < $150 MHz.
However, the VLBA observation shows a flat spectrum ($\rm \alpha^{4.99-8.42}_{VLBA}$ = 0.19). The change of spectral index could be related to the change of the emission structure in the core region.
Further multi-frequency observations are needed to confirm the spectrum change.
The 8.4 GHz VLBI data from the three epochs show a flux variability of about 25 per cent.
The central component C shows the most compact morphology, highest flux density and flat spectrum ($\rm \alpha^{4.87-8.37}_{VLBA}$ = 0.18, $\rm \alpha^{4.99-8.42}_{EVN}$ = 0.48) and is identified as the core.
Two components (E, W) are detected and denoted in a manner consistent with the previous results. We identify the components with a similar position relative to the central core as the corresponding component.
The observations show  that component E has a steep spectrum  ($\rm \alpha^{4.99-8.42}_{EVN}$ = $-$1.8) and the W1 component has an inverted spectrum ($\rm \alpha^{4.99-8.42}_{EVN}$ = 1.01).
Interestingly, the west component (W) has an optically thick spectral feature with $\rm \alpha$ of +1.01$\pm$0.36. One possibility is that W is strongly absorbed by the intervening ionised gas.
Using the core as a reference, a linear regression fit gives the proper motion velocities of the two components: 0.118$\pm$0.033 mas $\rm yr^{-1}$ for W  and 0.044$\pm$0.041 mas $\rm yr^{-1}$ for E.
These angular velocities correspond to $0.24\,c$ and $0.18\,c$ for W and E, respectively.

\subsection{J1230+4700}

The source J1230+4700 is associated with a galaxy at redshift $z = 0.039$ \citep{2004ApJ...607..202M}.
The source is unresolved in 150-MHz LOFAR observation and 1.4-GHz NVSS observation \citep{1998AJ....115.1693C,2020A&A...642A.107C}.
Our 5- and 8- GHz EVN images show a two-sided structure with jet extending in the northwest-southeast direction, consistent with the previous results \citep{2018ApJ...863..155C}. 
The source has a flat spectrum with a spectral index of $\rm \alpha$ = $-$0.14 between 150 MHz and 1.4 GHz \citep{2020A&A...642A.107C}, a spectrum with $\rm \alpha$ = $-$0.20 between 1.4 GHz and 5 GHz \citep{2019A&A...631A.176C}, and a flat spectrum with $\rm \alpha^{4.99-8.42}_{EVN}$ = $-$0.21.
Although the spectrum of the central component C shows a slightly steeper spectrum ($\rm \alpha^{4.87-8.37}_{VLBA}$ = 0.59, $\rm \alpha^{4.99-8.42}_{EVN}$ = $-$0.58), the compact size and highest flux density are identified as the core. 
Three previously identified components (N, S2, S1) are detected and have  similar positions relative to the central core. 
One new component S0 is detected at 10.6 mas at P.A. = 164$\degr$ at 5 GHz.

Using the core as a reference, the linear regression fitting gives the proper motion velocities of three components: 0.073$\pm$0.014 mas $\rm yr^{-1}$ for S2, $-$0.088$\pm$0.022 mas $\rm yr^{-1}$ for N, and 0.064$\pm$0.027 mas $\rm yr^{-1}$  for S1.
These angular velocities correspond to $0.22\,c$, $0.16\,c$, and $0.21\,c$ for S2, N and S1, respectively.
The low radial velocity obtained from proper motion measurements suggests an intrinsic low jet speed.

\subsection{J1246+1153}

The source J1246+1153 is at redshift $z = 0.047$ \citep{1985AJ.....90.1681B}.
J1246+1153 is unresolved in 150-MHz LOFAR observation and 1.4-GHz NVSS observation \citep{1998AJ....115.1693C,2020A&A...642A.107C}.
Our high resolution VLBI observations show a core-jet structure with the jet extending to the northwest direction, consistent with the previous results \citep{2018ApJ...863..155C}. 
The source has a flat spectrum with a spectral index of $\rm \alpha$ = 0.24 between 150 MHz and 1.4 GHz \citep{2020A&A...642A.107C}, a spectrum with $\rm \alpha$ = $-$0.34 between 1.4 GHz and 5 GHz \citep{2019A&A...631A.176C}, and a flat spectrum with $\rm \alpha^{4.99-8.42}_{EVN}$ = 0.12.
The component C shows the most compact size, highest flux density, and is located upstream of the jet base. It has a flat spectrum ($\rm \alpha^{4.87-8.37}_{VLBA}$ = 0.40, $\rm \alpha^{4.99-8.42}_{EVN}$ = 0.58) and is identified as the core. 
J1 is detected and denoted in a manner consistent with the previous results.
A new component J0 is detected at 5.5 mas in P.A. = $-$65$\degr$.
The flux density derived from the three epochs of 8.4 GHz VLBI data shows a very large variation.

Using the core as a reference, the linear regression fitting gives the proper motion velocities 0.128$\pm$0.101 mas $\rm yr^{-1}$ for J1, corresponding to an angular velocity of $0.39\,c$.

\begin{table*}
\caption{Model fitting parameters.}\label{modelfit}
%
\medskip
\centering
\renewcommand{\tabcolsep}{3.5mm}
\begin{tabular}{ccccccccc}
\hline\hline
Name & Array & $\nu$ & Comp. & $\rm S_{int}$ &  $\rm S_{peak}$      & $\rm R $  & $\rm P.A.$    & $\rm d$      \\
     &       & (GHz) &       &      (mJy) 	 & (mJy beam$^{-1}$) &    (mas)  &   $\rm \degr$ &  (mas)       \\
\hline
J0906+4124 & VLBA & 4.87 & C  & 38.63$\pm$3.86  & 36.21$\pm$3.62    &      ...          &   ...     & 0.55$\pm$0.08 \\
           &      &      & W3 & 9.12$\pm$0.91   & 27.33$\pm$2.73    & 0.96$\pm$0.04     & $-$59.17  & 0.18$\pm$0.03 \\
           &      &      & W2 & 8.02$\pm$0.80   & 13.31$\pm$0.13    & 1.75$\pm$0.21     & $-$58.10  & 1.04$\pm$0.16 \\
           &      &      & E2 & 8.25$\pm$0.83   & 9.62$\pm$0.96     & 1.57$\pm$0.27     & 120.18    & 1.37$\pm$0.25 \\
           &      &      & W1 & 4.59$\pm$0.46   & 2.43$\pm$0.24     & 5.40$\pm$0.31     & $-$65.40  & 1.56$\pm$0.23 \\
           &      &      & E1 & 3.91$\pm$0.39   & 2.14$\pm$0.21     & 5.53$\pm$0.33     & 119.92    & 1.63$\pm$0.24 \\
           & EVN  & 4.99 & C  & 48.72$\pm$4.87  & 54.71$\pm$5.47    &     ...           &   ...     & 0.21$\pm$0.03 \\
           &      &      & W3 & 9.33$\pm$1.53   & 14.02$\pm$1.40    & 1.12$\pm$0.05     & $-$61.35  & 0.27$\pm$0.04 \\
           &      &      & E2 & 8.07$\pm$0.81   & 22.11$\pm$2.21    & 1.59$\pm$0.11     & 118.15    & 0.57$\pm$0.09 \\
           &      &      & W2 & 7.27$\pm$0.73   & 10.35$\pm$1.04    & 2.03$\pm$0.27     & $-$63.88  & 1.37$\pm$0.21 \\
           &      &      & EN & 1.38$\pm$0.14   & 3.93$\pm$0.39     & 3.27$\pm$0.02     & 118.33    & 0.06$\pm$0.01 \\
           &      &      & W1 & 3.39$\pm$0.34   & 2.75$\pm$0.28     & 5.42$\pm$0.15     & $-$63.49  & 0.74$\pm$0.11 \\
           &      &      & E1 & 2.88$\pm$0.29   & 2.24$\pm$0.22     & 5.68$\pm$0.17     & 120.69    & 0.84$\pm$0.13 \\
           & VLBA & 8.37 & C  & 53.21$\pm$4.96  & 39.95$\pm$4.00    &     ...           &   ...     & 0.67$\pm$0.07 \\
           &      &      & W3 & 10.97$\pm$1.10  & 21.98$\pm$2.20    & 1.11$\pm$0.11     & $-$61.84  & 0.53$\pm$0.08 \\
           &      &      & E2 & 8.48$\pm$0.85   & 4.93$\pm$0.49     & 1.67$\pm$0.35     & 120.12    & 1.76$\pm$0.27 \\
           &      &      & W2 & 3.79$\pm$0.38   & 7.37$\pm$0.74     & 1.91$\pm$0.28     & $-$55.18  & 1.41$\pm$0.21 \\
           &      &      & W1 & 3.96$\pm$0.40   & 1.84$\pm$0.18     & 5.32$\pm$0.25     & $-$61.87  & 1.23$\pm$0.18 \\
           &      &      & E1 & 3.55$\pm$0.36   & 1.30$\pm$0.13     & 5.62$\pm$0.31     & 122.84    & 1.54$\pm$0.23 \\
           & EVN  & 8.42 & C  & 64.92$\pm$6.49  & 46.41$\pm$4.64    &    ...            &   ...     & 0.37$\pm$0.06 \\
           &      &      & E3 & 6.30$\pm$0.63   & 18.02$\pm$1.80    & 0.99$\pm$0.15     & 114.75    & 0.73$\pm$0.11 \\
           &      &      & W3 & 4.88$\pm$0.49   & 9.68$\pm$0.97     & 1.13$\pm$0.07     & 114.75    & 0.34$\pm$0.05 \\
           &      &      & W2 & 3.87$\pm$0.39   & 2.93$\pm$0.29     & 1.94$\pm$0.17     & $-$61.53  & 0.87$\pm$0.13 \\
           &      &      & E2 & 3.66$\pm$0.37   & 1.54$\pm$0.15     & 2.51$\pm$0.21     & 114.75    & 1.03$\pm$0.15 \\
           &      &      & E1 & 1.24$\pm$0.12   & 0.51$\pm$0.05     & 5.93$\pm$0.26     & $-$62.45  & 1.31$\pm$0.20 \\
           &      &      & W1 & 1.76$\pm$0.17   & 0.61$\pm$0.06     & 6.01$\pm$0.27     & 123.57    & 1.35$\pm$0.20 \\
J0909+1928 & VLBA & 4.87 & C  &118.54$\pm$11.85 & 110.49$\pm$11.05  &   ...             &   ...     & 0.21$\pm$0.03 \\
           &      &      & N2 & 2.44$\pm$0.24   & 1.85$\pm$0.19     & 3.15$\pm$0.14     & $-$3.33   & 0.69$\pm$0.10 \\
           &      &      & S3 & 1.71$\pm$0.17   & 1.96$\pm$0.20     & 3.64$\pm$0.19     &  165.34   & 0.93$\pm$0.14 \\
           &      &      & N1 & 1.69$\pm$0.17   & 1.02$\pm$0.10     & 6.99$\pm$0.29     &  $-$7.23  & 1.45$\pm$0.22 \\
           &      &      & S1 & 8.79$\pm$0.88   & 2.19$\pm$0.22     & 7.97$\pm$0.64     &  140.21   & 3.19$\pm$0.48 \\
           & EVN  & 4.99 & C  &121.07$\pm$12.11 & 103.11$\pm$10.31  &   ...             &   ...     & 0.13$\pm$0.02 \\
           &      &      & S3 & 1.52$\pm$0.15   & 5.67$\pm$0.57     & 3.81$\pm$0.01     &  167.38   & 0.01$\pm$0.01 \\
           &      &      & N2 & 3.75$\pm$0.38   & 3.07$\pm$0.31     & 3.65$\pm$0.13     &  $-$6.06  & 0.65$\pm$0.10 \\
           &      &      & S2 & 3.51$\pm$0.35   & 1.65$\pm$0.17     & 5.36$\pm$0.44     &  155.83   & 2.22$\pm$0.33 \\
           &      &      & N1 & 0.62$\pm$0.06   & 0.46$\pm$0.05     & 7.56$\pm$0.16     &  $-$5.47  & 0.86$\pm$0.12 \\
           &      &      & S1 & 7.04$\pm$0.79   & 2.33$\pm$0.23     & 8.37$\pm$0.53     &  138.09   & 2.65$\pm$0.40 \\ 
           & VLBA & 8.37 & C  & 137.95$\pm$13.80& 126.09$\pm$1.26   & ...               & ...       & 0.18$\pm$0.03 \\
           &      &      & N3 & 11.34$\pm$1.13  & 32.37$\pm$3.24    & 1.83$\pm$0.06     &  $-$2.66  & 0.31$\pm$0.06 \\
           &      &      & S1 & 4.69$\pm$0.47   & 0.66$\pm$0.07     & 7.97$\pm$0.60     &  143.22   & 3.01$\pm$0.45 \\
           & EVN  & 8.42 & C  & 102.47$\pm$10.25& 76.57$\pm$7.66    & ...               & ...       & 0.29$\pm$0.04 \\
           &      &      & N3 & 8.17$\pm$0.82   & 8.09$\pm$0.81     & 1.86$\pm$0.06     &  $-$4.39  & 0.29$\pm$0.04 \\
           &      &      & S4 & 1.46$\pm$0.15   & 1.45$\pm$0.15     & 2.44$\pm$0.05     &  174.39   & 0.24$\pm$0.04 \\
           &      &      & N1 & 1.87$\pm$0.19   & 4.81$\pm$0.48     & 7.21$\pm$0.25     &  $-$10.82 & 1.27$\pm$0.19 \\
           &      &      & S1 & 5.48$\pm$0.55   & 0.66$\pm$0.07     & 7.85$\pm$0.71     &  142.43   & 3.57$\pm$0.54 \\
J1025+1022 & VLBA & 4.87 & C  & 59.48$\pm$5.95  & 56.93$\pm$5.69    & ...               & ...       & 0.48$\pm$0.07 \\
           &      &      & W3 & 21.09$\pm$2.11  & 27.78$\pm$2.78    & 1.33$\pm$0.11     &  $-$91.85 & 0.52$\pm$0.08 \\
           &      &      & E2 & 18.23$\pm$1.82  & 11.98$\pm$1.20    & 4.82$\pm$0.26     &  131.79   & 1.29$\pm$0.19 \\
           &      &      & W2 & 10.68$\pm$1.07  & 4.78$\pm$0.48     & 5.69$\pm$0.40     &  $-$49.93 & 2.02$\pm$0.31 \\
           &      &      & E1 & 2.55$\pm$0.26   & 1.10$\pm$0.11     & 8.35$\pm$0.46     &  128.95   & 2.30$\pm$0.36 \\
           &      &      & W1 & 1.99$\pm$0.02   & 0.87$\pm$0.09     & 10.75$\pm$0.46    &  $-$59.07 & 2.29$\pm$0.34 \\
           & EVN  & 4.99 & C  & 53.97$\pm$5.40  & 50.67$\pm$5.07    & ...               & ...       & 0.49$\pm$0.07 \\
           &      &      & W3 & 22.74$\pm$2.27  & 30.53$\pm$3.05    & 1.35$\pm$0.13     &  $-$90.53 & 0.67$\pm$0.10 \\
           &      &      & E3 & 4.64$\pm$0.46   & 9.89$\pm$0.99     & 1.83$\pm$0.27     &  76.45    & 1.36$\pm$0.20 \\
           &      &      & W2 & 12.31$\pm$1.23  & 2.44$\pm$0.24     & 5.57$\pm$0.49     &  $-$58.16 & 2.43$\pm$0.36 \\
           &      &      & E2 & 17.88$\pm$1.79  & 10.19$\pm$1.02    & 4.78$\pm$0.24     &  130.01   & 1.22$\pm$0.18 \\
           &      &      & E1 & 1.49$\pm$0.15   & 1.51$\pm$0.15     & 7.87$\pm$0.02     &   131.09  & 0.12$\pm$0.02 \\
           &      &      & W1 & 1.56$\pm$0.16   & 1.02$\pm$0.10     & 11.63$\pm$0.18    &  $-$58.16 & 0.92$\pm$0.14 \\
           & VLBA & 8.37 & C  & 61.08$\pm$7.81  & 71.89$\pm$7.19    & ...               & ...       & 0.33$\pm$0.05 \\
           &      &      & W3 & 21.85$\pm$2.19  & 19.03$\pm$1.90    & 1.19$\pm$0.10     &  $-$93.84 & 0.51$\pm$0.08 \\
           &      &      & E2 & 6.18$\pm$0.62   & 7.33$\pm$0.73     & 4.92$\pm$0.26     &  132.05   & 1.32$\pm$0.19 \\

\hline
\end{tabular}
\end{table*}

\begin{table*}
\contcaption{}
%
\medskip
\centering
\renewcommand{\tabcolsep}{3.5mm}
\begin{tabular}{ccccccccc}
\hline\hline
Name & Array & $\nu$ & Comp. & $\rm S_{int}$ &  $\rm S_{peak}$      & $\rm R $  & $\rm P.A.$    & $\rm d$     \\
     &       & (GHz) &       &      (mJy) 	 & (mJy beam$^{-1}$) &    (mas)  &   $\rm \degr$ &  (mas)      \\
\hline
           &      &      & W2 & 7.41$\pm$0.74    & 2.55$\pm$0.26    & 5.72$\pm$0.32     &  $-$49.21 & 1.62$\pm$0.24 \\
           & EVN  & 8.42 & C  & 58.99$\pm$5.90   & 31.12$\pm$3.11   & ...               & ...       & 0.36$\pm$0.05 \\
           &      &      & E3 & 3.11$\pm$0.31    & 3.24$\pm$0.32    & 0.97$\pm$0.07     &  87.65    & 0.34$\pm$0.04 \\
           &      &      & W3 & 23.78$\pm$0.48   & 9.23$\pm$0.92    & 1.27$\pm$0.05     &  $-$91.70 & 0.23$\pm$0.03 \\
           &      &      & E2 & 15.01$\pm$1.50   & 2.45$\pm$0.25    & 4.96$\pm$0.32     &  130.66   & 1.60$\pm$0.25 \\
           &      &      & W2 & 5.76$\pm$0.58    & 1.51$\pm$0.15    & 5.25$\pm$0.23     &  $-$49.87 & 1.15$\pm$0.17 \\
J1037+4335 & VLBA & 4.87 & C  & 18.66$\pm$1.87   & 11.16$\pm$1.12   & ...               & ...       & 1.57$\pm$0.24 \\
           &      &      & J1 & 3.24$\pm$0.32    & 1.58$\pm$0.16    & 8.15$\pm$0.40     & $-$28.81  & 1.98$\pm$0.29 \\
           & EVN  & 4.99 & C  & 12.13$\pm$1.21   & 11.21$\pm$1.12   & ...               & ...       & 0.27$\pm$0.04 \\
           &      &      & J3 & 5.12$\pm$0.51    & 1.75$\pm$0.18    & 3.16$\pm$0.18     & $-$20.34  & 0.92$\pm$0.14 \\
           &      &      & J1 & 5.31$\pm$0.53    & 0.62$\pm$0.06    & 8.62$\pm$0.24     & $-$24.96  & 5.52$\pm$0.84 \\
           & VLBA & 8.37 & C  & 12.79$\pm$1.38   & 13.24$\pm$1.32   & ...               & ...       & 0.22$\pm$0.03 \\
           &      &      & J3 & 3.57$\pm$0.36    & 1.96$\pm$0.20    & 3.05$\pm$0.21     & $-$18.67  & 1.05$\pm$0.16 \\
           &      &      & J2 & 2.47$\pm$0.25    & 1.58$\pm$0.16    & 6.57$\pm$0.18     & $-$30.01  & 0.89$\pm$0.13 \\
           & EVN  & 8.42 & C  & 11.09$\pm$1.11   & 10.85$\pm$1.09   & ...               & ...       & 0.11$\pm$0.02 \\
           &      &      & J3 & 3.21$\pm$0.32    & 0.94$\pm$0.09    & 3.15$\pm$0.19     & $-$24.08  & 2.03$\pm$0.31 \\
           &      &      & J2 & 2.37$\pm$0.24    & 0.37$\pm$0.04    & 6.71$\pm$0.23     & $-$24.93  & 2.31$\pm$0.35 \\
J1205+2031 & VLBA & 4.87 & C  & 31.85$\pm$3.19   & 20.74$\pm$2.07   & ...               & ...       & 0.07$\pm$0.01 \\
           &      &      & SW2& 1.65$\pm$0.17    & 10.82$\pm$1.08   & 1.55$\pm$0.12     & $-$117.04 & 0.62$\pm$0.10 \\
           &      &      & NE & 3.79$\pm$0.38    & 2.03$\pm$0.20    & 15.39$\pm$1.06    & 50.40     & 5.29$\pm$0.81 \\
           & EVN  & 4.99 & C  & 24.71$\pm$2.47   & 18.41$\pm$1.84   & ...               & ...       & 0.81$\pm$0.12 \\
           &      &      & SW1& 2.75$\pm$0.28    & 1.12$\pm$0.11    & 2.64$\pm$0.39     & $-$114.27 & 1.94$\pm$0.28 \\
           &      &      & NE & 7.21$\pm$0.72    & 0.88$\pm$0.09    & 15.55$\pm$0.80    & 47.99     & 4.01$\pm$0.61 \\
           & VLBA & 8.37 & C  & 37.13$\pm$4.21   & 32.58$\pm$3.26   & ...               & ...       & 0.78$\pm$0.12 \\
           &      &      & SW2& 1.05$\pm$0.11    & 5.94$\pm$0.59    & 1.25$\pm$0.01     & $-$148.71 & 0.03$\pm$0.01 \\
           &      &      & SW1& 0.67$\pm$0.07    & 5.41$\pm$0.54    & 2.77$\pm$0.25     & $-$123.84 & 1.26$\pm$0.19 \\
           &      &      & NE & 4.86$\pm$0.49    & 1.31$\pm$0.13    & 16.09$\pm$0.54    & 46.85     & 2.69$\pm$0.40 \\
           & EVN  & 8.42 & C  & 27.67$\pm$2.77   & 18.07$\pm$1.81   & ...               & ...       & 0.38$\pm$0.06 \\
           &      &      & SW2& 3.76$\pm$0.38    & 4.73$\pm$0.47    & 1.27$\pm$ 0.01    & $-$136.24 & 0.07$\pm$0.01 \\
           &      &      & SW1& 1.52$\pm$0.15    & 0.71$\pm$0.07    & 2.71$\pm$0.16     & $-$118.40 & 0.81$\pm$0.12 \\
           &      &      & NE & 1.23$\pm$0.12    & 0.65$\pm$0.07    & 15.56$\pm$0.14    & 43.66     & 0.69$\pm$0.11 \\
J1213+5044 & VLBA & 4.87 & C  & 25.37$\pm$2.54   & 27.42$\pm$2.74   & ...               & ...       & 1.34$\pm$0.14 \\
           &      &      & E  & 18.12$\pm$1.81   & 25.38$\pm$2.54   & 1.42$\pm$0.27     & 88.87     & 0.94$\pm$0.20 \\
           &      &      & W  & 4.06$\pm$0.41    & 15.44$\pm$1.54   & 1.92$\pm$0.24     & $-$96.31  & 1.21$\pm$0.18 \\
           & EVN  & 4.99 & C  & 23.10$\pm$2.31   & 20.48$\pm$2.05   & ...               & ...       & 0.75$\pm$0.11 \\
           &      &      & E  & 14.95$\pm$1.50   & 15.36$\pm$1.54   & 1.46$\pm$0.13     & 95.11     & 0.63$\pm$0.09 \\
           &      &      & W  & 7.53$\pm$0.75    & 4.88$\pm$0.49    & 1.89$\pm$0.37     & $-$74.64  & 1.84$\pm$0.28 \\
           & VLBA & 8.37 & C  & 27.97$\pm$2.80   & 24.91$\pm$2.49   & ...               & ...       & 0.50$\pm$0.07 \\
           &      &      & W  & 23.24$\pm$2.12   & 15.50$\pm$1.55   & 1.67$\pm$0.17     & $-$89.30  & 0.85$\pm$0.13 \\
           &      &      & E  & 4.75$\pm$0.48    & 5.32$\pm$0.53    & 1.89$\pm$0.19     & 90.01     & 0.97$\pm$0.15 \\
           & EVN  & 8.42 & C  & 18.75$\pm$2.98   & 14.56$\pm$1.46   & ...               & ...       & 0.65$\pm$0.10 \\
           &      &      & W  & 8.76$\pm$1.28    & 6.53$\pm$0.65    & 1.85$\pm$0.09     & $-$81.39  & 0.47$\pm$0.07 \\
           &      &      & E  & 5.67$\pm$0.30    & 0.85$\pm$0.09    & 1.93$\pm$0.30     & 96.88     & 1.49$\pm$0.22 \\
J1230+4700 & VLBA & 4.87 & C  & 17.45$\pm$1.74   & 26.18$\pm$2.62   & ...               & ...       & 0.04$\pm$0.02 \\
           &      &      & S2 & 13.09$\pm$1.31   & 6.86$\pm$0.69    & 1.29$\pm$0.02     & 169.13    & 0.05$\pm$0.02 \\
           &      &      & N  & 4.35$\pm$0.44    & 9.34$\pm$0.93    & 2.38$\pm$0.22     & $-$11.44  & 1.10$\pm$0.16 \\
           &      &      & S1 & 3.22$\pm$0.32    & 0.83$\pm$0.08    & 3.12$\pm$0.19     & 161.33    & 0.95$\pm$0.14 \\
           &      &      & S0 & 1.21$\pm$0.12    & 0.45$\pm$0.05    & 10.69$\pm$0.28    & 164.20    & 1.41$\pm$0.21 \\
           & EVN  & 4.99 & C  & 14.08$\pm$1.41   & 24.49$\pm$2.45   & ...               & ...       & 0.15$\pm$0.02 \\
           &      &      & S2 & 13.32$\pm$1.33   & 9.15$\pm$0.92    & 1.21$\pm$0.04     & 173.86    & 0.18$\pm$0.03 \\
           &      &      & N  & 4.65$\pm$0.47    & 11.24$\pm$1.12   & 2.12$\pm$0.14     & $-$13.08  & 0.70$\pm$0.10 \\
           &      &      & S1 & 3.24$\pm$0.32    & 0.67$\pm$0.07    & 3.31$\pm$0.21     & 160.98    & 1.04$\pm$0.16 \\
           &      &      & S0 & 3.30$\pm$0.33    & 0.54$\pm$0.05    & 10.71$\pm$0.71    & 163.65    & 3.55$\pm$0.53 \\
           & VLBA & 8.37 & C  & 24.03$\pm$2.40   & 28.16$\pm$2.82   & ...               & ...       & 0.07$\pm$0.01 \\
           &      &      & S2 & 11.06$\pm$1.11   & 18.01$\pm$1.80   & 1.48$\pm$0.11     & 169.63    & 0.55$\pm$0.08 \\
           &      &      & N  & 5.66$\pm$0.57    & 8.71$\pm$0.87    & 1.98$\pm$0.17     & $-$12.21  & 0.87$\pm$0.13 \\
           &      &      & S1 & 1.86$\pm$0.19    & 1.64$\pm$0.16    & 3.51$\pm$0.20     & 160.91    & 0.98$\pm$0.15 \\
           & EVN  & 8.42 & C  & 13.39$\pm$0.94   & 13.03$\pm$1.30   & ...               & ...       & 0.18$\pm$0.02 \\
           &      &      & S2 & 11.30$\pm$0.73    & 10.62$\pm$1.06   & 1.31$\pm$0.01     & 172.93   & 0.02$\pm$0.01 \\
           &      &      & N  & 3.36$\pm$0.33    & 4.43$\pm$0.44    & 2.03$\pm$0.10     & $-$9.58   & 0.49$\pm$0.07 \\
           &      &      & S1 & 2.16$\pm$0.22    & 1.66$\pm$0.17    & 3.41$\pm$0.20     & 163.97    & 0.99$\pm$0.15 \\
\hline\end{tabular}
\end{table*}

\begin{table*}
\contcaption{}
%
\medskip
\centering
\renewcommand{\tabcolsep}{3.5mm}
\begin{tabular}{ccccccccc}\hline\hline
Name & Array & $\nu$ & Comp. & $\rm S_{int}$ &  $\rm S_{peak}$      & $\rm R $  & $\rm P.A.$    & $\rm d$    \\
     &       & (GHz) &       &      (mJy) 	 & (mJy beam$^{-1}$) &    (mas)  &   $\rm \degr$ &  (mas)     \\
\hline
J1246+1153 & VLBA & 4.87 & C  & 6.73$\pm$0.67    & 5.71$\pm$0.57    & ...               & ...       & 0.94$\pm$0.14 \\
           &      &      & J1 & 1.30$\pm$0.13    & 1.32$\pm$0.13    & 3.06$\pm$0.14     & $-$61.88  & 0.67$\pm$0.11 \\
           & EVN  & 4.99 & C  & 4.93$\pm$0.41    & 3.88$\pm$0.39    & ...               & ...       & 0.41$\pm$0.06 \\
           &      &      & J1 & 1.42$\pm$0.19    & 0.56$\pm$0.06    & 2.78$\pm$0.25     & $-$60.02  & 1.27$\pm$0.19 \\
           &      &      & J0 & 0.48$\pm$0.05    & 0.65$\pm$0.07    & 5.52$\pm$0.01     & $-$65.69  & 0.01$\pm$0.01 \\
           & VLBA & 8.37 & C  & 8.38$\pm$0.84    & 8.05$\pm$0.81    & ...               & ...       & 0.23$\pm$0.02 \\
           & EVN  & 8.42 & C  & 6.69$\pm$0.67    & 6.10$\pm$0.61    & ...               & ...       & 0.28$\pm$0.03 \\
           &      &      & J1 & 0.47$\pm$0.05    & 0.16$\pm$0.02    & 2.93$\pm$0.21     & $-$81.44  & 1.05$\pm$0.16 \\
\hline
\end{tabular}
\begin{flushleft}
Notes: column description: (1) sources name; (2) VLBI array; (3) observing frequency; (4) component label; (5) integrated flux density of the fitted Gaussian model component; (6) peak intensity; (7) radial distance from the core; (8) position angle respect to the core, measured from north through the east; (9) component size
\end{flushleft}
\end{table*}

\begin{table*}
\centering
    \caption{Proper Motions of 8 FR 0s}
	\begin{tabular}{ccccc}
	\hline\hline
Source & Comp. & $\rm \mu$  & $\rm \beta_{app}$ & Freq.    \\
       &       & (mas yr$^{-1}$) &       (c)         &  (GHz)   \\
\hline
J0906+4124 & W2 & 0.178$\pm$0.029 & 0.32$\pm$0.05  & 8     \\
           & E2 & 0.208$\pm$0.029 & 0.37$\pm$0.05  & 8     \\
           & W1 & 0.198$\pm$0.030 & 0.35$\pm$0.05  & 8     \\
           & E1 & 0.182$\pm$0.046 & 0.32$\pm$0.08  & 8     \\
J0909+1928 & S2 & 0.104$\pm$0.037 & 0.18$\pm$0.07  & 5     \\
           & N1 & 0.219$\pm$0.017 & 0.39$\pm$0.03  & 5     \\
     & S1 & $-$0.005$\pm$0.054 & $-$0.08$\pm$0.09  & 5     \\
J1025+1022 & W3 & 0.106$\pm$0.024 & 0.31$\pm$0.07  & 8     \\
           & W2 & 0.173$\pm$0.041 & 0.51$\pm$0.12  & 8     \\
           & E2 & 0.045$\pm$0.051 & 0.13$\pm$0.15  & 8     \\
J1037+4335 & J3 & 0.178$\pm$0.465 & 0.30$\pm$0.78  & 8     \\ 
           & J2 & 0.249$\pm$0.520 & 0.42$\pm$0.87  & 8     \\
           & J1 & 0.783$\pm$0.991 & 1.32$\pm$1.67  & 5     \\
J1205+2031 & SW2& 0.047$\pm$0.009 & 0.07$\pm$0.01  & 8     \\
           & SW1& 0.028$\pm$0.021 & 0.04$\pm$0.03  & 8     \\
           & NE & 0.130$\pm$0.114 & 0.21$\pm$0.18  & 8     \\
J1213+5044 & W  & 0.118$\pm$0.033 & 0.24$\pm$0.07  & 8     \\
           & E  & 0.044$\pm$0.041 & 0.18$\pm$0.04  & 8     \\
J1230+4700 & S2 & 0.073$\pm$0.014 & 0.22$\pm$0.06  & 8     \\
           & N  & 0.088$\pm$0.022 & 0.16$\pm$0.04  & 8     \\
           & S1 & 0.064$\pm$0.027 & 0.16$\pm$0.07  & 8     \\
J1246+1153 & J1 & 0.128$\pm$0.101 & 0.39$\pm$0.31  & 5,8   \\
           
\hline
	\end{tabular}
\label{proper motion}
\begin{flushleft}
Notes. Column description: (1) source name; (2) component label; (3) proper motion; (4) apparent speed of the component; (5) the frequency data we used.
\end{flushleft}
\end{table*}

\bsp	
\label{lastpage}

\end{document}